\documentclass[pra,superscriptaddress,twocolumn,showpacs,nofootinbib]{revtex4}
\usepackage{graphicx}
\usepackage{amssymb}
\usepackage{tikz} 
\usepackage{graphicx,epsf,natbib}
\usepackage{hyperref} 
\usepackage{bm} 
\usepackage{amsbsy,amssymb,amsmath,bm}
\usepackage{amsfonts}

\def\be{\begin{equation}}
\def\ee{\end{equation}}
\def\bea{\begin{eqnarray}}
\def\eea{\end{eqnarray}}
\def\bdm{\begin{displaymath}}
\def\edm{\end{displaymath}}
\def\ba{\begin{array}}
\def\ea{\end{array}}
\def \bde {\begin{description}}
\def \ede {\end{description}}
\def \nn {\nonumber}
\def\bes{\begin{subequations}}
\def\ees{\end{subequations}}

\def\bq{{\bf q}}
\def\ba{\bar{a}}
\def\bj{\bar{\bf j}}

\def \a {\alpha}

\def \m {\mu}

\def \lam {\lambda}

\def \dag {\dagger}
\def\e{\rm e}

\newcommand{\abs}[1]{{\left|{#1}\right|}} 
\newcommand{\ket}[1]{\vert{#1}\rangle} 
\newcommand{\bra}[1]{\langle{#1}\vert} 

\newcommand{\eq}[1]{(\ref{#1})}

\begin{document}

\title{Refined Characterization of Lattice Chern Insulators by Bulk Entanglement Spectrum}

\author{Dah-Wei Chiou}
\email{dwchiou@gmail.com}
\affiliation{Department of Physics, National Taiwan Normal University, Taipei 11677, Taiwan}
\affiliation{Center for Condensed Matter Sciences, National Taiwan University, Taipei 10617, Taiwan}
\author{Hsien-Chung Kao}
\thanks{Correspondence}
\email{hckao12@gmail.com}
\affiliation{Department of Physics, National Taiwan Normal University, Taipei 11677, Taiwan}
\author{Feng-Li Lin}
\thanks{Correspondence}
\email{fengli.lin@gmail.com}
\affiliation{Department of Physics, National Taiwan Normal University, Taipei 11677, Taiwan}


\begin{abstract}
We have studied extensively the band crossing patterns of the bulk entanglement spectrum (BES) for various lattice Chern insulators. We find that only partitions with dual symmetry can have either stable nodal-lines or nodal-points in the BES when the system is in the topological phase of a nonzero Chern number. By deforming the Hamiltonian to lift the accidental symmetry, one can see that only nodal points are robust. They thus should bear certain topological characteristics of the BES. By studying the band crossing patterns in details we conclude that the topological characteristics of the BES are inherited from the topological order of the underlying Chern insulators and the former can have more refined topological structures. We then propose the conjecture that the sum of the vorticities in the BES in a properly chosen reduced Brillouin zone equals the Chern number of the underlying Chern insulator. This relation is beyond the usual classification scheme of topological insulators/superconductors.
\end{abstract}

\pacs{73.43.Cd,03.67.Mn,73.43.Nq}

\maketitle

\tableofcontents

\section{Introduction}

   Entanglement is an intrinsic quantum characteristic with no classical analogue. Thus, quantum states without good mean-field (classical) description, such as the (symmetry-protected)  topological ordered states, should have nontrivial patterns of quantum entanglement \cite{spt}. Given a reduced density matrix, one may defined the corresponding entanglement spectrum (ES), which is a natural and more refined measure to characterize the patterns of quantum entanglement compared to the entanglement entropy.  A nice demonstration for the above idea was provided in \cite{Li} showing the degeneracy pattern of entanglement spectrum is related to that of the edge-mode spectrum for some fractional quantum Hall states. It is worthwhile to mention that in \cite{Ryu1} the ES for the Chern insulators was first calculated to evaluate the entanglement entropy even before the term ES was coined \cite{Li}.  Similar considerations and demonstrations have been extended to the topological insulators/superconductors \cite{Pollman,Spain1,Turner,Fidkowski,Arovas,Qi-Ludwig}. For a nice review on the recent progress of the subject please see \cite{PYChang}.

   All the ES's considered in the aforementioned works are limited to the case that the system is partitioned into two parts with a single domain-wall like boundary. Recently, an extensive partition, as illustrated in Fig.\ \ref{fig-checkerboard-1} and Fig.\ \ref{fig-checkerboard-2}, has been proposed \cite{MIT1}.  (See also \cite{BES-1,BES-2} for related considerations.). The corresponding ES is dubbed the bulk entanglement spectrum (BES). One of the advantages of using an extensive partition is that certain subset of the translational symmetry of the original lattice may be preserved.  As a result, the corresponding BES can be found easily by invoking the Bloch theorem.

    In \cite{MIT1} the authors found that band crossings in BES occur for a free fermion lattice model when the partition is in the checkerboard form. These band crossings are robust and signify the topological characteristics of BES.  This is reminiscent of the zero energy edge states which appear on the boundary of topological insulators/superconductors. They are believed to reflect the topological orders of the underlying ground states.  The authors further argued that the band crossing pattern of BES can be manipulated by tuning the partition geometry or the entanglement temperature, and it can be used to probe the topological phase transition. This scenario has been demonstrated to be generic by the subsequent works on the AKLT spin chain \cite{Wan2,MIT2,Santos-1,Santos-2}, integer quantum Hall states \cite{Wan1,BES-QHE}, random partitions \cite{MIT3,Wan3} and topological insulators \cite{Fukui,Fukui1}.

     In this work we will extend the previous studies of BES to explore the topological connection between BES and the quantum state of the underlying system. For this purpose, we will study the BES of the Chern insulators based on \cite{Haldane} and its extensions \cite{Kitaev-per,AIP-1,SPT-cls,Hasan2010,Qi2011,Furusaki}, whose ground states are characterized by the Chern number associated with the Berry connection \cite{TKNN}. As the Chern insulators are free fermion lattice systems, we can follow the prescription given in \cite{Peschel,Cheong,Chung1} to evaluate their BES via the restricted correlation function. Moreover, these Chen insulators can have different topological phases which can be accessed by varying some dynamical parameters rather than the geometry or entanglement temperature as suggested in \cite{MIT1}. One can then use the band crossing patterns of these BES to probe the topological phase transitions, and find out a definite relation between BES criticality and quantum criticality, i.e., a relation between the total vorticity of the nodal points in the BES and the Chern number of the underlying quantum state.

    Our work is an extension of previous studies on BES, and our results clarify some important issues, which are summarized as follows.
\begin{enumerate}

   \item When the lattice is partitioned into sub-lattices $L_A$ and $L_B$ for evaluating its BES, the band crossings in the BES occur when there is a dual symmetry between $L_A$ and $L_B$ (also referred to as ``A-B symmetry''). Whether there is percolation of the of sites in $L_A$ or $L_B$ is irrelevant. This could be related to the emerging ``chiral symmetries" of the entanglement Hamiltonian when the partition is A-B symmetric \cite{PYChang,Turner},   see section \ref{sec IID} for more detailed discussions.

  \item There could be band crossings due to some accidental symmetries of the underlying Hamiltonian, which can be lifted by deforming the Hamiltonian to break these accidental symmetries. More specifically, we find that some nodal-line crossings are of this type.

  \item There are nodal point band crossings which remain robust under deformation.  The robustness of these nodal points suggests that the BES may have inherited certain characteristic of the topological order from the underlying Chern insulator.

  After an extensive study of the vortex structure of all the robust nodal points in the BES, we arrive at the following conjecture up to a sign ambiguity:
      \be\label{master rel}
      \mathtt{C} = \pm \sum_{\mathbf{q}_i\in\widetilde{\mathbf{BZ}}_\mathbf{q}}v_{\mathbf{q}_i},
      \ee
      where $\widetilde{\mathbf{BZ}}_\mathbf{q}$ is a properly-chosen reduced Brillouin zone (from the full Brillouin zone denoted by ${\bf BZ_k}$) and $v_{\mathbf{q}_i}$ is the \emph{vorticity} of the $i$-th nodal point $\mathbf{q}_i$.   Note that the vorticity of the nodal point is defined through the Berry connection of the entanglement Hamiltonian's eigenstates.  Under arbitrary deformation, this identity \emph{always} and \emph{continuously} admits existence of $\widetilde{\mathbf{BZ}}_\mathbf{q}$.\footnote{See Sec.\ \ref{Section-remarks on subtleties} for the subtlety on choosing $\widetilde{\mathbf{BZ}}_\mathbf{q}$.} The unavoidable sign ambiguity is a reflection of the emerging chiral symmetries of the entanglement Hamiltonian for the A-B symmetric partition of the lattice.

      As will be discussed in section II D, this relation is beyond the usual classification schemes of topological insulators/superconductors.

\end{enumerate}

The remaining of the paper is organized as follows. In the next section, we will spell out the framework used to evaluate BES for free fermion lattice systems. We will consider both checkerboard and stripe partitions, and discuss how to deform the Hamiltonian to lift the accidental symmetries leading to the nodal lines in the BES.  We also discuss the emerging chiral symmetries, their implication to the robustness of band-crossing of BES and beyond. In section \ref{BES for CI} we will first delineate the four Chern insulators and their phase diagrams for which we will evaluate the corresponding BES. Then, we plot the relevant band crossing patterns of the BES, from which we collect the evidences which support the conjectured relation \eq{master rel}. Finally, we will conclude with some discussions in \ref{conclusion}. In Appendix \ref{Appendix-dispersion} we show the details of how the dispersion relation and the vorticity of the nodal point in the BES are evaluated.   In Appendix \ref{Appendix-nodal lines} we investigate the physical significance of accidental and nontrivial nodal lines.

\section{Formalism of evaluating BES}\label{formulation}

\subsection{Eigenvalue equation}
The Hamiltonian of a two-band lattice fermion system in two-dimensions (2D) takes the following form
\be\label{Hd}
 H={\bf d(k)} \cdot \boldsymbol{\sigma}.
\ee
In our notation, a bold-faced letter denotes a vector, e.g., ${\bf d}=(d_x,d_y,d_z)$. Here $\bf d(k)$ is a function of momentum ${\bf k}=(k_x,k_y)$ and $\boldsymbol{ \sigma}=(\sigma_x,\sigma_y,\sigma_z)$ are the Pauli matrices. There are two non-degenerate eigenstates for each $\bf k$. When the system is in the ground state, the lower band is completely filled and the upper band empty, i.e.
\be
|\Phi_0\rangle:=\prod_{\bf k \in BZ_{\bf k}} \prod_{s=\uparrow,\downarrow} \chi_-^s({\bf k}) c_{{\bf k},s}^{- \dagger} |0\rangle
\ee
where $c_{{\bf k},s}^{- \dagger}$ is the creation operator of a particle carrying momentum $\bf k$ and spin $s$ which is in the lower band (labeled by the superscript $-$) with $\chi_-^s({\bf k})$ the corresponding wavefunction. $\bf k$ is restricted to the first Brillouin zone of the original lattice, which is denoted as  ${\bf BZ_k}:=(-\pi, \pi]\otimes (-\pi, \pi]$.

The corresponding correlation function in the $\bf k$ space is given by
\be\label{Ck1}
({\bf C}_{\bf k})_{s,s'}:=\sum_{\a,\a'} \langle \Phi_0|  c_{{\bf k},s}^{\a \dagger}  c^{\a'}_{{\bf k},s'} |\Phi_0\rangle,
\ee
with $s,s'$ and $\a, \a'$ the spin and band indices.  Since the upper band is empty, only the states in the lower band contribute so that
\be\label{Ck2}
({\bf C}_{\bf k})_{s,s'}=\chi^{s*}_-({\bf k})\; \chi_-^{s'}({\bf k}),
\ee
where the lower band eigenfunctions are
\be
\chi^{\uparrow}_-={d_z- |{\bf d}| \over \sqrt{2|{\bf d}|(|{\bf d}|-d_z})}\;,\;\quad \chi^{\downarrow}_-={d_x+id_y \over \sqrt{2|{\bf d}|(|{\bf d}|-d_z})}\;.
\ee
For simplicity, we will always set the lattice spacing to be unity.   Moreover, it is easy to see that for a given $\bf k$ the eigenvalues  of ${\bf C}_{\bf k}$ are either zero or one as expected.

Due to the translational symmetry of the total lattice, the spatial correlation function can be obtained through the following Fourier transform:
\be\label{Cfourier}
(\tilde{\bf C}_{{\bf j},{\bf j'}})_{s,s'}={1\over N_T} \sum_{\bf k \in BZ_k} e^{-i {\bf k}\cdot ({\bf j}-{\bf j'})} ({\bf C}_{\bf k})_{s,s'}\;,
\ee
where $N_T$ is the total number of lattice sites.

Let's divide the whole lattice into two sub-lattices  $L_A$ and $L_B$ so that we may define the entanglement spectrum.  To retain some degree of translational symmetry, we consider the case that the sites in $L_A$ may be described by a superlattice with a basis, $\bar{L}_A=\{\bar{\bf j}_{\bar{a}}\}, \bar{a}= 1, \dots, \bar{N}_A$.  The  corresponding primitive vectors are given by $\vec{\ell}_x=\ell_x \hat{\bf x},\ \vec{\ell}_y=\ell_y \hat{\bf y}$ and we will refer $\bar{\bf j}_{\bar{a}}$ as the orbital indices which specify the locations of all the A sites in a superlattice primitive cell. See Fig.\ \ref{fig-checkerboard-1} and Fig.\ \ref{fig-checkerboard-2} for examples.  From the definition of the reciprocal vector $\bf G$ of the superlattice, we have
\be\label{superG}
{\bf G}\cdot {\bf j}={\bf G}\cdot \bar{\bf j} + 2\pi n, \qquad n \in \mathbb{Z}\;,
\ee
where ${\bf G}=2\pi\left( \frac{m_x}{\ell_x}\hat{\bf x}+ \frac{m_y}{\ell_y}\hat{\bf y} \right)$, with $m_x =0,1, \dots, \ell_x-1$ and $m_y =0,1, \dots, \ell_y-1$.

The reduce density matrix $\rho_A$ can be found by tracing out the degree of freedoms coming from $L_B$, which is related to the entanglement Hamiltonian by
\be
\rho_A={\cal K} \e^{{\cal H}_e},
\ee
with ${\cal K}$ a normalization constant, and
\be
{\cal H}_e = \sum_{{\bf j}, {\bf j}'} (H_e)_{{\bf j j}'}c^\dag_{\bf j} c_{{\bf j}'}.
\ee
In the above equation, we consider ${\cal H}_e, c^\dag_{\bf j}$ and $c_{{\bf j}'}$ matrices in the spinor space and thus the spin indices are suppressed to simplify the expression.
It has been shown that in a free fermion system, the entanglement Hamiltonian can be expressed in terms of the restricted correlation function \cite{Peschel,Cheong,Chung1}
\be\label{HeC}
H_e^T= \ln {1-\tilde{\bf C}_{{\bf j},{\bf j'}}|_{L_A} \over\tilde{\bf C}_{{\bf j},{\bf j'}}|_{L_A}}.
\ee
Here, $\tilde{\bf C}_{{\bf j},{\bf j'}}\big|_{L_A}$ is the correlation matrix in the real space with ${\bf j},{\bf j'}$ restricted to the sites in the sub-lattice $L_A$.  In light of this relation, finding the entanglement spectrum  is equivalent to solving the following eigenvalue equation
\be\label{eigenC}
\sum_{{\bf j' \in L_A}} (\tilde{\bf C}_{{\bf j}, {\bf j'}}) \Psi ({\bf j'})=\lambda \Psi ({\bf j})\;,   \qquad  {\bf j} \in L_A \;.
\ee

 Because of the residual translational symmetry, eigenfunction $\Psi({\bf j})$ takes the following form upon using Bloch's theorem:
\be\label{BlochC}
\Psi({\bf j})=e^{-i {\bf q}\cdot {\bf j}}\; \psi_{\bf q}(\bar{\bf j}).
\ee
 Plugging (\ref{Cfourier}) and (\ref{BlochC}) into (\ref{eigenC}) and using (\ref{superG}), we arrive at the following simplified form of eigenvalue equation:
\be
 \sum_{{\bf G}}\sum_{\bar{a}'=1}^{\bar{N}_A} \frac{\e^{-i{\bf G}\cdot(\bar{\bf j}_{\bar{a}} - \bar{\bf j}_{\bar{a}'} )} }{\ell_x \ell_y} \left({\bf C}_{{\bf q} + {\bf G}} \right) \psi_{\bf q}(\bj_{\ba'}) =\lam_{\bf q} \psi_{\bf q}(\bj_{\ba}),
\ee
where $\bar{a}= 1, \dots, \bar{N}_A$.  By convention, we here let $\bf q$ belong to the principal reduced Brillouin zone of the superlattice denoted by ${\bf BZ_q}:=(-{\pi\over \ell_x},{\pi\over \ell_x}]\otimes (-{\pi\over \ell_y},{\pi\over \ell_y}]$.  To find the eigenvalues $\lam_{\bq},$ all we have to do now is to diagonalize the following $(2\bar{N}_A) \times(2\bar{N}_A)$ matrix:
\be\label{masterM}
[{\bf M(q)}]_{\ba,s; \ba',s'} =\sum_{{\bf G}} \frac{\e^{-i{\bf G}\cdot(\bar{\bf j}_{\bar{a}} - \bar{\bf j}_{\bar{a}'} )} }{\ell_x \ell_y} \left({\bf C}_{{\bf q} + {\bf G}} \right)_{s, s'}.
\ee
Note that in deriving the above result we have used the relation:
\be
\sum_{{\bf J}} e^{i{\bf (k-q)\cdot J}}={1\over \ell_x \ell_y}\sum_{\bf G} \delta_{\bf k, q+G}
\ee
where ${\bf J}$ is a lattice vector of the superlattice such that ${\bf J\cdot G} \in 2\pi \mathtt{Z}$.

\subsection{Various Partitions}

   Let's first consider a few simple partition patterns by choosing  $(\ell_x, \ell_y)=(2, 2)$,  and find out the corresponding BES.  For these partitions, the Bloch wave vector ${\bf q} \in (-{\pi \over 2}, {\pi\over 2}]\otimes (-{\pi \over 2}, {\pi\over 2}]$.

\begin{figure}
\begin{tikzpicture}
\begin{scope}[scale=0.5]

 \draw[help lines] (-1.3,-1.3) grid (4.3,4.3);
 \draw [thin,->] (-1.3,0) -- (4.5,0);
 \node at (4.7,0) {$x$};
 \draw [thin,->] (0,-1.3) -- (0,4.5);
 \node at (-0.3,4.5) {$y$};

 \foreach \i in {0,1,2}{
  \foreach \j in {0,1,2}{
   \begin{scope}[shift={(2*\i,2*\j)}]
   \draw [fill] (0,0) circle [radius=0.1];
   \end{scope}
  }
 }
\draw [blue] (-0.5,-0.5) rectangle (1.5,1.5);
\end{scope}
\end{tikzpicture}
\caption{Partition type {\bf a}: A superlattice primitive cell is specified by the enclosed region, and the corresponding primitive vectors are $\vec{\ell}_x=2 \hat{\bf x},\ \vec{\ell}_y=2 \hat{\bf y}$. The principal reduced Brillouin zone is ${\bf BZ_q}=(-{\pi\over 2},{\pi\over 2}]\otimes (-{\pi\over 2},{\pi\over 2}]$. The dotted sites are those belong to sub-lattice $L_A$, and the undotted ones to $L_B$. Note that there is no dual symmetry between $L_A$ and $L_B$.}
\label{fig-checkerboard-1}
\end{figure}
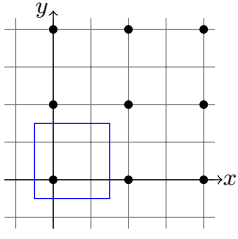

\begin{figure}
 \begin{tikzpicture}
\begin{scope}[scale=0.5]

 \draw[help lines] (-1.3,-1.3) grid (4.3,4.3);
 \draw [thin,->] (-1.3,0) -- (4.5,0);
 \node at (4.7,0) {$x$};
 \draw [thin,->] (0,-1.3) -- (0,4.5);
 \node at (-0.3,4.5) {$y$};

 \foreach \i in {0,1}{
  \foreach \j in {0,1}{
   \begin{scope}[shift={(2*\i,2*\j)}]
   \draw [fill] (0,0) circle [radius=0.1];
   \draw [fill] (-1,-1) circle [radius=0.1];
   \draw [fill] (0,2) circle [radius=0.1];
   \draw [fill] (2,0) circle [radius=0.1];
     \draw [fill] (1,1) circle [radius=0.1];
   \draw [fill] (2,2) circle [radius=0.1];
   \draw [fill] (1,-1) circle [radius=0.1];
   \draw [fill] (-1,1) circle [radius=0.1];
    \end{scope}
  }
 }

\draw [blue] (-0.5,-0.5) rectangle (1.5,1.5);
\end{scope}
\end{tikzpicture}
\caption{Partition type {\bf b}: In the checkerboard partition, the superlattice primitive cell is the same as the one in Fig.\ \ref{fig-checkerboard-1}, however, one more site $(1,1)$ of the superlattice primitive cell is included in $L_A$ so that there is dual symmetry between $L_A$ and $L_B$.}
\label{fig-checkerboard-2}
\end{figure}
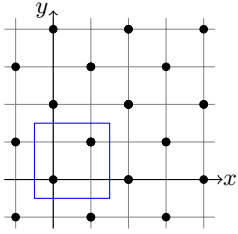

    The explicit form of the matrix ${\bf M(q)}$ depends on what sites in the superlattice primitive cell are included in $\bar{L}_A$.    Here, we consider five types of partitions labelled by ${\bf a}$, ${\bf b}$, ${\bf c}$, ${\bf d}$ and ${\bf e}$. However, later on we will focus especially on partition ${\bf b}$ and ${\bf c}$ for which the BES can be solved analytically and there are band crossings because of the A-B symmetry. Let us now discuss them one by one:

\begin{itemize}

\item{Partition type \bf{a}}\hfil\break
   Let's begin with the case that there is only one site in $\bar{L}_A$, which can be chosen to be $\bj_1=(0,0)$ without loss of generality (Fig.\ \ref{fig-checkerboard-1}).  It is then straightforward to see that
\be\label{M1site}
{\bf M(q)}={1\over 4} ({\bf C}_{0,0}+{\bf C}_{0,1}+{\bf C}_{1,0}+{\bf C}_{1,1})\;.
\ee
Here, ${\bf C}_{m_x, m_y}:={\bf C}_{{\bf q}+{\bf G}}$ with ${\bf G} = \pi(m_x \hat{\bf x} + m_y \hat{\bf y})$.
In this case, $\bf M(q)$ is only a 2 by 2 matrix, as there is only one orbital degree of freedom.

\item{Partition type \bf{b}}\hfil\break
   When there are two sites in $\bar{L}_A$, there are two inequivalent possibilities. Let's first consider the so-called checkerboard partition, in which we choose $\bj_1=(0,0),\ \bj_2=(1,1)$ (Fig.\ \ref{fig-checkerboard-2}). We thus have a 4 by 4 matrix $\bf M(q)$ taking the form as follows:
\be\label{M2sites}
{\bf M(q)}=\left(
\begin{matrix}
{\bf M}_{1,1} & {\bf M}_{1,2} \cr
{\bf M}_{2,1} &{\bf M}_{2,2} \cr
\end{matrix}
\right).
\ee
Here, ${\bf M}_{1,1}={\bf M}_{2,2}={1\over 4} ({\bf C}_{0,0}+{\bf C}_{0,1}+{\bf C}_{1,0}+{\bf C}_{1,1}), $ and ${\bf M}_{1,2}={\bf M}_{2,1}={1\over 4} ({\bf C}_{0,0}-{\bf C}_{0,1}-{\bf C}_{1,0}+{\bf C}_{1,1}).$
Introducing $P_{\pm}={1\over 2} (1\pm \tau_x)$ with $\vec \tau$ the Pauli matrices for the orbital degrees of freedom, we may rewrite ${\bf M(q)}$  as
\bea
{\bf M(q)}&=&{1\over 2}({\bf C}_{0,0}+{\bf C}_{1,1})\otimes P_+  \nonumber \\
&& \mbox{} + {1\over 2} ({\bf C}_{0,1}+{\bf C}_{1,0}) \otimes P_-.
\eea
Since $P_{\pm}$ are commuting projection operators, they can be diagonalized simultaneously. Moreover, as $P_+$ and $P_-$ are orthogonal to each other, the eigenvalues of $\bf M(q)$ are given by those of  ${1\over 2}({\bf C}_{0,0}+{\bf C}_{1,1})$ and ${1\over 2}({\bf C}_{0,1}+{\bf C}_{1,0})$, giving rise to 4 bands.  In contrast to the first case, there is a dual symmetry between $L_A$ and $L_B$.  As we shall see later, this turns out to be the key property which determines whether band crossing patterns would appear in the BES when the system is in the topological phase.

Moreover, we can solve the eigenvectors and  eigenvalues  for $\bf M(q)$ explicitly  in this case.  The eigenvectors are
\begin{subequations}\label{eigenvec b}
\begin{eqnarray}
&&\quad\psi_{1,2}^s(\mathbf{q})\\
&&\propto \left\{ \sqrt{\alpha_1(\mathbf{q})}\,\chi_-^s(\mathbf{q})  \pm \sqrt{\alpha_1(\mathbf{q})^*}\,\chi_-^s(\mathbf{q}_1) \right\} \otimes |+\rangle,
\nonumber\\
&&\quad\psi_{3,4}^s(\mathbf{q})\\
&&\propto \left\{ \sqrt{\alpha_2(\mathbf{q})}\,\chi_-^s(\mathbf{q}_3)
  \pm \sqrt{\alpha_2(\mathbf{q})^*}\,\chi_-^s(\mathbf{q}_2) \right\} \otimes |-\rangle.
  \nonumber
\end{eqnarray}
\end{subequations}
Here, $P_{\pm}|\pm\rangle=|\pm \rangle$ and
\begin{eqnarray}\label{bfqi}
&&{\bf q}_1:={\bf q}+(\pi,\pi), \quad {\bf q}_2:={\bf q}+(0,\pi),\nonumber\\
&&{\bf q}_3:={\bf q} + (\pi,0),
\end{eqnarray}
and
\begin{subequations}\label{alpha1 alpha2}
\bea\label{alpha1}
\alpha_1(\mathbf{q})&:=&\sum_{s=\uparrow,\downarrow} \chi_-^s(\mathbf{q})^* \chi_-^s(\mathbf{q}_1)\;,\\
\alpha_2(\mathbf{q})&:=&\sum_{s=\uparrow,\downarrow} \chi_-^s(\mathbf{q}_3)^* \chi_-^s(\mathbf{q}_2)\;. \label{alpha2}
\eea
\end{subequations}
The corresponding eigenvalues are
\begin{subequations}\label{eigenval b}
\begin{eqnarray}
\lambda_{1,2}(\mathbf{q}) &=& \frac{1}{2}\left\{1\pm |\alpha_1(\mathbf{q})|\right\}\\
\lambda_{3,4}(\mathbf{q}) &=& \frac{1}{2}\left\{1\pm |\alpha_2(\mathbf{q})|\right\}\;.
\end{eqnarray}
\end{subequations}

These informations will be useful when we analyze the band crossing patterns.

\item{Partition type \bf{c}}\hfil\break
   Another interesting partition is the stripe partition in which we choose $\bj_1=(0,0),\ \bj_2=(0,1)$.   Here, there is also a dual symmetry between $L_A$ and $L_B$. This partition is in fact equivalent to the case that $(\ell_x, \ell_y)=(2, 1)$ with one site in $\bar{L}_A$ specified by $\bj_1=(0,0)$. Since the translational symmetry along y-direction is intact, the Bloch wave vector $\bq \in (-{\pi \over 2}, {\pi \over 2}]\otimes (-\pi, \pi ]$ (see Fig.\ \ref{fig-stripe 1}) . The corresponding 2-band matrix ${\bf M(q)}$ is given by
\be\label{M-c}
{\bf M(q)}={1\over 2} ({\bf C}_0+ \bf{C}_1)\;.
\ee
Here, ${\bf C}_{m_x}:={\bf C}_{\bq+{\bf G}}$ with ${\bf G} = m_x \pi  \hat{\bf x}$ and $m_x=0,1$.

 Similarly, in this case we can explicitly solve the eigenvectors and eigenvalues for $\bf M(q)$. The eigenvectors are
\begin{equation}\label{eigenvec c}
\psi_{1,2}^s(\mathbf{q}) \propto \sqrt{\alpha_3(\mathbf{q})}\,\chi_-^s(\mathbf{q})
\pm \sqrt{\alpha_3(\mathbf{q})^*}\,\chi_-^s(\mathbf{q}_3)
\end{equation}
with the corresponding eigenvalues
\begin{equation}\label{eigenval c}
\lambda_{1,2}(\mathbf{q}) = \frac{1}{2}\left(1\pm |\alpha_3(\mathbf{q})| \right)\;.
\end{equation}
Note that ${\bf q}_3$ is defined in \eq{bfqi} and
\be\label{alpha3}
\alpha_3({\bf q}):=\sum_{s=\uparrow,\downarrow} \chi_-^s(\mathbf{q})^* \chi_-^s(\mathbf{q}_3)\;.
\ee

\begin{figure}
\begin{tikzpicture}

\begin{scope}[shift={(0,0)},scale=0.5]

 \draw[help lines] (-1.3,-1.3) grid (6.3,5.3);
 \draw [->] (-1.3,0) -- (6.5,0);
 \node at (6.9,0) {$x$};
 \draw [->] (0,-1.3) -- (0,5.8);
 \node at (-0.3,6.2) {$y$};

 \foreach \i in {0,1,2,3}{
  \foreach \j in {-1,0,1,2,3,4,5}{
   \begin{scope}[shift={(2*\i,\j)}]
   \draw [fill] (0,0) circle [radius=0.1];
   \end{scope}
  }
 }

 \draw [blue] (-0.5,-0.5) rectangle (1.5,0.5);

\end{scope}

\end{tikzpicture}
\caption{Partition type {\bf c}: A stripe partition with the superlattice primitive vector $\vec{\ell}_x=2 \hat{\bf x}$ so that ${\bf BZ_q}:=(-{\pi \over 2}, {\pi \over 2}]\otimes(-\pi, \pi]$. Note that there is a dual symmetry between $L_A$ and $L_B$.}\label{fig-stripe 1}
\end{figure}
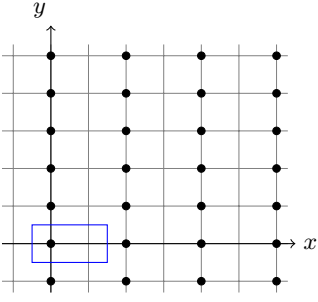

\item{Partition type \bf{d}}\hfil\break
   To make comparison with partition {type \bf c}, let's introduce another stripe partition in which  $(\ell_x, \ell_y)=(4, 1)$. There is only one site in $\bar{L}_A$ which is specified by $\bj_1=(0,0)$, and the partition is not A-B symmetric.  The Bloch wave vector $\bq \in (-{\pi \over 4}, {\pi \over 4}]\otimes (-\pi, \pi ]$ (see Fig.\ \ref{fig-partition d}). The corresponding 2-band matrix ${\bf M(q)}$ is given by
\be\label{M-d}
{\bf M(q)}={1\over 4} ({\bf D}_0+ \bf{D}_1+{\bf D}_2+ \bf{D}_3)\;.
\ee
Here, ${\bf D}_{m_x}:={\bf C}_{\bq+{\bf G}}$ with ${\bf G} = (m_x \pi/2) \hat{\bf x}$ and $m_x=0,1,2,3$.   The BES can not be simplified analytically.  Using numerical calculation, we see that there is no robust band crossing for this partitions as there is no dual symmetry between A and B sublattices.

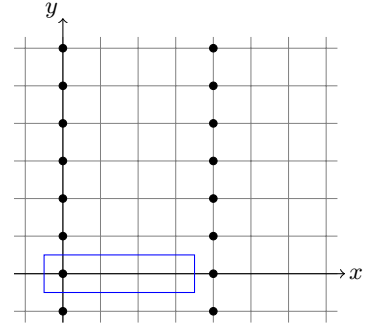
\begin{figure}
\begin{tikzpicture}

\begin{scope}[shift={(0,0)},scale=0.5]

\draw[help lines] (-1.3,-1.3) grid (7.3,6.3);
 \draw [->] (-1.3,0) -- (7.5,0);
 \node at (7.8,0) {$x$};
 \draw [->] (0,-1.3) -- (0,6.8);
 \node at (-0.3,7) {$y$};

 \foreach \i in {0,4}{
  \foreach \j in {-1,0,1,2,3,4,5,6}{
   \begin{scope}[shift={(\i,\j)}]
   \draw [fill] (0,0) circle [radius=0.1];
   \end{scope}
  }
 }

 \draw [blue] (-0.5,-0.5) rectangle (3.5,0.5);

\end{scope}

\end{tikzpicture}
\caption{Partition type {\bf d}: A stripe partition with the superlattice primitive vector $\vec{\ell}_x=4 \hat{\bf x}$ so that ${\bf BZ_q}:=(-{\pi \over 4}, {\pi \over 4}]\otimes(-\pi, \pi]$. Note that there is no dual symmetry between $L_A$ and $L_B$.}\label{fig-partition d}
\end{figure}

\item{Partition type \bf{e}} \hfil\break
     To further confirm that band crossing occurs if and only if the partition is A-B symmetric and the system is in the topological phase, let's introduce the double stripe partition in which  $(\ell_x, \ell_y)=(4, 1)$. There are two sites in $\bar{L}_A$, which are specified by $\bj_1=(0,0)$ and $\bj_2=(1,0)$, so the partition is A-B symmetric.  The Bloch wave vector $\bq \in (-{\pi \over 4}, {\pi \over 4}]\otimes (-\pi, \pi ]$ (see Fig.\ \ref{fig-partition e}). The corresponding $4\times4$ matrix ${\bf M(q)}$ takes the form
\be\label{M-e}
{\bf M(q)}=\left(
\begin{matrix}
{\bf M}_{1,1} & {\bf M}_{1,2} \cr
{\bf M}_{2,1} &{\bf M}_{2,2} \cr
\end{matrix}
\right),
\ee
where ${\bf M}_{1,1}={\bf M}_{2,2}={1\over 4} ({\bf D}_0+{\bf D}_1+{\bf D}_2+{\bf D}_3)$, ${\bf M}_{1,2}={\bf M}_{2,1}^\dag={1\over 4} ({\bf D}_0+i{\bf D}_1-{\bf D}_2-i{\bf D}_3)$.
Again, the BES can not be simplified analytically. Nevertheless, numerical calculation confirms that the BES has band crossing as long as the tight-binding Hamiltonian exhibits a nonzero Chern number.

\begin{figure}
\begin{tikzpicture}

\begin{scope}[shift={(0,0)},scale=0.5]

\draw[help lines] (-1.3,-1.3) grid (7.3,6.3);
 \draw [->] (-1.3,0) -- (7.5,0);
 \node at (7.8,0) {$x$};
 \draw [->] (0,-1.3) -- (0,6.8);
 \node at (-0.3,7) {$y$};

 \foreach \i in {0,4}{
  \foreach \j in {-1,0,1,2,3,4,5,6}{
   \begin{scope}[shift={(\i,\j)}]
   \draw [fill] (0,0) circle [radius=0.1];
   \draw [fill] (1,0) circle [radius=0.1];
   \end{scope}
  }
 }

 \draw [blue] (-0.5,-0.5) rectangle (3.5,0.5);

\end{scope}

\end{tikzpicture}
\caption{Partition type {\bf e}: A double stripe partition with the superlattice primitive vector $\vec{\ell}_x=4 \hat{\bf x}$ so that ${\bf BZ_q}:=(-{\pi \over 4}, {\pi \over 4}]\otimes(-\pi, \pi]$. Note that there is a dual symmetry between $L_A$ and $L_B$.}\label{fig-partition e}
\end{figure}
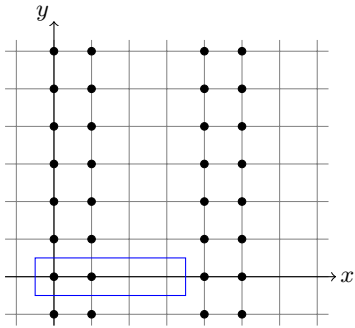

\end{itemize}

 Although each individual ${\bf C}_{m_x, m_y}$,  ${\bf C}_{m_x}$ and  ${\bf D}_{m_x}$ have either zero or unity eigenvalues for all $\bf q$, it is generally not the case for their linear combinations which appear in (\ref{M1site}),  (\ref{M2sites}), (\ref{M-c}), (\ref{M-d}) and (\ref{M-e}).  We may also evaluate the BES for $L_B$ for partition {\bf a} shown in Fig.\ \ref{fig-checkerboard-1}. The most evident distinction is that the sites in $L_B$ percolate the whole lattice.  The corresponding $\bf M(q)$ is 6 by 6, and numerical method must be used to find the BES.  The results show that there is no robust band crossing for this kind of partition. This tells us that percolation is not a determining factor for the existence of band crossing in the BES.

\subsection{Deformations}

\begin{figure}

\centering
  \begin{minipage}[b]{0.23\textwidth}
    \includegraphics[width=\textwidth]{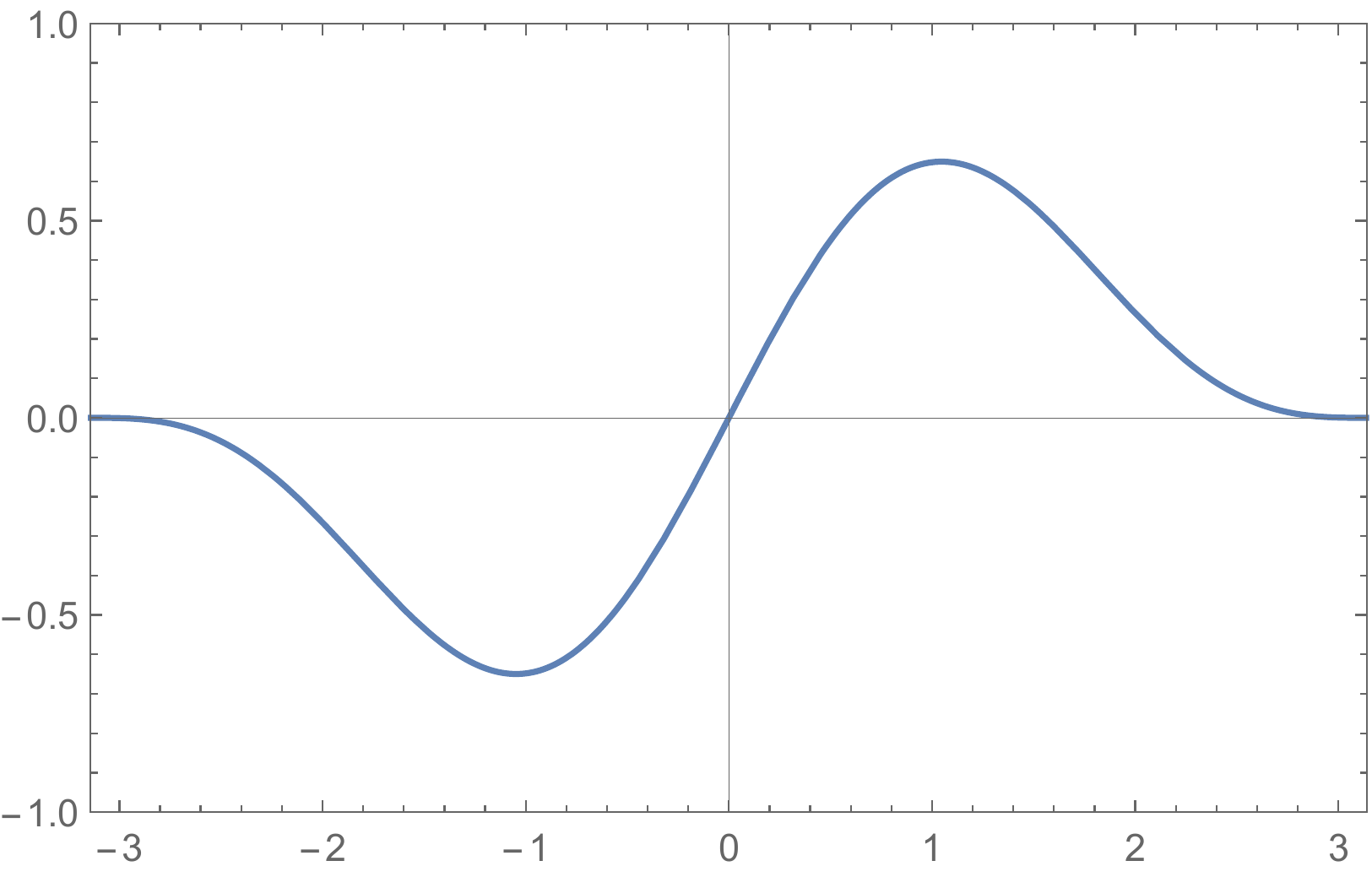}
  \end{minipage}
  \hspace{0.1cm} 
  \begin{minipage}[b]{0.23\textwidth}
    \includegraphics[width=\textwidth]{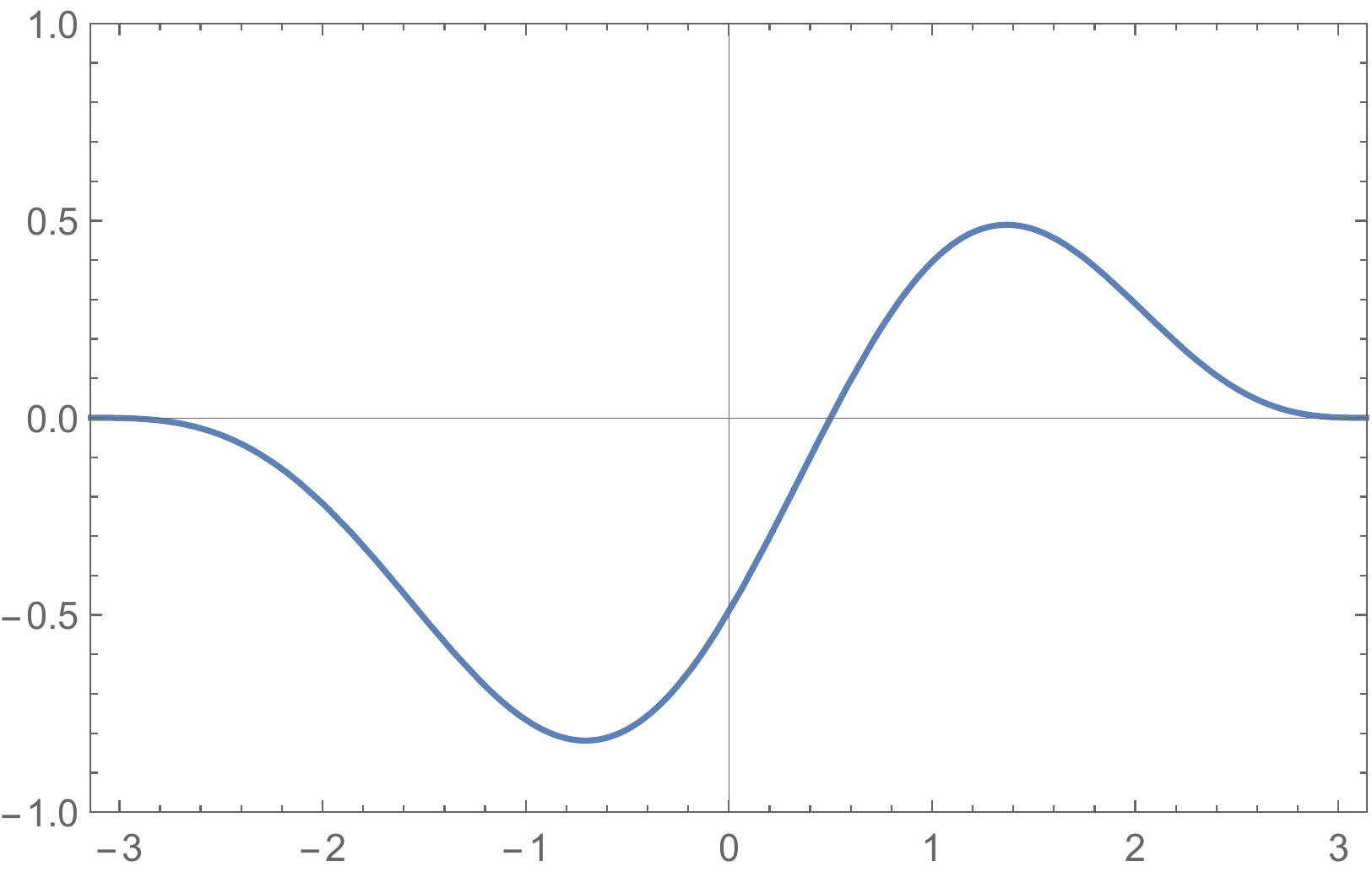}
  \end{minipage}

\caption{The function $g(x;a,b)$ in $(-\pi,\pi]$. Left: $a=1$, $b=0$. Right: $a=1$, $b=0.5$.}
\label{fig-deforming function}
\end{figure}

  In our study we find that there are accidental band crossings which can not be attributed to symmetries that protect the topological phases. Thus, we believe these accidental band crossings are not robust under perturbation, and we show that they do get lifted when we deform the Hamiltonian. Instead of introducing an additional perturbation, we deform the original Hamiltonian by relabeling the momentum using the following map
\be\label{deformk}
k \longrightarrow k'=f(k):=k+ g(k;a,b)
\ee
with
\be
g(x;a,b):=a \sin\left[ {b \over \pi^2 -b^2} (k^2 - \pi^2) +k\right] \cos^2 \left[ {k\over 2}\right].
\ee
Here, $k$ can be either $k_x$ or $k_y$, and $a, b$ are free parameters which are arbitrarily chosen as long as they are small enough to ensure the $k\rightarrow k'$ map is injective. In particular, this map preserves the boundary conditions at the zone boundary as required for a physical band structure, i.e., $f(k=\pm \pi)=\pm \pi$ and ${d f(k)\over d k}|_{k=\pm \pi}=1$.  Examples of $g(x;a,b)$
are shown in Fig.\ \ref{fig-deforming function}.\footnote{The deformation via a homeomorphism $\mathbf{k}\rightarrow\mathbf{k}'$ is highly restricted. Generically, the topological deformation in a tight-binding model is described by a homotopy (i.e., a homotopic deformation) between the two maps, $\mathbf{d}:\mathbf{k}\mapsto\mathbf{d}(\mathbf{k})$ and $\mathbf{d}':\mathbf{k}'\mapsto\mathbf{d}'(\mathbf{k}')$. However, for our purpose, we focus only on the restricted kind.}

\subsection{Chiral symmetry, its implications and beyond}\label{sec IID}
  In the above five types of partitions, we see that type $\bf b$, $\bf c$ and $\bf e$ are A-B symmetric. We now briefly sketch the emerging chiral symmetry from the A-B symmetric partition and discuss its relevance to our results. We do not aim for a rigorous proof but provide a heuristic argument to support the findings of our case studies.

  Under the A-B partition, the correlation function matrix may be written as
\be
{\bf C} =
\left(\begin{matrix}
{\bf C}_A & {\bf C}_{AB} \cr
{\bf C}_{AB}^{\dagger} &{\bf C}_B \cr
\end{matrix}\right)
\ee
where the subscripts denote the region on which the corresponding matrix acts.  Since the eigenvalue of $\bf C$ is either zero or one, we have ${\bf C}^2={\bf C}$.  The condition can then be turned into the following form:
\bea
&\; \hskip -2.0cm &  {\bf C}_A^2+{\bf C}_{AB}{\bf C}^{\dagger}_{AB}={\bf C}_A\;, \nn \\
&\; \hskip -2.0cm &  {\bf C}_B^2+{\bf C}^{\dagger}_{AB}{\bf C}_{AB}={\bf C}_B\;, \nn \\
&\; \hskip -2.0cm & {\bf C}_A{\bf C}_{AB}+{\bf C}_{AB}{\bf C}_B={\bf C}_{AB}\;.\nn
\eea

 If the partition is A-B symmetric, then we have ${\bf C}_A={\bf C}_B$, and the above conditions leads to the following commuting relation:
\be
 \left\{ {\bf C}_A - \frac{1}{2}{\bf I}, {\bf C}_{AB} \right\} =0\;.
\ee
The above algebra implies that for the ``entanglement Hamiltonian" ${\bf C}_A - \frac{1}{2}{\bf I}$ there is emerging  chiral symmetry generated by ${\bf C}_{AB}$.  From the cases studied in this paper we find that the nodal points of BES are robust in the A-B symmetric partitions but not in the A-B nonsymmetric ones. We believe that the emerging chiral symmetry in the A-B symmetric cases should be responsible for the observed robustness, however, a more rigorous argument for this connection similar to the one for the usual symmetry-protected topological insulators is needed.

   Moreover, the emerging chiral symmetry also leads to the fact that the ``entanglement Hamiltonian"  ${\bf C}_A - \frac{1}{2}{\bf I}$ should have trivial topological properties. That is, if we associate some vorticity with each robust nodal point of BES, then the overall vorticity should be zero.  From the classification of topological insulators/superconductors \cite{Kitaev-per,AIP-1,SPT-cls}, the 2D Chern insulators considered here belong to class $\textrm{A}$, which do not have time-reversal, particle-hole or chiral symmetry, and are characterized by integer Chern number. Therefore, the reduced correlation function ${\bf C}_A$ inherits no symmetry from the underlying Chern insulator but possesses emerging chiral symmetries. According to \cite{Kitaev-per,AIP-1,SPT-cls}, the ``entanglement Hamiltonian"  ${\bf C}_A - \frac{1}{2}{\bf I}$  is classified as $\textrm{AIII}$ for which there is no nontrivial topology in 2D. This is consistent with the facts that the overall vorticity is zero.

    Now comes a puzzle: if the entanglement Hamiltonian has trivial topology, how can it match the Chern insulators of integer class via the relation \eq{master rel}? This is because the trivial topology of the entanglement Hamiltonian is defined with respect to the full Brillouin zone ${\bf BZ_k}$, i.e., the overall vorticity inside ${\bf BZ_k}$ should be zero. However, our \eq{master rel} is relating the Chern number to the overall vorticity inside the reduced Brillouin zone ${\bf BZ_q}$ or the appropriately chosen one  $\widetilde{\mathbf{BZ}}_\mathbf{q}$. In this sense, our relation \eq{master rel} is nontrivial since it goes beyond the usual classification scheme of topological insulators/superconductors.
    Empirically, its validity is supported by all of our case studies as the reduced Brillouin zone $\widetilde{\mathbf{BZ}}_\mathbf{q}$ can always and continuously be chosen in responde to continuous change of tuning parameters and deformation (see the discussions related to Fig. \ref{fig-non-principal BZ}).

   Besides, one may expect that it is always possible to choose some $\widetilde{\mathbf{BZ}}_\mathbf{q}$ so that the sign ambiguity in \eq{master rel} can be lifted. However, our case studies show that this is not true. There are some cases for which no \emph{continuous} adaption of the reduced Brillouin zone can undo the sign change of the overall vorticity.  This leads to a new kind of topological phase transition of BES at which the sign of the overall vorticity flips and one or more nontrivial nodal lines appear. This also reveals the refined topological structure of BES which is manifest only in the BES but not in the spectrum of the underlying Chern insulator. For more details, please see the related discussion about Fig.\ \ref{fig-model I (refined diagram)}. This refined structure and the sign ambiguity of \eq{master rel} reinforces the fact that the \eq{master rel} is nontrivial and can not be inferred simply from  the standard classification scheme of topological insulators/superconductors with the emerging chiral symmetries.

\section{BES of Chern Insulators} \label{BES for CI}

  The main purpose of this work is to explore the connection between the topological nature of the underlying lattice fermion states and the band-crossing patterns of the corresponding BES. Therefore, we will consider the Chern insulators which admit topological phases characterized by the Chern number \cite{TKNN}:
\be\label{chern n}
\mathtt{C}={1\over 4 \pi} \int_{\bf BZ_k} d^2k \;  {\bf \hat{d}} \cdot ( \partial_{k_x} {\bf \hat{d}} \times \partial_{k_y}{\bf \hat{d}})
\ee
where ${\bf \hat{d}}:={\bf d(k) \over |d(k)|}$ with ${\bf d(k)}$ defined in \eq{Hd}.

\subsection{Topological states considered}

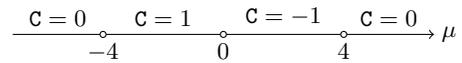
\begin{figure}
\begin{tikzpicture}

\begin{scope}[shift={(0,0)},scale=0.4]

 \draw [->] (-7,0) -- (7,0);
 \node at (7.5,0) {$\mu$};

 \draw [fill=white] (0,0) circle [radius=0.1];
 \node [below] at (0,0) {$0$};

 \draw [fill=white] (4,0) circle [radius=0.1];
 \node [below] at (4,0) {$4$};

 \draw [fill=white] (-4,0) circle [radius=0.1];
 \node [below] at (-4,0) {$-4$};

 \node [above] at (2,0) {$\mathtt{C}=-1$};
 \node [above] at (-2,0) {$\mathtt{C}=1$};
 \node [above] at (5.5,0) {$\mathtt{C}=0$};
 \node [above] at (-5.5,0) {$\mathtt{C}=0$};

\end{scope}

\end{tikzpicture}
\caption{The phase diagram of $\bf CI_1$ for various $\mu$.  Each phase is characterized by a specific Chern number \eq{model-1}.}\label{fig-model I}
\end{figure}

$\bf CI_1$: The first Chern insulator (denoted by $\bf CI_1$) we consider is given by the following $\bf d(k)$ \cite{TI-model}:
\bea
d_x &=& d^{(1)}_x :=   \sin k_x, \nn \\
d_y &=& d^{(1)}_y :=  \sin k_y, \nn \\
d_z & =& -\mu - 2 \cos k_x-  2 \cos k_y \;. \label{model-1}
\eea
This model has four special points in ${\bf BZ_k}$ at $(k_x,k_y)=(0,0),(0,\pi),(\pi,0)$ and $(\pi,\pi),$ which are also the time-reversal invariant (TRI) momenta.  In the topological phase,  the sign of  $d_z$ at $(0,0)$ (or $(\pi, \pi)$) is different from the other three points.  As a result, we need to use two different gauges around these points to calculate the Berry connection $\mathbf{A}:=-i \langle \chi^- | \boldsymbol{\nabla}_{\bf k} | \chi^- \rangle$.  Making use of the Stoke theorem, we may calculate the Chern number $\mathtt{C}$ by doing the integration of the Berry connection along a contour that separates the two gauge patches.  The result is equivalent to \eq{chern n}.  The phase diagram of this model by varying $\m$ is shown in Fig.\ \ref{fig-model I}, in which the phases are characterized by the Chern number.  For $|\m|>4$, the phase is topologically trivial with $\mathtt{C}=0$. On the other hand, there are topologically nontrivial phases characterized by $\mathtt{C}=1$ for $-4<\m<0$ and $\mathtt{C}=-1$ for $0<\m<4$. Thus, there are three topological phase transition points at $\m=-4,0,4$.

\bigskip

\begin{figure}
\begin{tikzpicture}

\begin{scope}[shift={(0,0)},scale=0.4]

 \draw [->] (-7,0) -- (7,0);
 \node at (7.5,0) {$\mu$};

 \draw [fill=white] (4,0) circle [radius=0.1];
 \node [below] at (4,0) {$4$};

 \draw [fill=white] (-4,0) circle [radius=0.1];
 \node [below] at (-4,0) {$-4$};

 \node [above] at (0,0) {$\mathtt{C}=2$};
 \node [above] at (5.5,0) {$\mathtt{C}=0$};
 \node [above] at (-5.5,0) {$\mathtt{C}=0$};

\end{scope}

\end{tikzpicture}
\caption{The phase diagram of $\bf CI_2$ for various $\mu$.  Each phase is characterized by a specific Chern number \eq{model-2}.}\label{fig-model II}
\end{figure}
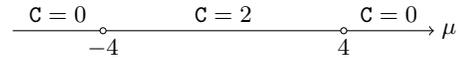

   $\bf CI_2$: Next we consider the second type of Chern insulator (denoted by $\bf CI_2$) with the following $\bf d(k)$:
\bea
d_x &=& d^{(2)}_x :=\cos k_y-\cos k_x, \nn \\
d_y &=& d^{(2)}_y := \sin k_x \sin k_y, \nn \\
d_z & =& -\mu-2\cos k_x-2\cos k_y\;. \label{model-2}
\eea
This model has two special points in $\bf BZ_k$ at $(k_x,k_y)=(0,0)$ and $(\pi,\pi)$.  As shown in Fig.\ \ref{fig-model II}, various phases show up when we vary $\mu$ in the model. There are two phases in the phase diagram.  For $|\m|<4,$ $\mathtt{C}=2$ in contrast to the previous model.  For $|\m|>4$, the system is in the trivial phase. Again, the relative sign of $d_z$ at the two special points in $\bf BZ_k$ can be used to distinguish the topological and trivial phases.

\bigskip
\bigskip

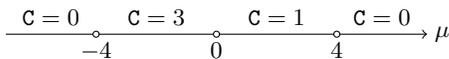
\begin{figure}
\begin{tikzpicture}

\begin{scope}[shift={(0,0)},scale=0.4]

 \draw [->] (-7,0) -- (7,0);
 \node at (7.5,0) {$\mu$};

 \draw [fill=white] (0,0) circle [radius=0.1];
 \node [below] at (0,0) {$0$};

 \draw [fill=white] (4,0) circle [radius=0.1];
 \node [below] at (4,0) {$4$};

 \draw [fill=white] (-4,0) circle [radius=0.1];
 \node [below] at (-4,0) {$-4$};

 \node [above] at (2,0) {$\mathtt{C}=1$};
 \node [above] at (-2,0) {$\mathtt{C}=3$};
 \node [above] at (5.5,0) {$\mathtt{C}=0$};
 \node [above] at (-5.5,0) {$\mathtt{C}=0$};

\end{scope}

\end{tikzpicture}
\caption{The phase diagram of $\bf CI_3$ for various $\mu$.  Each phase is characterized by a specific Chern number \eq{model-3}.}\label{fig-model III}
\end{figure}

 $\bf CI_3$: It is desirable to check  Chern insulators with $\mathtt{C}=3$ to ensure the generality of our proposed relation \eq{master rel}. Since the Chern number is dictated by the Berry phase  $\phi({\bf k}):=-\arg(d_x+i d_y)$ near the special points $(0,0),(0,\pi),(\pi,0)$ and $(\pi,\pi),$ we can construct a model with $\mathtt{C}=3$ by multiplying $d^{(1)}_x+i d^{(1)}_y$ and $d^{(2)}_x+i d^{(2)}_y$. The third Chern insulator to be consider in this paper (denoted by $\bf CI_3$) is explicitly described by the following ${\bf d(k)}$:
\bea
d_x &=& \sin k_x\left(\cos k_y-\cos k_x\right) -\sin k_x \sin^2k_y, \nn \\
d_y &=& \sin k_y\left(\cos k_y-\cos k_x\right) +\sin k_y \sin^2k_x,\nn \\
d_z &=& -\mu - 2 \cos k_x - 2 \cos k_y. \label{model-3}
\eea
Its phase diagram characterized by $\mathtt{C}$ is shown in Fig.\ \ref{fig-model III}, from which one can read out the details as before. Note that there is a phase with $\mathtt{C}=3$ for $-4<\mu<0$.
\\

\begin{figure}
\begin{tikzpicture}
\begin{scope}[shift={(0,0)},scale=0.4]

 \path [fill=orange] (-2,-1) -- (2,0) -- (-2,1);
 \path [fill=pink] (-2,1) -- (2,0) -- (5,3/4) -- (5,5) -- (-2,5);
 \path [fill=pink] (-2,-1) -- (2,0) -- (5,-3/4) -- (5,-5) -- (-2,-5);
 \path [fill=lime] (-2,1) -- (-2,5) -- (-5,5) -- (-5,7/4);
 \path [fill=lime] (-2,-1) -- (-2,-5) -- (-5,-5) -- (-5,-7/4);

 \draw[help lines] (-5,-5) grid (5,5);

 \draw [thick,blue] (-2,-5) -- (-2,5);
 \draw [thick,blue,domain=-5:5] plot (\x, {-0.5+0.25*\x});
 \draw [thick,blue,domain=-5:5] plot (\x, {0.5-0.25*\x});

 \node [below] at (-4,-5) {$-4$};
 \node [below] at (-2,-5) {$-2$};
 \node [below] at (0,-5) {$0$};
 \node [below] at (2,-5) {$2$};
 \node [below] at (4,-5) {$4$};
 \node [below right] at (5,-5) {$\mu$};

 \node [left] at (-5,-4) {$-4$};
 \node [left] at (-5,-2) {$-2$};
 \node [left] at (-5,0) {$0$};
 \node [left] at (-5,2) {$2$};
 \node [left] at (-5,4) {$4$};
 \node [above left] at (-5,5) {$t$};

 \node [] at (-0.7,0) {$\mathtt{C}=-2$};
 \node [] at (4,0) {$\mathtt{C}=0$};
 \node [] at (1.5,2.5) {$\mathtt{C}=-1$};
 \node [] at (1.5,-2.5) {$\mathtt{C}=-1$};
 \node [] at (-3.5,3) {$\mathtt{C}=1$};
 \node [] at (-3.5,-3) {$\mathtt{C}=1$};
 \node [] at (-3.5,0) {$\mathtt{C}=0$};

\end{scope}

\end{tikzpicture}
\caption{The phase diagram of $\bf CI_4$ in the $\mu$-$t$ plane. The phase are characterized by the Chern number\eq{model-4}.}\label{fig-model 4}
\end{figure}
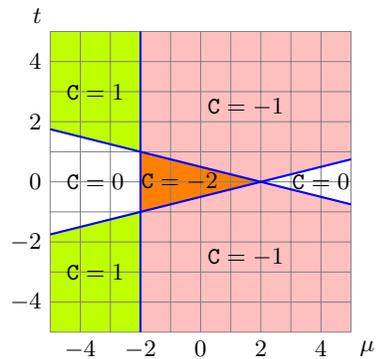

   $\bf CI_4$: Finally, for generality we also consider the BES of a more complicated model of the Chern insulator (denoted by $\bf CI_4$) with the following ${\bf d(k)}$ \cite{Simon-4}:
\begin{eqnarray}
d_x &=& \sin k_x,\nn \\
d_y &=&  \sin k_y, \nn \\
d_z &=& -\mu- 2t\left(\cos k_x+\cos k_y\right)+ 2\cos(k_x+k_y). \label{model-4}
\end{eqnarray}
There are two physical parameters in this system which can be tuned to alter the phase of the system. This model also has a phase with $|\mathtt{C}|=2$ as in $\bf CI_2$. However, the Chern number here receives contribution from two of the TRI momenta rather than one. We will see that they have different band crossing pattern in the BES. The full phase diagram is shown in Fig.\ \ref{fig-model 4}. Compared to the previous cases, this phase diagram is 2-dimensional and more complicated. Therefore, it serves as a stringent check on the the correspondence between the existence of band crossing in BES and the phase of the original system.  In Fig.\ \ref{fig-model 4}, one can see that there are three phase transition ``lines'' $\m=-2$, $\m+4t=2$ and $\m-4t=2$ which divide the phase diagram into seven regions with one of them having $\mathtt{C}=2$, and three pairs having $\mathtt{C}=0$, $1$ and $-1$, respectively.

\subsection{Band crossing patterns of BES}

We now discuss the band crossing patterns in the BES for the four Chern insulators under different partitions.  For convenience, we will use the notation like $\bf CI_{1a}$ to refer to the Chern insulator model $\bf CI_1$ under partition type $\bf a$. As discussed, the pattern of band-crossing is robust if and only if the the partition is A-B symmetric so that in the following discussions we will mainly focus on the cases with partition type $\bf b$ and $\bf c$.

\subsubsection{$\bf CI_{1c}$}

  This is the simplest case.  Because of the relation between the entanglement Hamiltonian and the restricted correlation function in eq.\ \eq{HeC}, band crossing in the BES occurs when
\be
{\rm det}[{\bf M(q)}-{\bf I}/2] = 0.
\ee
The condition is an algebraic equation in terms of $\cos q_x$ and $\cos q_y$, which has solution only for  ${\bf q}=(0,0), (0,\pi)$.
For ${\bf q}=(0,0)$
\be
{\rm det}[{\bf M} (0,0)-{\bf I}/2] = \frac{-\m(\m+4)-|\m(\m+4)| }{8|\m(\m+4)|}.
\ee
The partition is A-B symmetric and indeed the above determinant vanishes for $-4<\m<0$ and thus band crossing in the BES occurs at ${\bf q}=(0,0)$ only in the $\mathtt{C}=1$ phase. On the other hand, for ${\bf q}=(0,\pi)$ one has
\be
{\rm det}[{\bf M}(0,\pi)-{\bf I}/2] = \frac{-\m(\m-4)-|\m(\m-4)| }{8|\m(\m-4)|}.
\ee
The determinant vanishes for $0<\m<4,$ and therefore band crossing in the BES  occurs at ${\bf q}=(0,\pi)$ only in the $\mathtt{C}=-1$ phase.

\begin{figure}
\centering
  \begin{minipage}[b]{0.2\textwidth}
    \includegraphics[width=\textwidth]{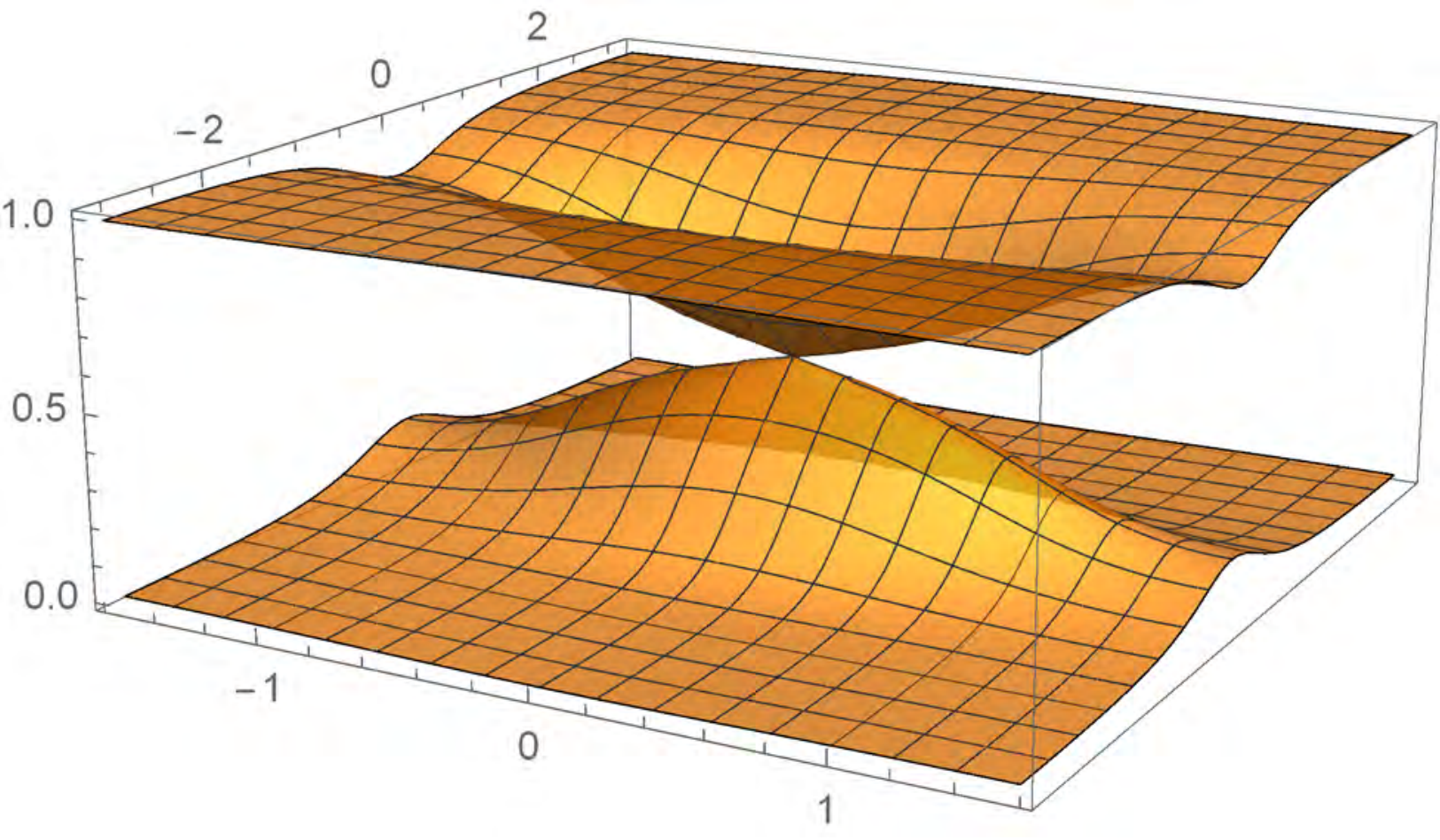}
  \end{minipage}
  \hspace{0.7cm} 
  \begin{minipage}[b]{0.2\textwidth}
    \includegraphics[width=\textwidth]{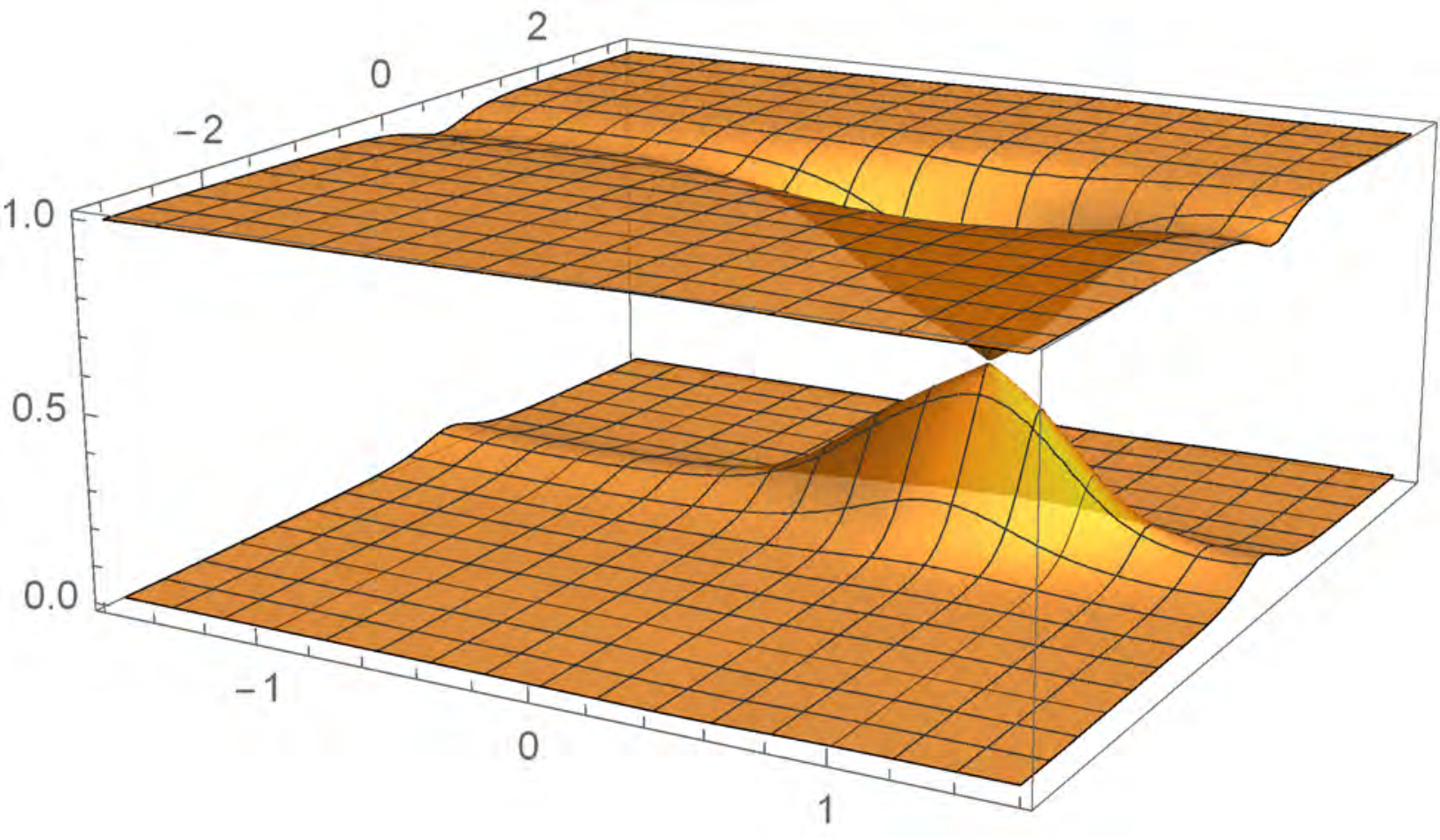}
  \end{minipage}
\caption{BES for $\bf CI_{1c}$ for $\mu=-3$ ($\mathtt{C}=1$). Left:  without deformation. Right: with deformation $a_x=b_x=a_y=b_y=1$.}\label{fig-CI1c-I}
\end{figure}

\begin{figure}
\centering
  \begin{minipage}[b]{0.2\textwidth}
    \includegraphics[width=\textwidth]{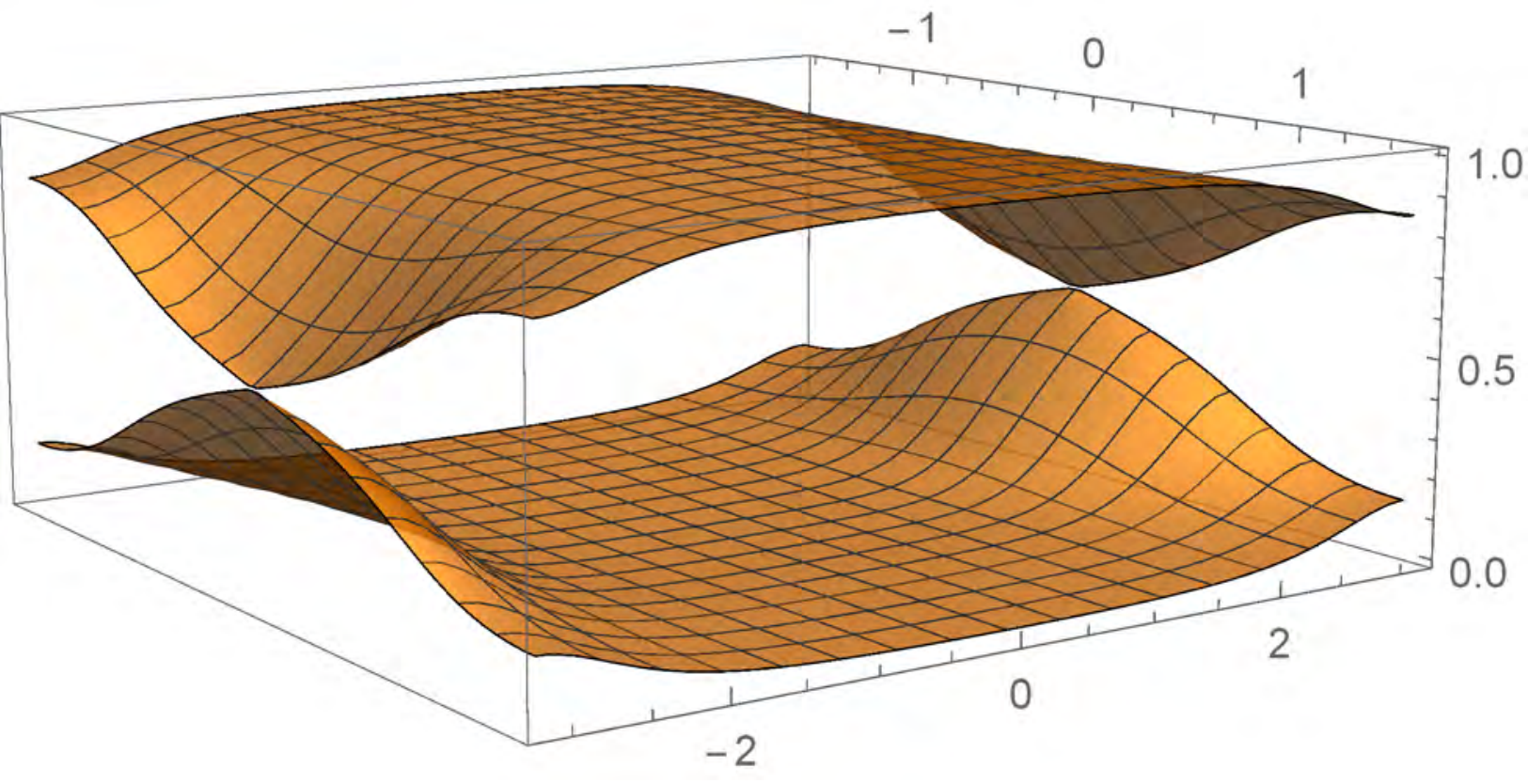}
  \end{minipage}
  \hspace{0.7cm} 
  \begin{minipage}[b]{0.2\textwidth}
    \includegraphics[width=\textwidth]{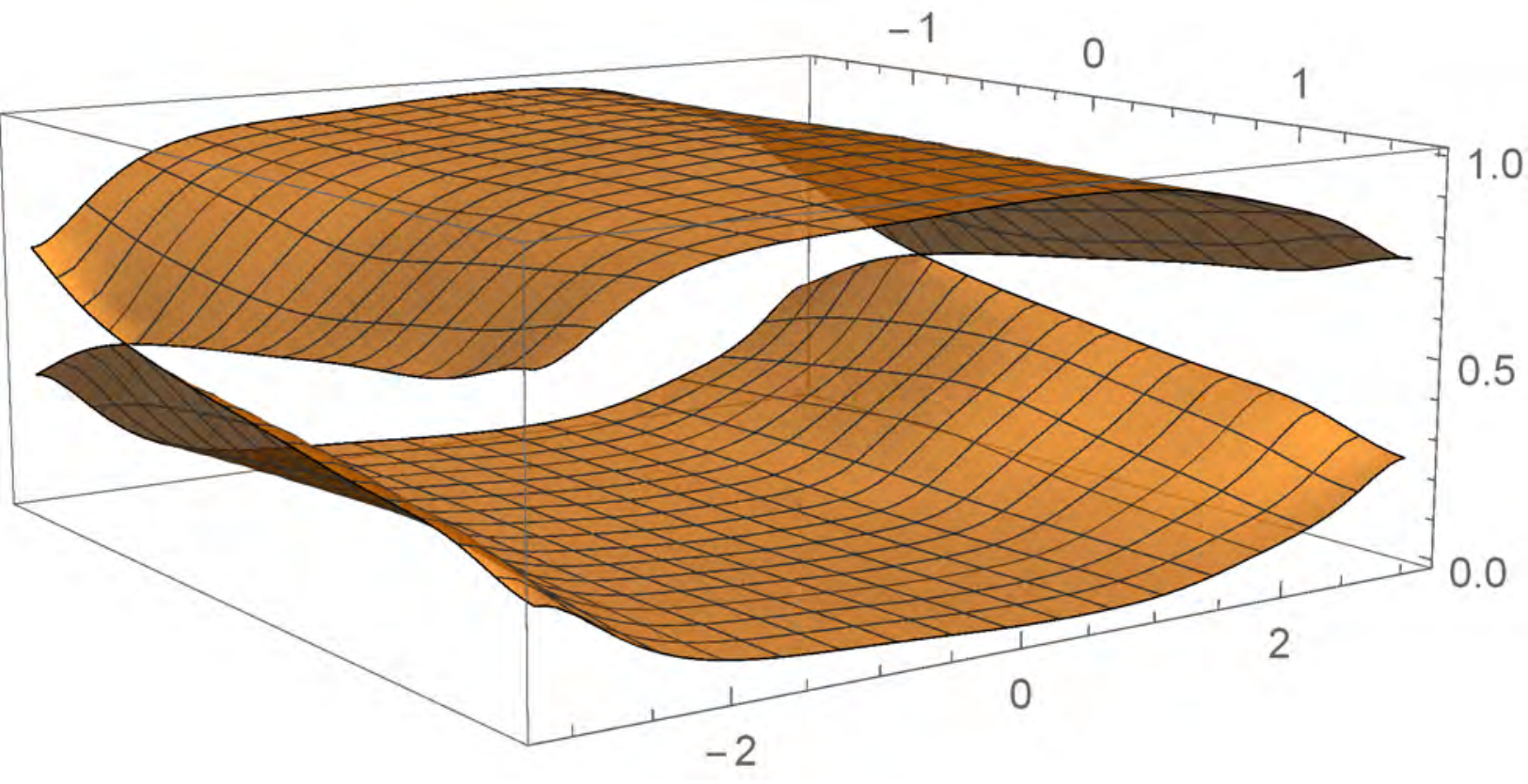}
  \end{minipage}
\caption{BES for $\bf CI_{1c}$ for $\mu=3$ ($\mathtt{C}=-1$). Left:  without deformation. Right: with deformation $a_x=b_x=a_y=b_y=1$.}\label{fig-CI1c-II}
\end{figure}

It is interesting to note that the band crossing pattern in the topological phase varies with $\m$'s. For example, setting $\mu=-3$ so that $\mathtt{C}=1$, we see a Dirac nodal point at ${\bf q}=(0,0)$ in the left panel of Fig.\ \ref{fig-CI1c-I}. In contrast, setting $\mu=3$ so that $\mathtt{C}=-1$, the Dirac point is now at ${\bf q}=(0, \pi)$ in the left panel of Fig.\ \ref{fig-CI1c-II}. When we introduce the deformation defined in eq.\ \eq{deformk}, there is still a Dirac point in the BES although its location gets shifted. See the right panels of Fig.\ \ref{fig-CI1c-I} and Fig.\ \ref{fig-CI1c-II}.  In Appendix \ref{Appendix-dispersion}, the details of checking the dispersion relation around the Dirac nodal point are given.   In general, we see there is a nodal point as shown in Fig.\ \ref{fig-CI1c-I} and Fig.\ \ref{fig-CI1c-II} for $|\mu| < 4$ except for $|\mu| = 2$.

\begin{figure*}
\centering
  \begin{minipage}[b]{0.2\textwidth}
    \includegraphics[width=\textwidth]{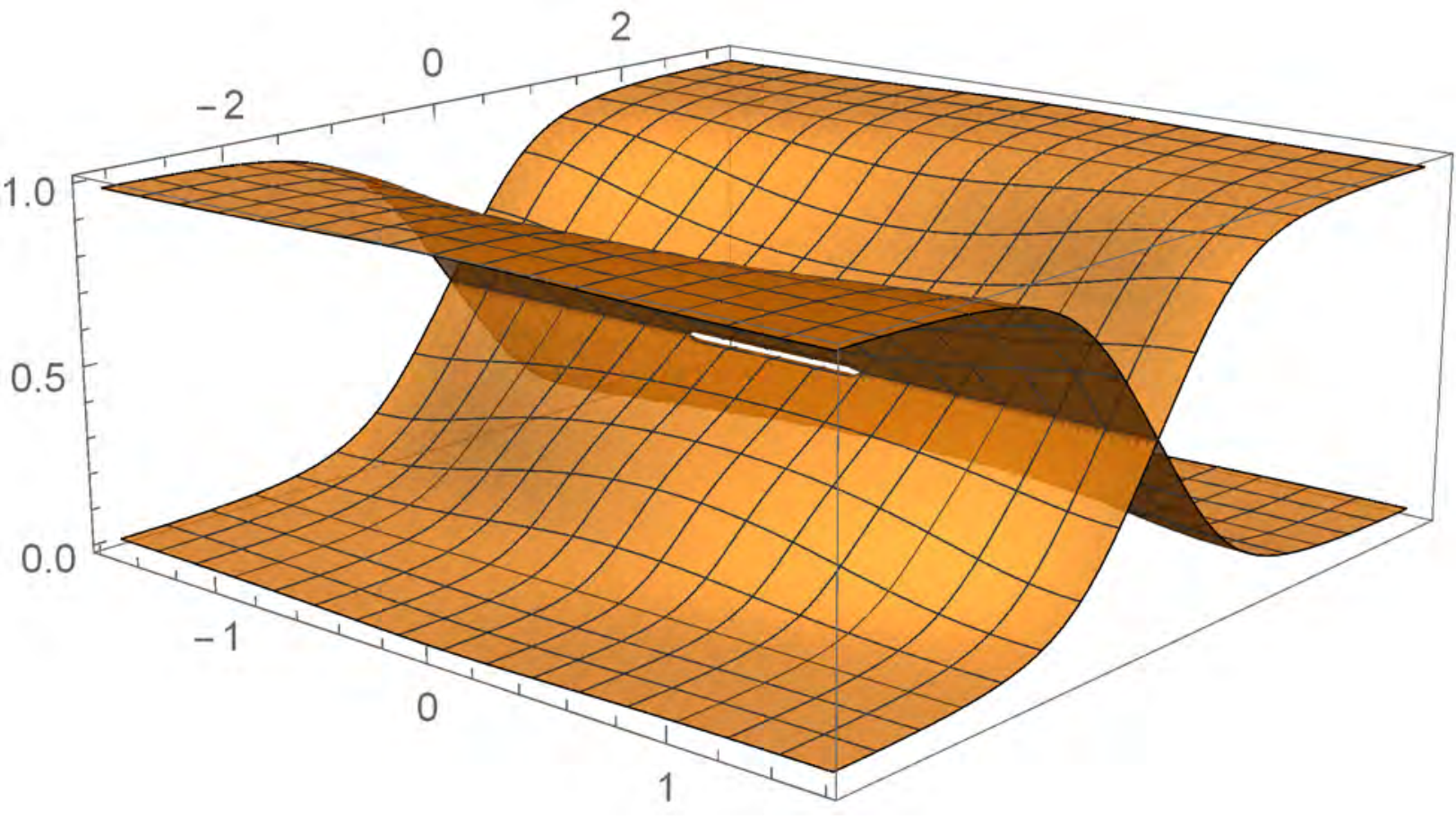}
  \end{minipage}
  \hspace{1cm} 
  \begin{minipage}[b]{0.2\textwidth}
    \includegraphics[width=\textwidth]{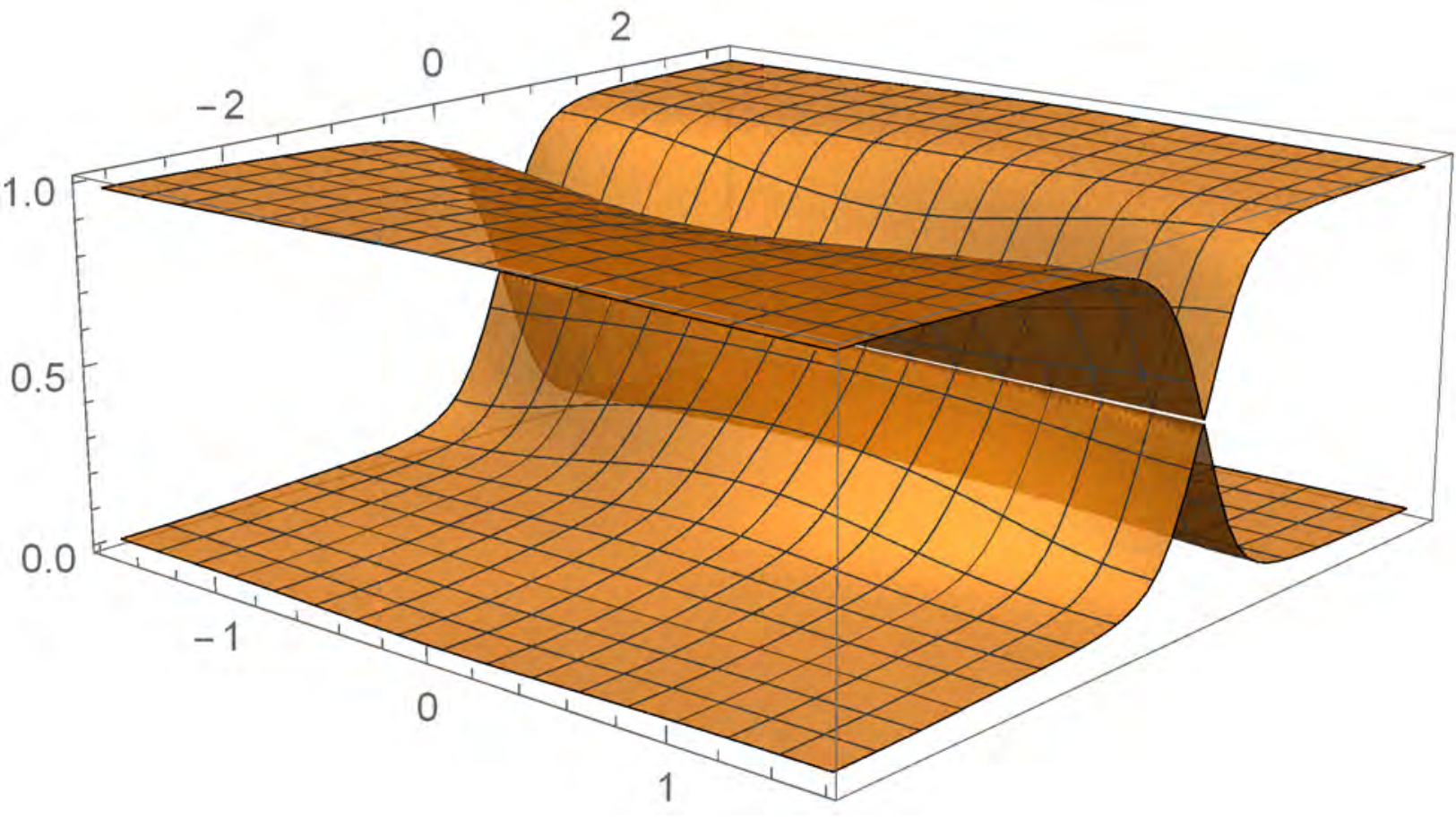}
  \end{minipage}
  \hspace{1cm} 
  \begin{minipage}[b]{0.2\textwidth}
    \includegraphics[width=\textwidth]{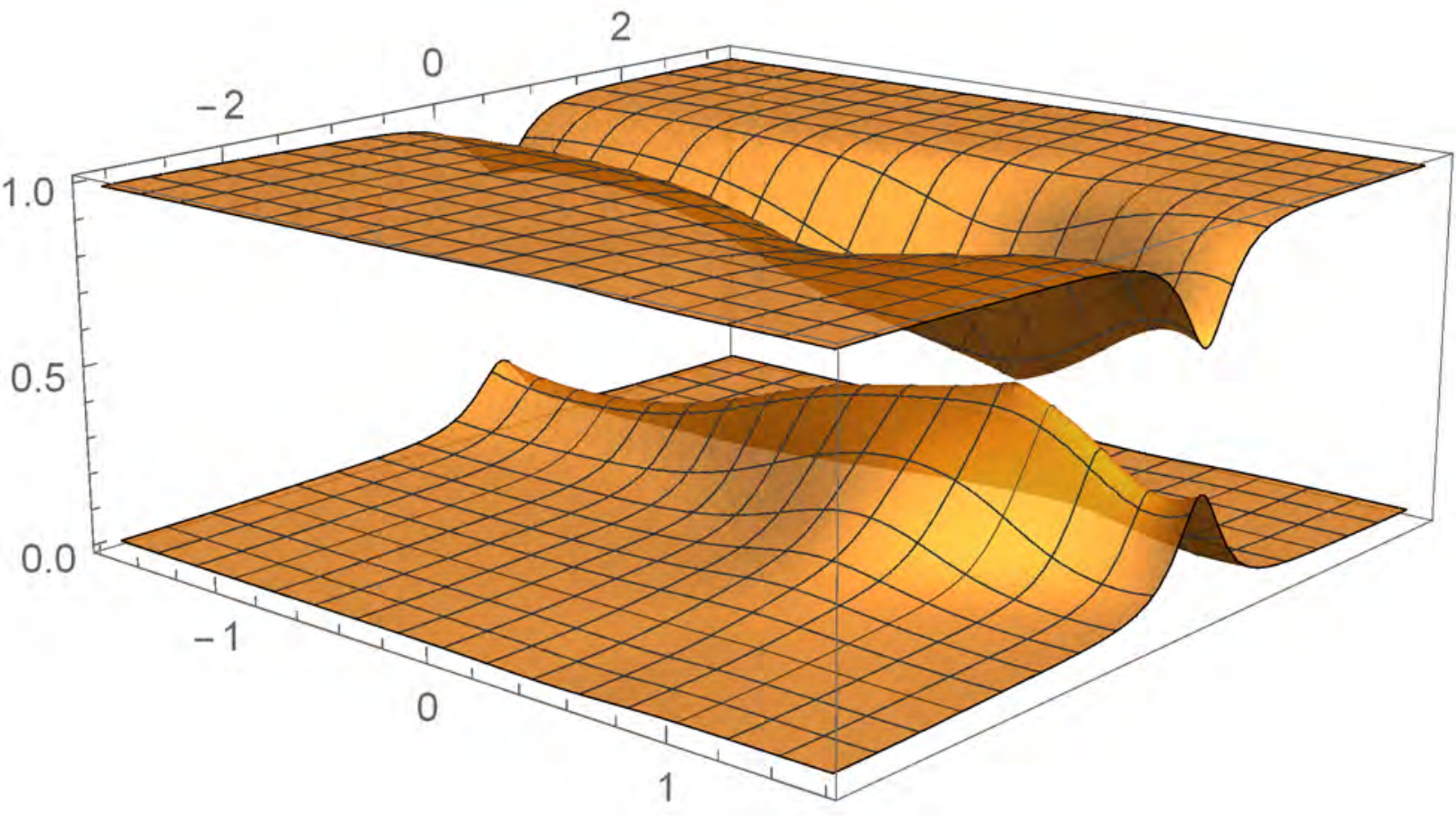}
  \end{minipage}
\caption{BES for $\bf CI_{1c}$ for $\mu=-2$ ($\mathtt{C}=1$). Left:  without deformation. Middle: with deformation $a_x=b_x=0$, $a_y=b_y=1$. Right: with deformation $a_x=b_x=a_y=b_y=1$.}\label{fig-CI1c-III}
\end{figure*}

For $\mu=-2$ ($\mathtt{C}=1$) in particular, we notice that peculiarly there is a nodal line along $q_y=0$ in the BES, shown in the left panel of Fig.\ \ref{fig-CI1c-III}.  Similar pattern remains even if we deform the BES with parameters $a_x=b_x=0$, $a_y=b_y=1$, as one can see in the middle one of Fig.\ \ref{fig-CI1c-III}.  With further deformation e.g.\  $a_x=b_x=1=a_y=b_y=1$, the nodal line reduces to a point as shown in the right one of Fig.\ \ref{fig-CI1c-III}.  Similar phenomenon can also be observed for the case that $\mu=2$ ($\mathtt{C}=-1$), where the nodal line is given by $q_y= \pi$. The topological properties of such kind of nontrivial nodal lines are further discussed in Appendix \ref{Appendix-nodal lines}. It is interesting to see that a single phase in the physical Hamiltonian is further differentiated into two sub-phases in regard to the BES. This suggests that the BES not only inherits the topology of the Chern insulator but, in some particular A-B symmetric partitions, also manifests a more refined topological structure which can not be seen from the spectrum of the physical Hamiltonian.

\subsubsection{$\bf CI_{1d}$}\label{Section-CI1d}
The partition is not A-B symmetric, and we generally do not see any band crossing in the BES except for $\mu =\pm 2$.  When $\mu =2$, again we see a nodal line at $q_y =0$.  In contrast to the $\bf CI_{1c}$ case,  the nodal line is completely lifted after we introduce deformation, as shown in Fig.\ \ref{fig-CI1c-III}.  Therefore, we think the nodal line here is unstable under perturbation and its origin has nothing to do with the topology of the system.

\begin{figure*}
\centering
  \begin{minipage}[b]{0.2\textwidth}
    \includegraphics[width=\textwidth]{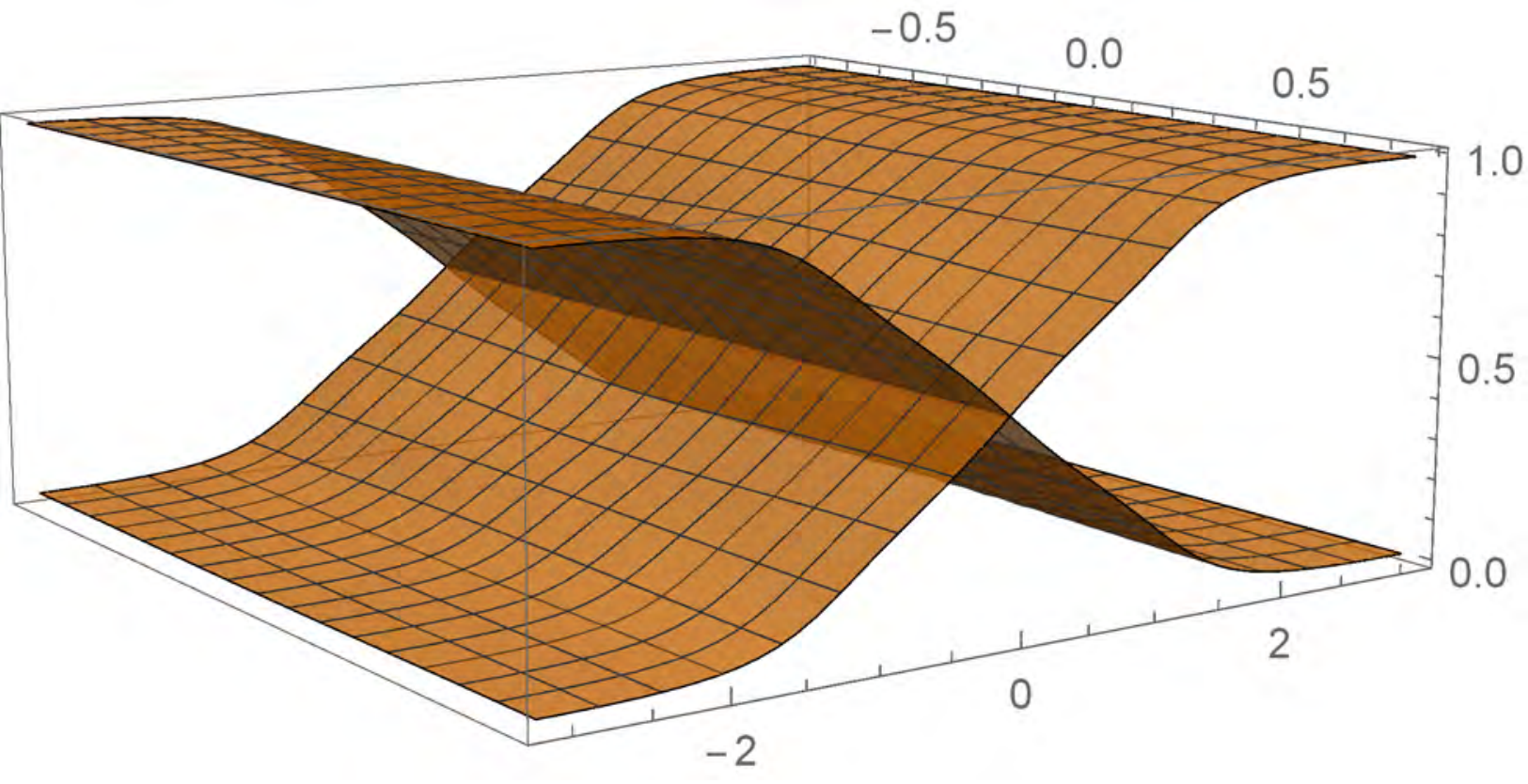}
  \end{minipage}
  \hspace{1cm} 
  \begin{minipage}[b]{0.2\textwidth}
    \includegraphics[width=\textwidth]{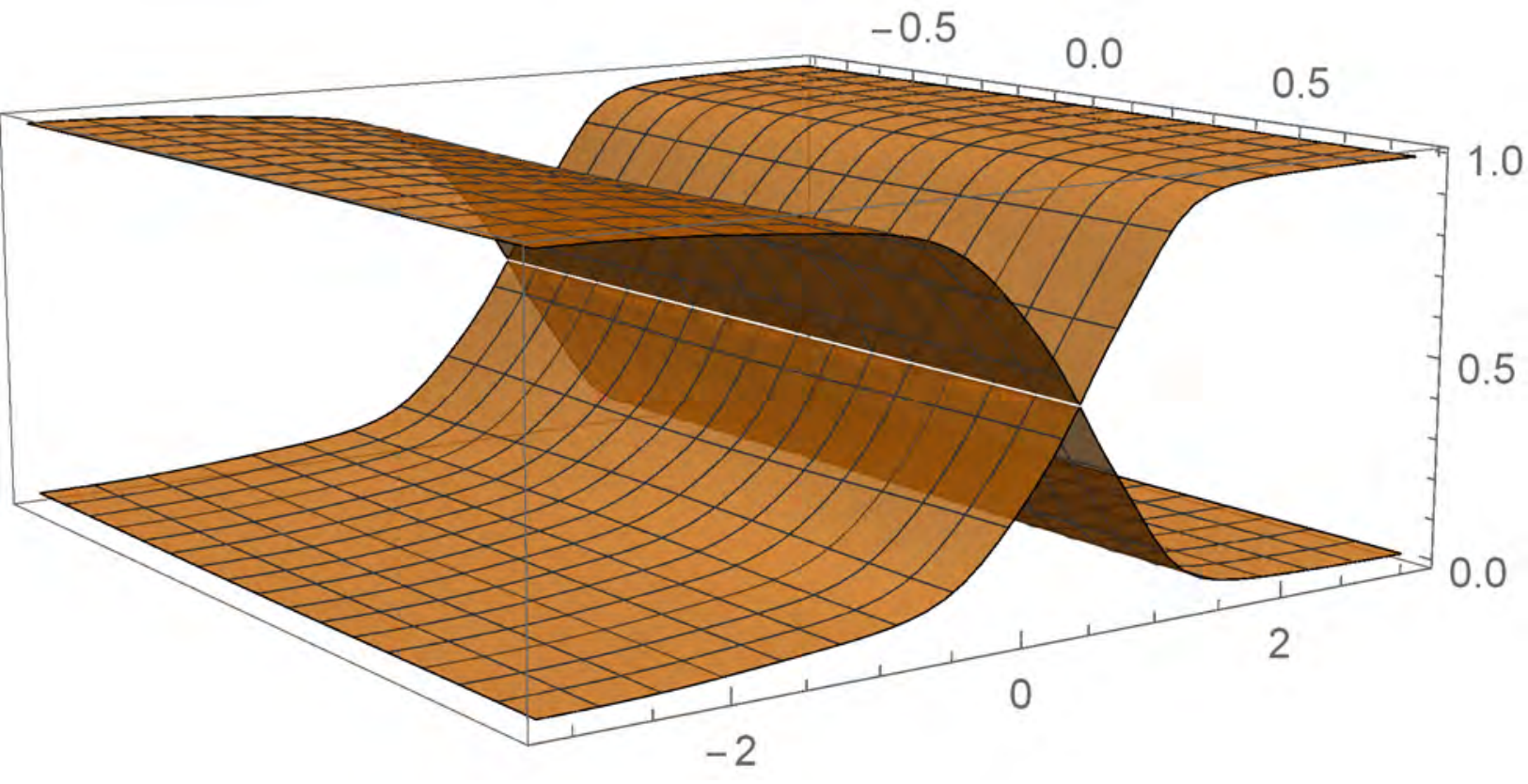}
  \end{minipage}
  \hspace{1cm} 
  \begin{minipage}[b]{0.2\textwidth}
    \includegraphics[width=\textwidth]{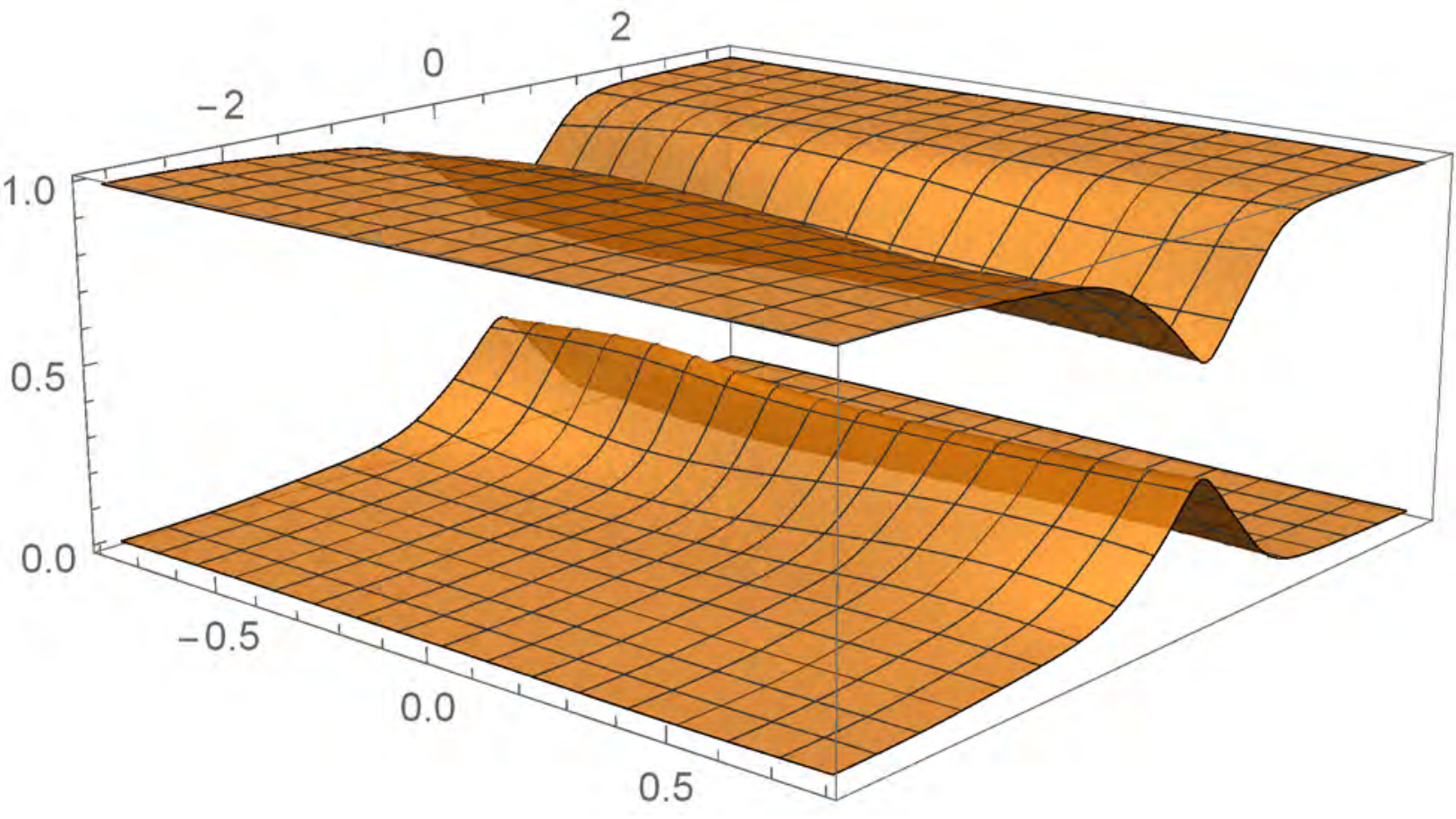}
  \end{minipage}
\caption{BES for $\bf CI_{1d}$ for $\mu=-2$ ($\mathtt{C}=1$). Left:  without deformation. Middle: with deformation $a_x=b_x=0$, $a_y=b_y=1$. Right: with deformation $a_x=b_x=a_y=b_y=1$.}\label{fig-CI1d}
\end{figure*}

\subsubsection{$\bf CI_{1e}$}
\begin{figure}
\centering
  \begin{minipage}[b]{0.2\textwidth}
    \includegraphics[width=\textwidth]{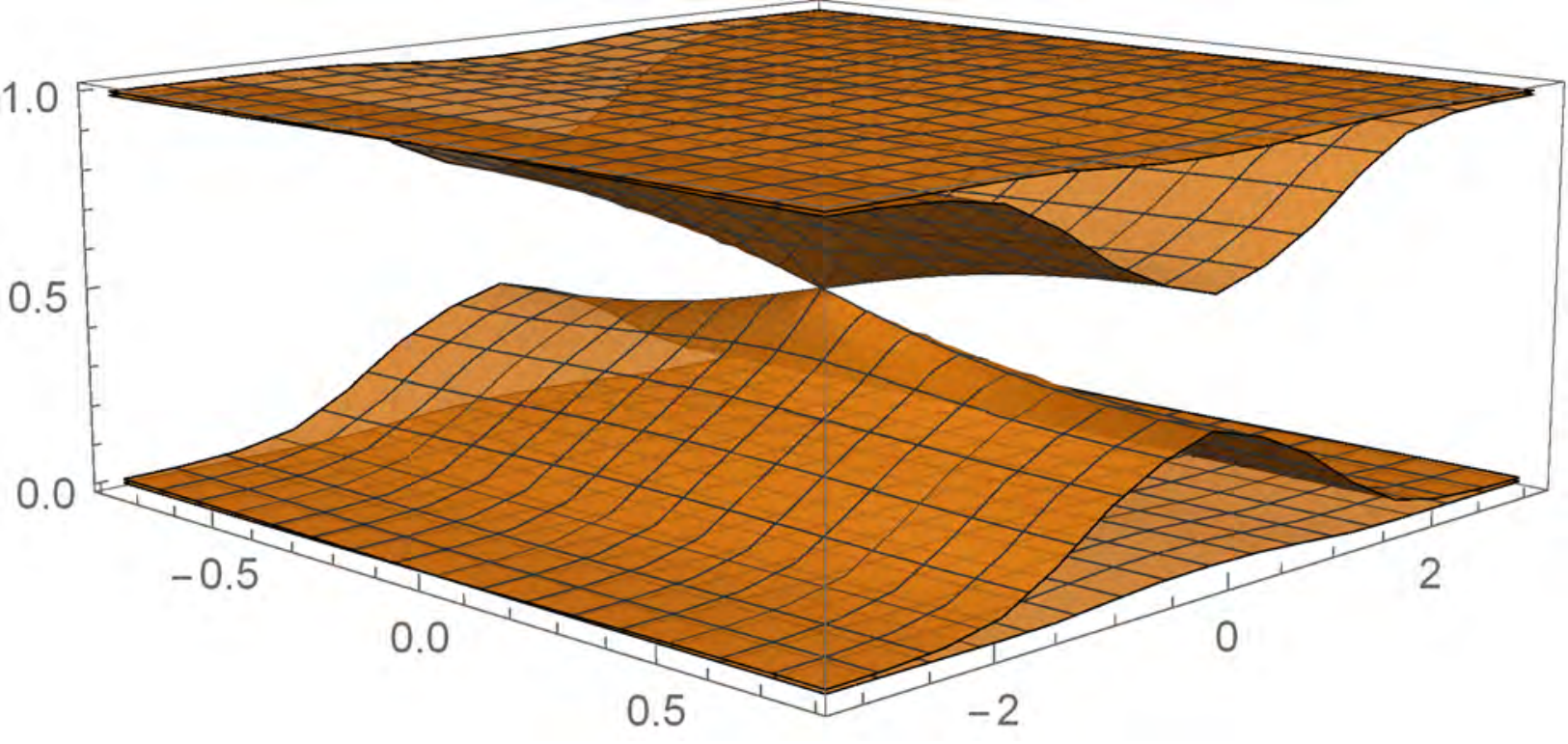}
  \end{minipage}
  \hspace{0.7cm} 
  \begin{minipage}[b]{0.2\textwidth}
    \includegraphics[width=\textwidth]{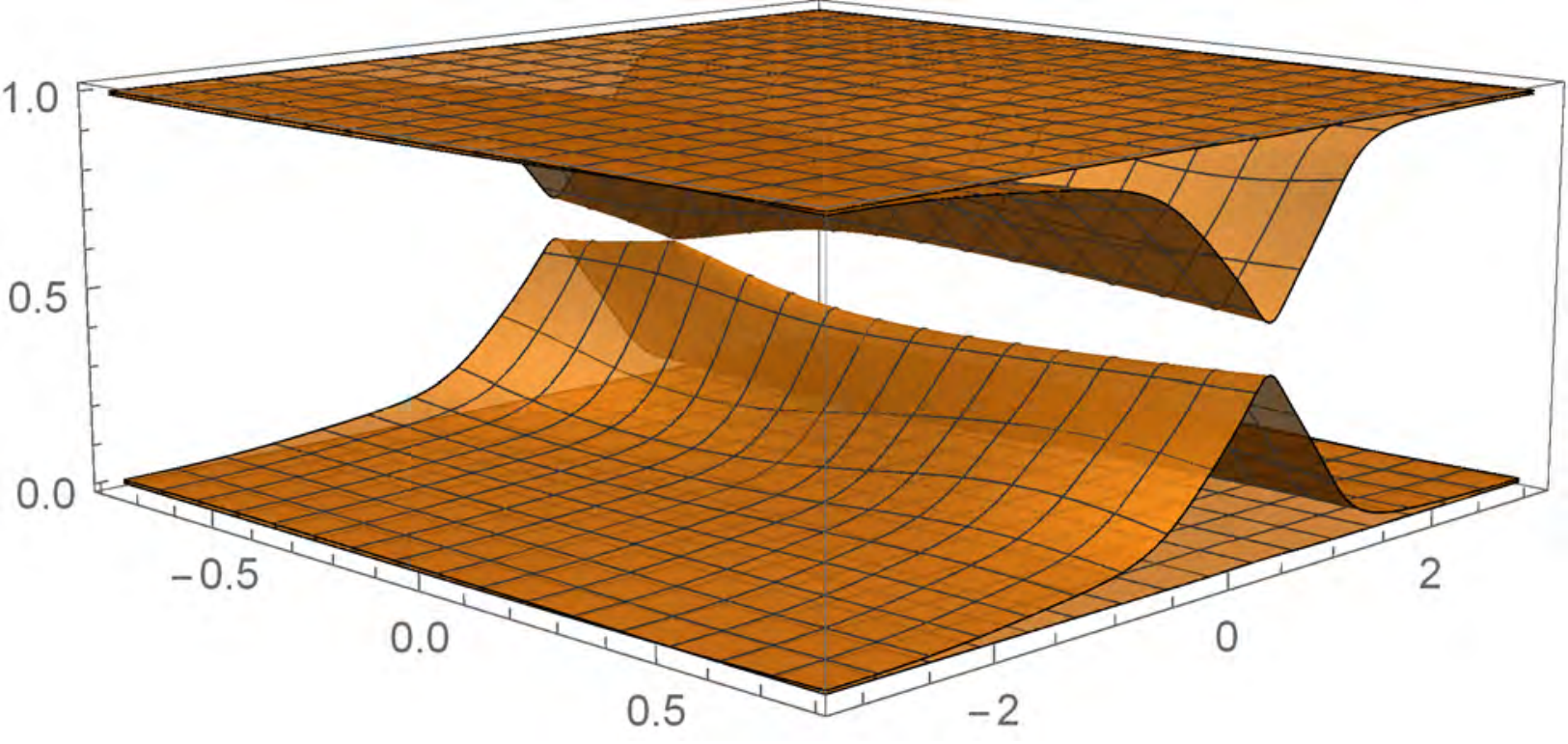}
  \end{minipage}
\caption{BES for $\bf CI_{1e}$ for $\mu=-2$ ($\mathtt{C}=1$). Left:  without deformation. Right: with deformation $a_x=b_x=a_y=b_y=1$.}\label{fig-CI1e-I}
\end{figure}

\begin{figure}
\centering
  \begin{minipage}[b]{0.2\textwidth}
    \includegraphics[width=\textwidth]{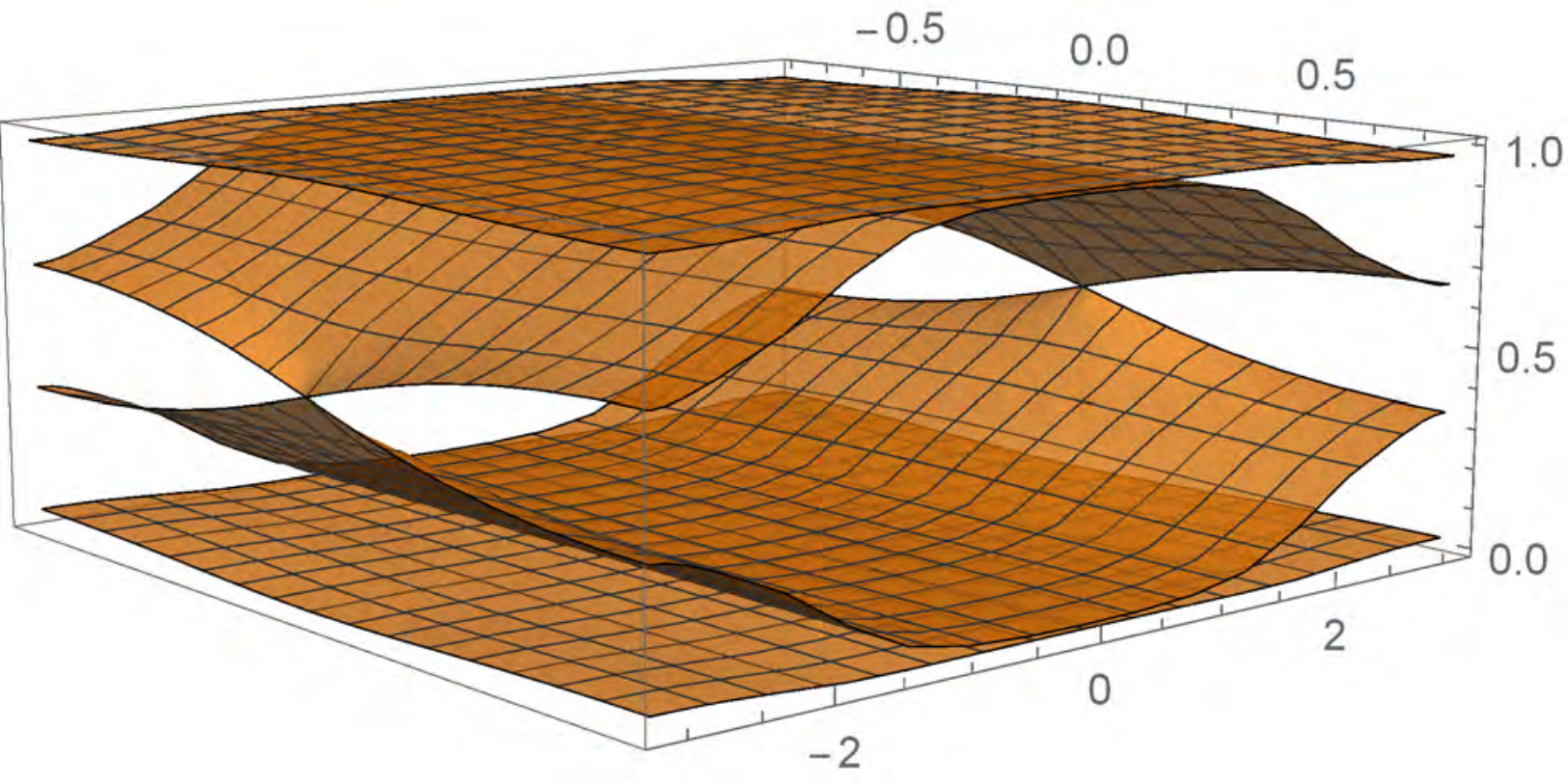}
  \end{minipage}
  \hspace{0.7cm} 
  \begin{minipage}[b]{0.2\textwidth}
    \includegraphics[width=\textwidth]{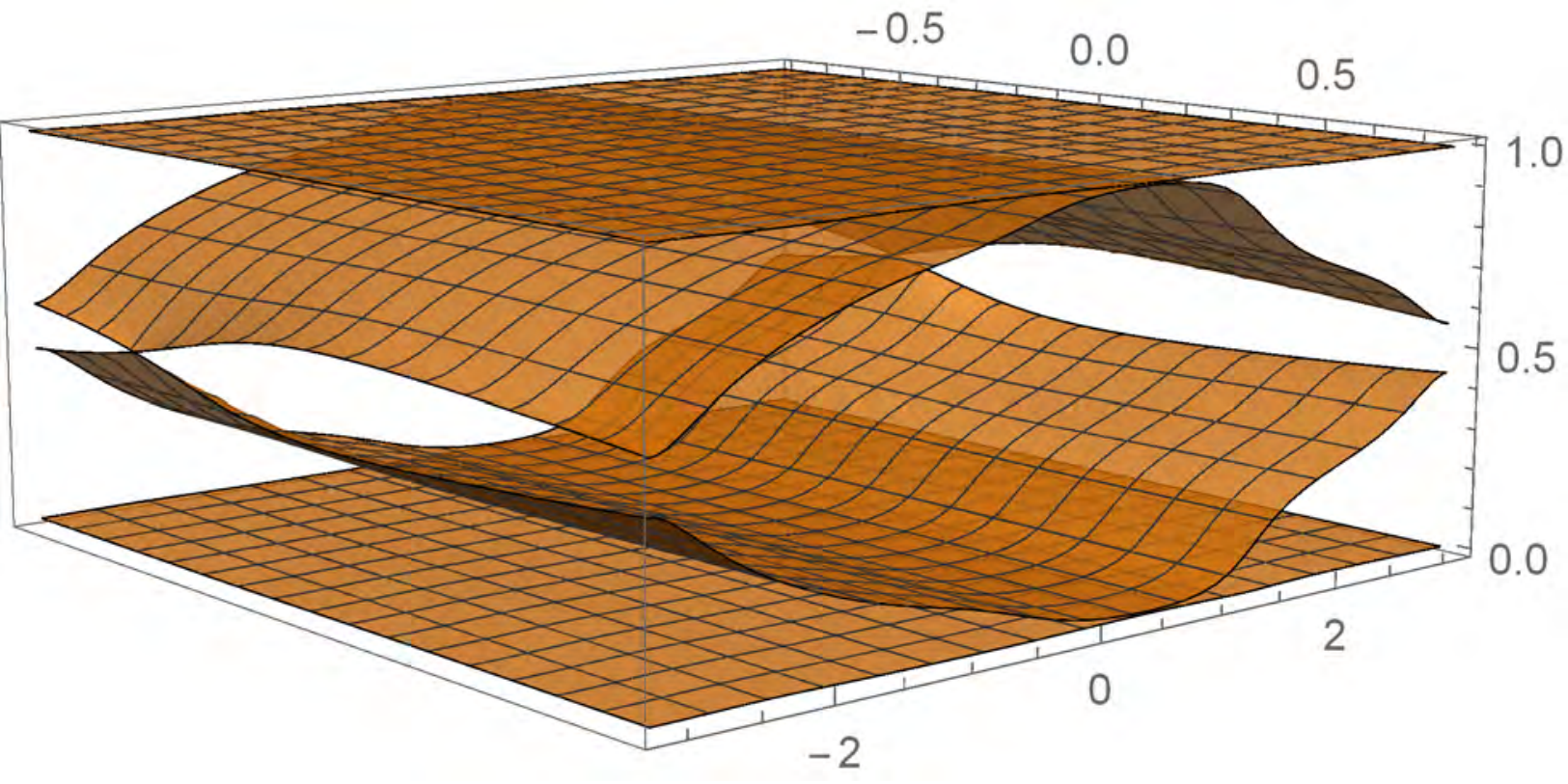}
  \end{minipage}
\caption{BES for $\bf CI_{1e}$ for $\mu=2$ ($\mathtt{C}=-1$). Left:  without deformation. Right: with deformation $a_x=b_x=a_y=b_y=1$.}\label{fig-CI1e-II}
\end{figure}

The partition is also A-B symmetric. Even though we do not have an analytical solution, numerical calculation shows that the BES has band crossing whenever the Chern insulator exhibits non-vanishing $\mathtt{C}$. Setting $\mu=-3$ so that $\mathtt{C}=1$, we see a Dirac nodal point at ${\bf q}=(0,0)$ in the left panel of Fig.\ \ref{fig-CI1e-I}. In contrast, setting $\mu=3$ so that $\mathtt{C}=-1$, the Dirac point is now at ${\bf q}=(0,\pi)$ in the left panel of Fig.\ \ref{fig-CI1e-II}. When we introduce the deformation defined in e.q.\ \eq{deformk}, there is still a Dirac point in the BES although its location gets shifted, which are shown in the right panels of Fig.\ \ref{fig-CI1e-I} and Fig.\ \ref{fig-CI1e-II}.

\subsubsection{$\bf CI_{1a}$ and $\bf CI_{1b}$}
\begin{figure}
\centering
  \begin{minipage}[b]{0.2\textwidth}
    \includegraphics[width=\textwidth]{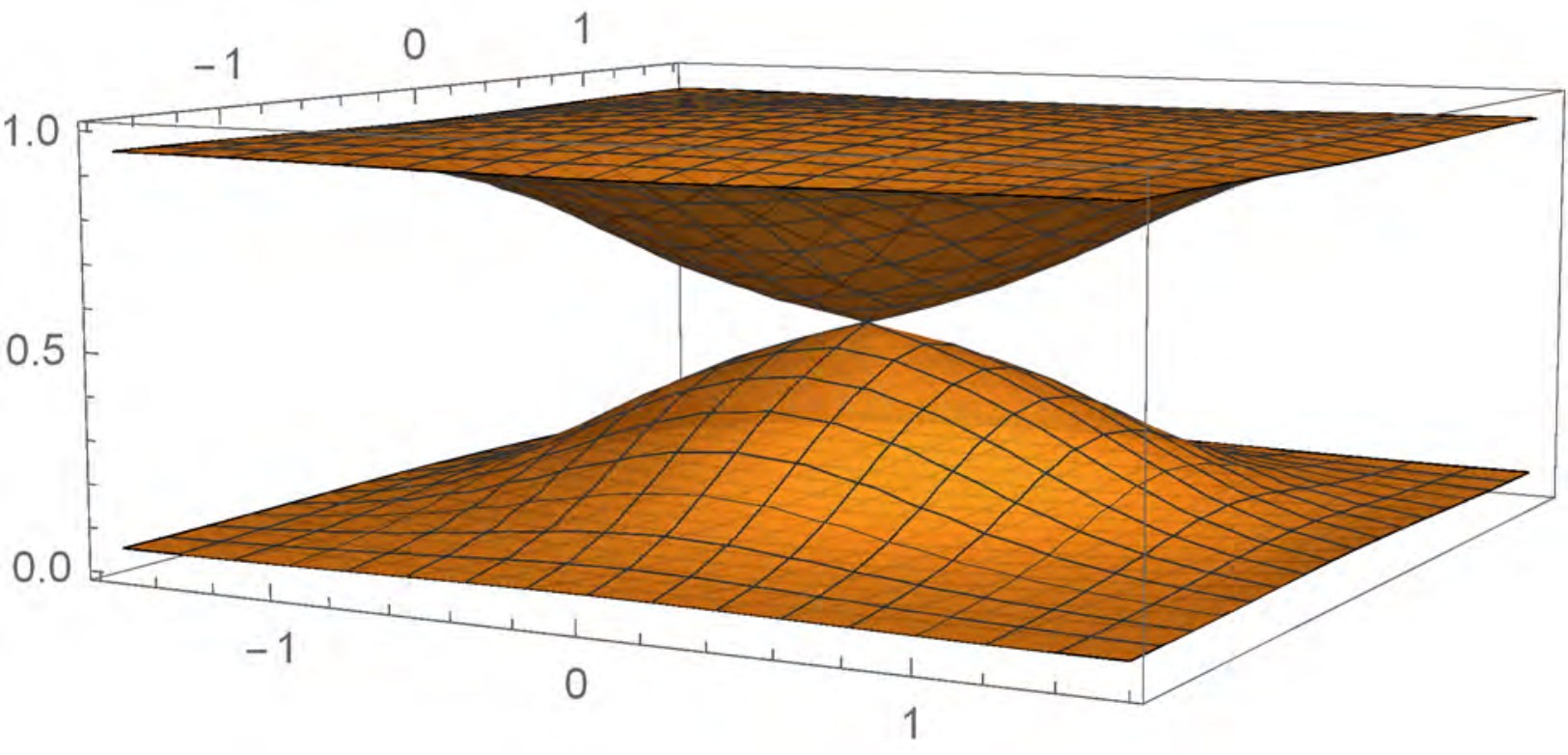}
  \end{minipage}
  \hspace{0.7cm} 
  \begin{minipage}[b]{0.2\textwidth}
    \includegraphics[width=\textwidth]{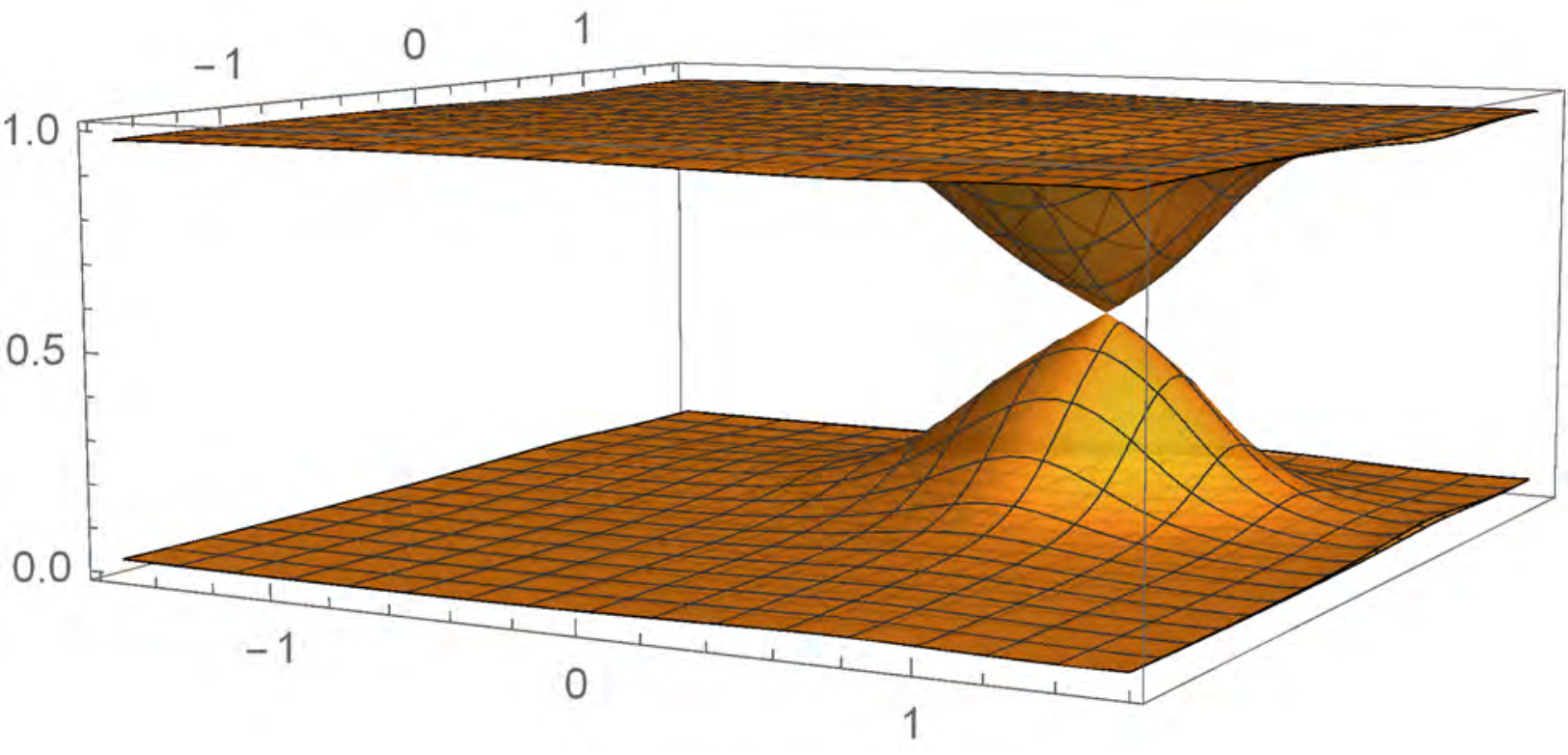}
  \end{minipage}
\caption{BES for ${\bf CI_{1b}}$ for $\mu=-3$  ($\mathtt{C}=+1$). Left: without deformation. Right: with deformation $a_x=b_x=1$, $a_y=b_y=1$.}\label{fig-CI1b}
\end{figure}

A similar story repeats for partition $\bf a$ (A-B asymmetric) and $\bf b$ (A-B symmetric), except that here there is no nodal line in the BES. For the A-B symmetric case $\bf CI_{1b}$, there is a nodal point at ${\bf q}=(0,0)$ which is similar to the ones shown in Fig.\ \ref{fig-CI1c-I} for both phases with $\mathtt{C}=\pm 1$  as shown in Fig.\ \ref{fig-CI1b}. This band crossing condition can also be seen from the determinant
\be
{\rm det}[{\bf M}(0,0)-{\bf I}/2] = \frac{(\m-4)(\m+4)+|(\m-4)(\m+4)| }{32(\m-4)(\m+4)},
\ee
which vanishes for $|\m|<4$.

In contrast, there is no band crossing at all in the case of $\bf CI_{1a}$ for any value of $\mu$. This case is similar to ${\bf CI_{1d}}$ as discussed in Sec.\ \ref{Section-CI1d} in the sense that there is no \emph{robust} band crossing at all even when the physical Hamiltonian gives a nonzero Chern number. In addition to the two simplest cases of $\bf CI_{1a}$ and ${\bf CI_{1d}}$, we have also numerically tested other more complicated A-B asymmetric partitions. All of our results indicate that the BES does not have any robust band crossing against generic deformation. Our empirical evidence strongly suggests that the A-B symmetry is essential for the BES to reflect the nontrivial topology of the tight-binding model. The importance of the A-B symmetry is also underscored by the emerging chiral symmetry as argued before.

\subsubsection{$\bf CI_{2b}$ and $\bf CI_{2c}$}\label{Section-CI2b and CI2c}

In all the cases we have studied so far, robust nodal points of BES only occur for A-B symmetric partitions. Thus, from now on we will just focus on A-B symmetric partitions.

\begin{figure}
\centering
  \begin{minipage}[b]{0.2\textwidth}
    \includegraphics[width=\textwidth]{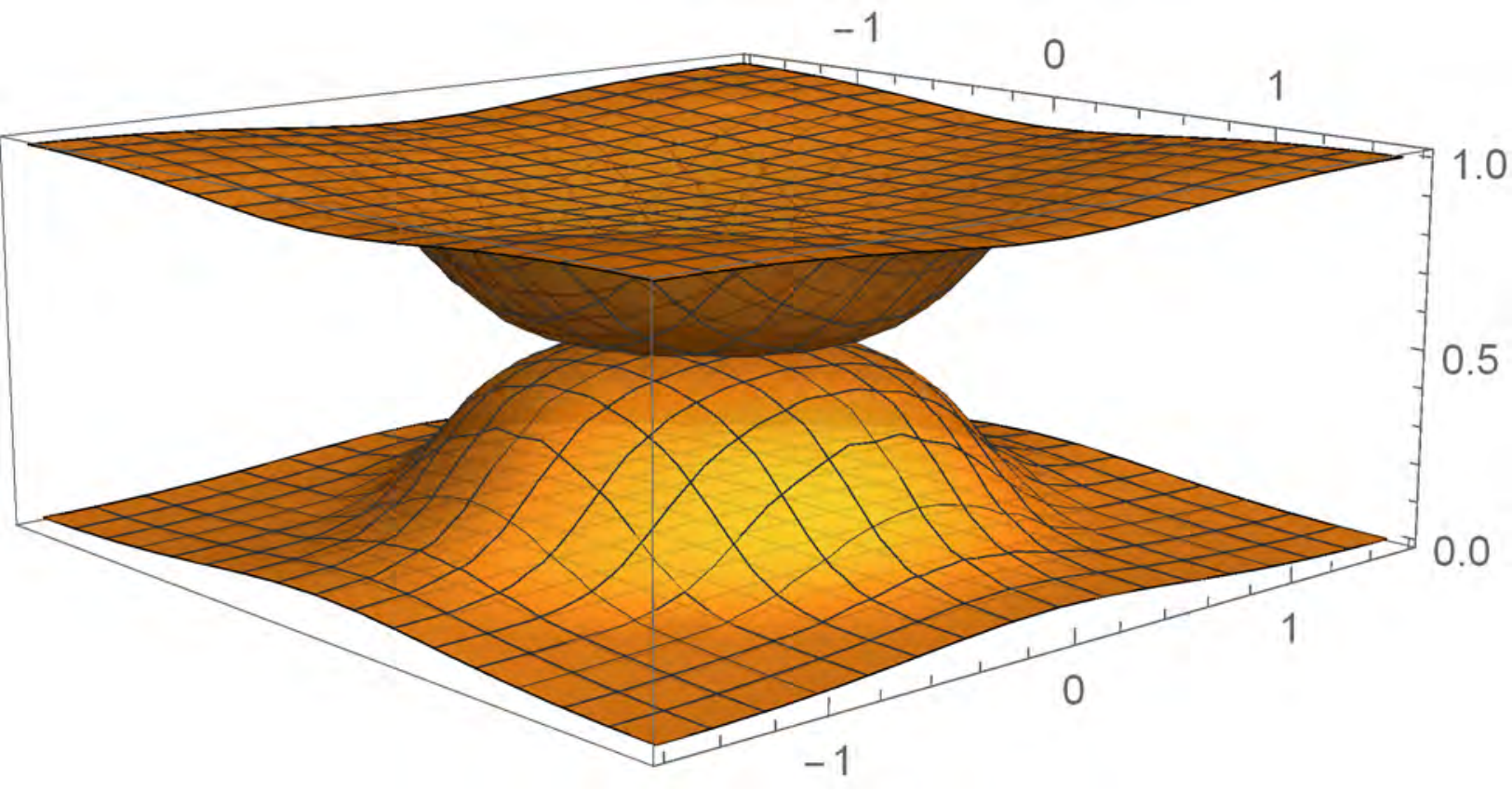}
  \end{minipage}
  \hspace{0.7cm} 
  \begin{minipage}[b]{0.2\textwidth}
    \includegraphics[width=\textwidth]{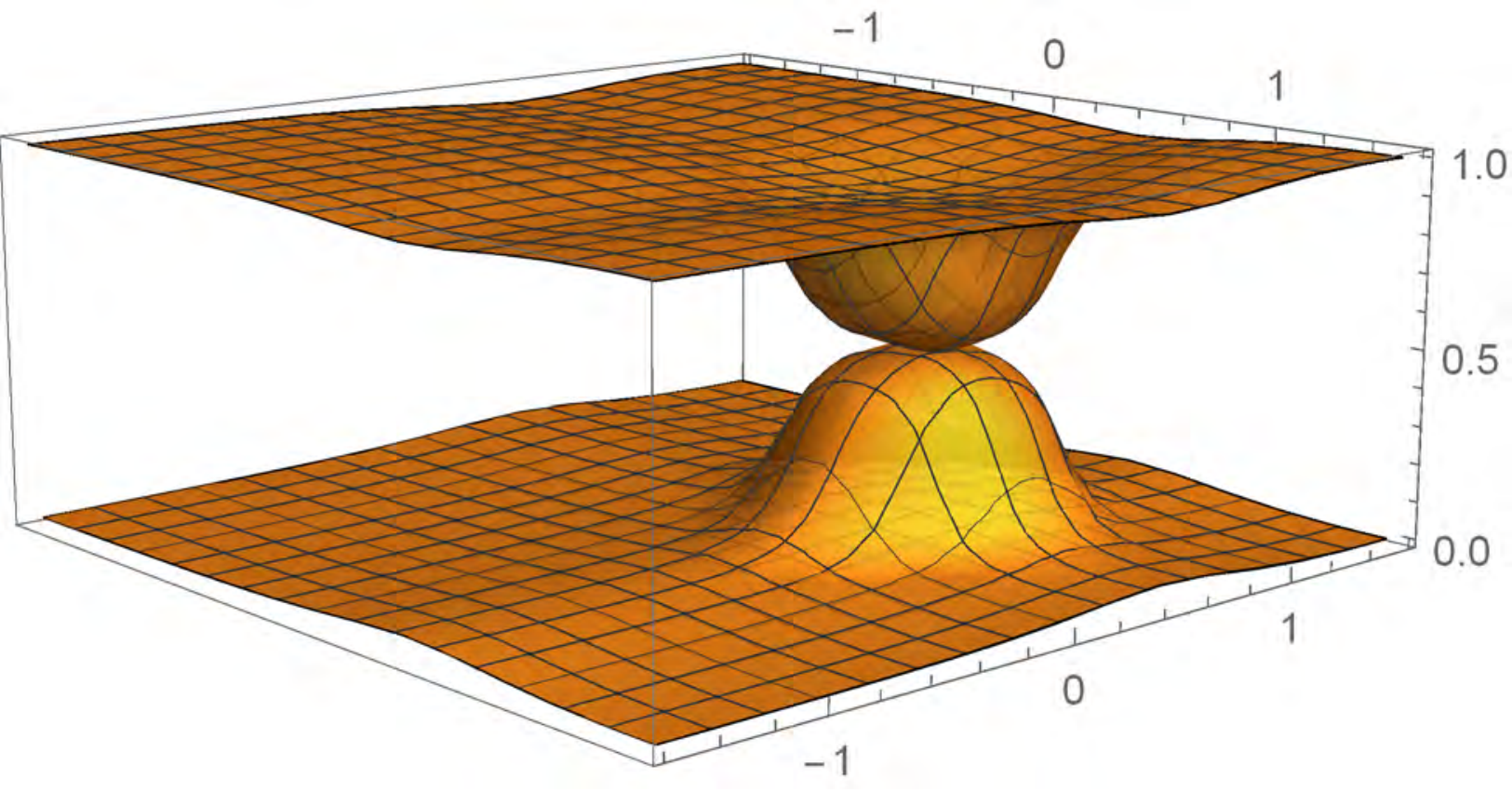}
  \end{minipage}
\caption{BES for $\bf CI_{2b}$ for $\mu=-3$ ($\mathtt{C}=2$). Left:  without deformation. Right: with deformation $a_x=b_x=a_y=b_y=1$.}\label{fig-CI2a}
\end{figure}

The band crossing condition for $\bf CI_{2b}$ can be seen from
\be
{\rm det}[{\bf M}(0,0)-{\bf I}/2] = \frac{\m^2\left[(\m-4)(\m+4)+|(\m-4)(\m+4)| \right] }{32(\m-4)(\m+4)(\m^2+4)}
\ee
which vanishes for $|\m|<4$, just like in the ${\bf CI_1}$ case. Its band crossing patterns as shown in Fig.\ \ref{fig-CI2a} do show a quadratic nodal point, which is expected since $\mathtt{C}=2$ in this topological phase \cite{BES-QHE}.  The details of demonstrating the nodal point to be quadratic can be found in Appendix \ref{Appendix-dispersion}. At the special value $\m=0$, there are two nodal lines along $q_x=0$ and $q_y=0$. Under generic deformation \eq{deformk}, the two nodal lines are lifted but their intersection point remains a quadratic nodal point. (Also see Appendix \ref{Appendix-nodal lines}.)

\begin{figure}
\centering
  \begin{minipage}[b]{0.2\textwidth}
    \includegraphics[width=\textwidth]{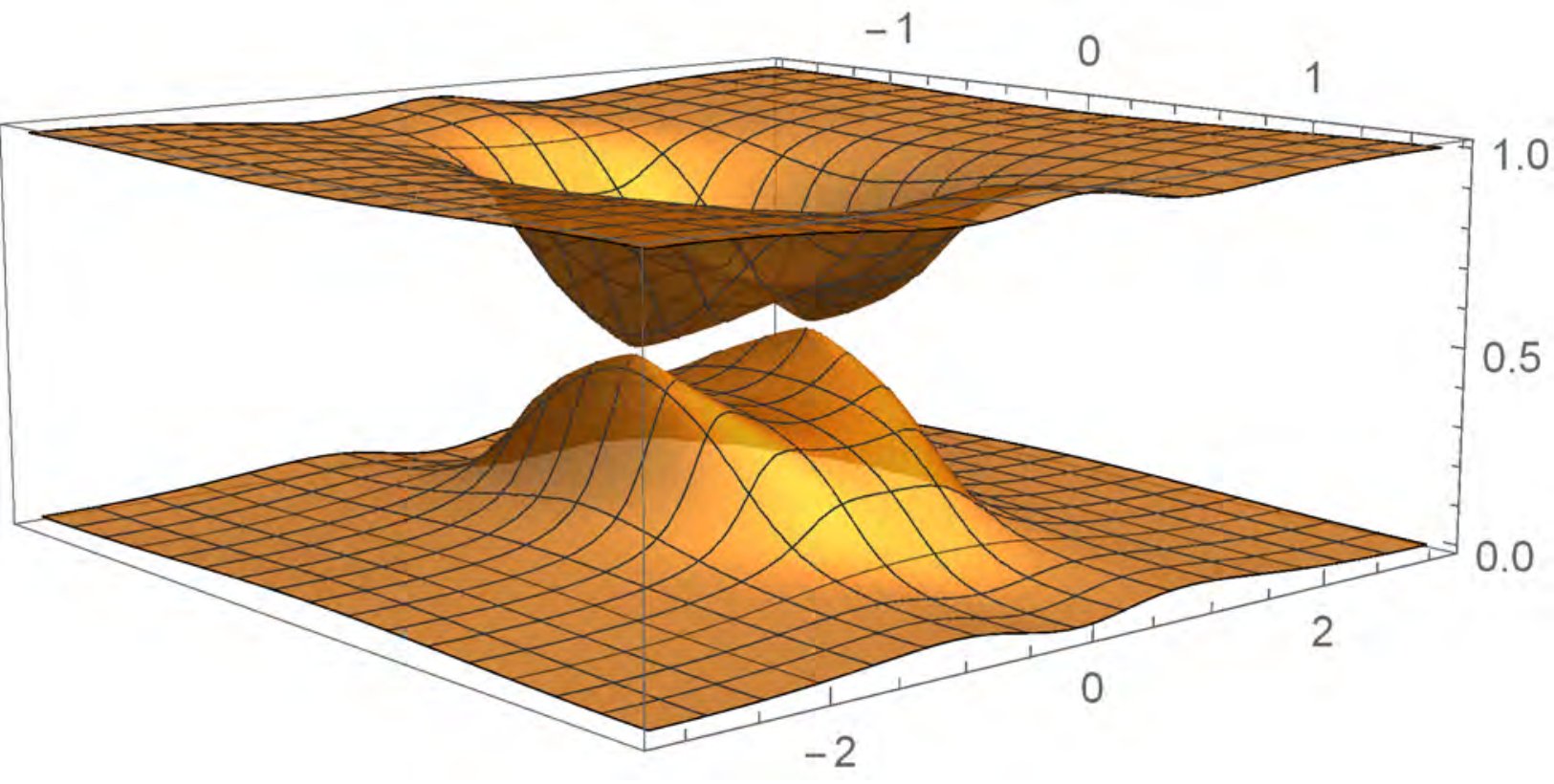}
  \end{minipage}
  \hspace{0.7cm} 
  \begin{minipage}[b]{0.2\textwidth}
    \includegraphics[width=\textwidth]{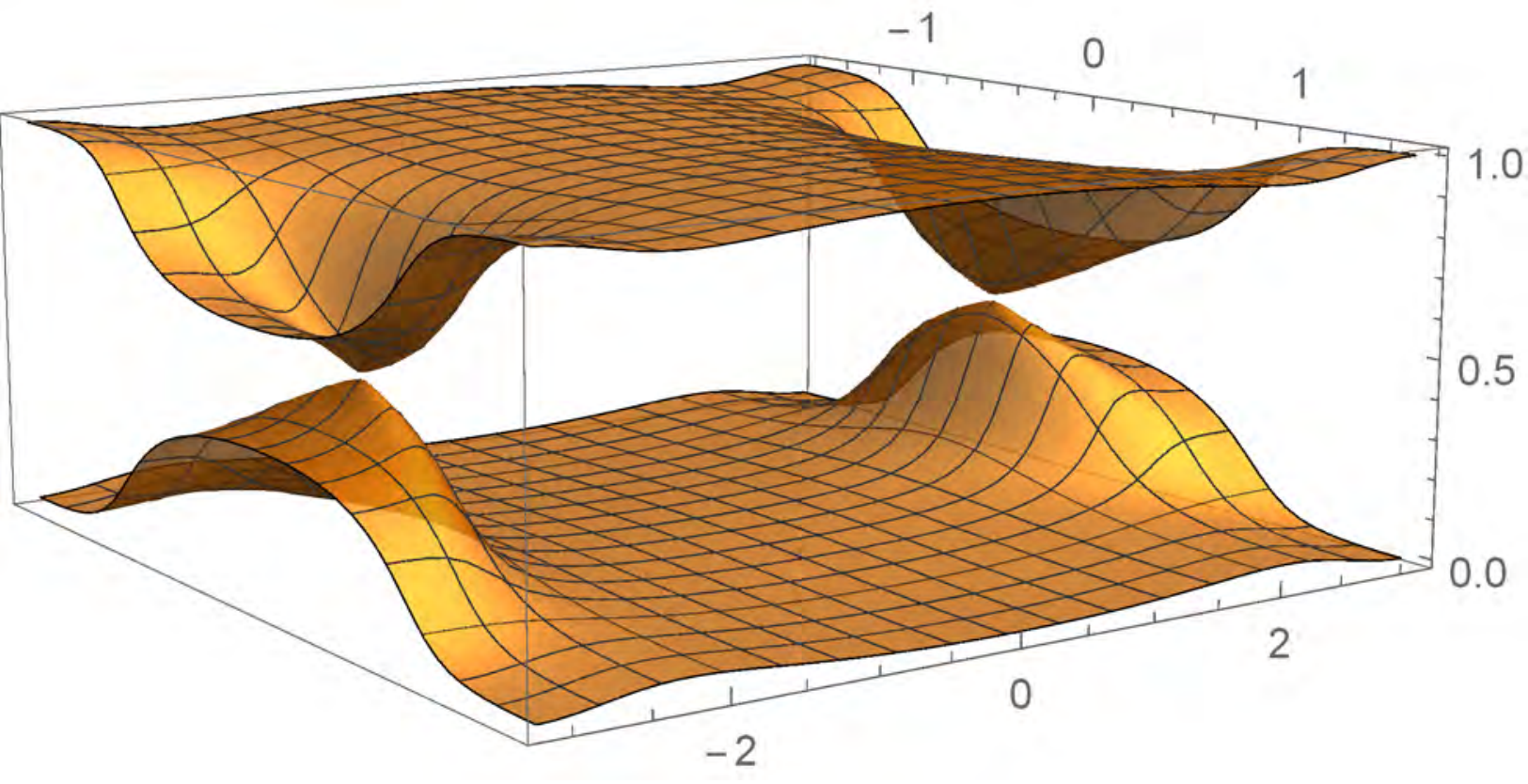}
  \end{minipage}
\caption{BES for $\bf CI_{2c}$ without deformation. Left: for $\mu=-3$ ($\mathtt{C}=2$). Right: for $\mu=3$ ($\mathtt{C}=2$).  The generic deformation \eq{deformk} will just shift the positions of the Dirac nodal points.}\label{fig-CI2c}
\end{figure}

On the other hand, for the stripe partition $\bf CI_{2c}$, which is A-B symmetric, we find that the condition \hfil\break
${\rm det}[{\bf M}(\bq)-{\bf I}/2] =0$ yields band crossing at
\be
 {\bf q}=(0,\pm\cos^{-1}\frac{-\mu}{4}), \quad   \mbox{for}\; -4<\m<4.
\ee
The band crossing patterns show two robust Dirac nodal points, see Fig.\ \ref{fig-CI2c}.  This is in sharp contrast to the quadratic nodal points in Fig.\ \ref{fig-CI2a} for the checkerboard partition. Again, we find that at the special point $\m=0$ there are two nodal lines along $q_y=\pm\pi/2$, which reduce to two Dirac nodal points under the deformation \eq{deformk}. (Also see Appendix \ref{Appendix-nodal lines}.)

For all the band crossing patterns of BES considered so far, they are also consistent with the following relation:
\be\label{not master}
|\mathtt{C}|= \text{sum of degrees of robust nodal points},
\ee
where the degree is counted as one for a Dirac node, two for a quadratic node, three for a cubic node, etc. Although the relation \eq{not master} is quite intuitive, we will see it is violated in the next subsection.   The correct relation should be \eq{master rel}, for which we need to define the vorticity of the nodal point through the Berry connection of the entanglement Hamiltonian's eigenstates as defined in section \ref{sec4}.

\subsubsection{$\bf CI_{3b}$ and $\bf CI_{3c}$}

In the third Chern insulator model, there are three phases with $\mathtt{C}=0, 1, 3$.

Let's first consider the stripe partition, i.e.\ $\bf CI_{3c}$. In the $\mathtt{C}=3$ ($-4<\mu<0$) phase, there are five points in the $\bf q$-space on which the band crossings may occur.
They are
\begin{eqnarray}
&&\mathbf{q}_\mathrm{O} := (0,0)\;, \quad
\mathbf{q}_\mathrm{N,S} := \left(0,\pm \cos^{-1}\frac{-\mu}{4}\right)\;, \nn
\\
&&\mathbf{q}_\mathrm{E,W} := \left(0,\pm \cos^{-1}\sqrt{\frac{-2-\mu}{2}}\right)\;. \label{5 nodal-1}
\end{eqnarray}
Note the first and second sub-index refer to the ``$+$" and "$-$", respectively.

\begin{figure*}

\centering

  \begin{minipage}[b]{0.2\textwidth}
    \includegraphics[width=\textwidth]{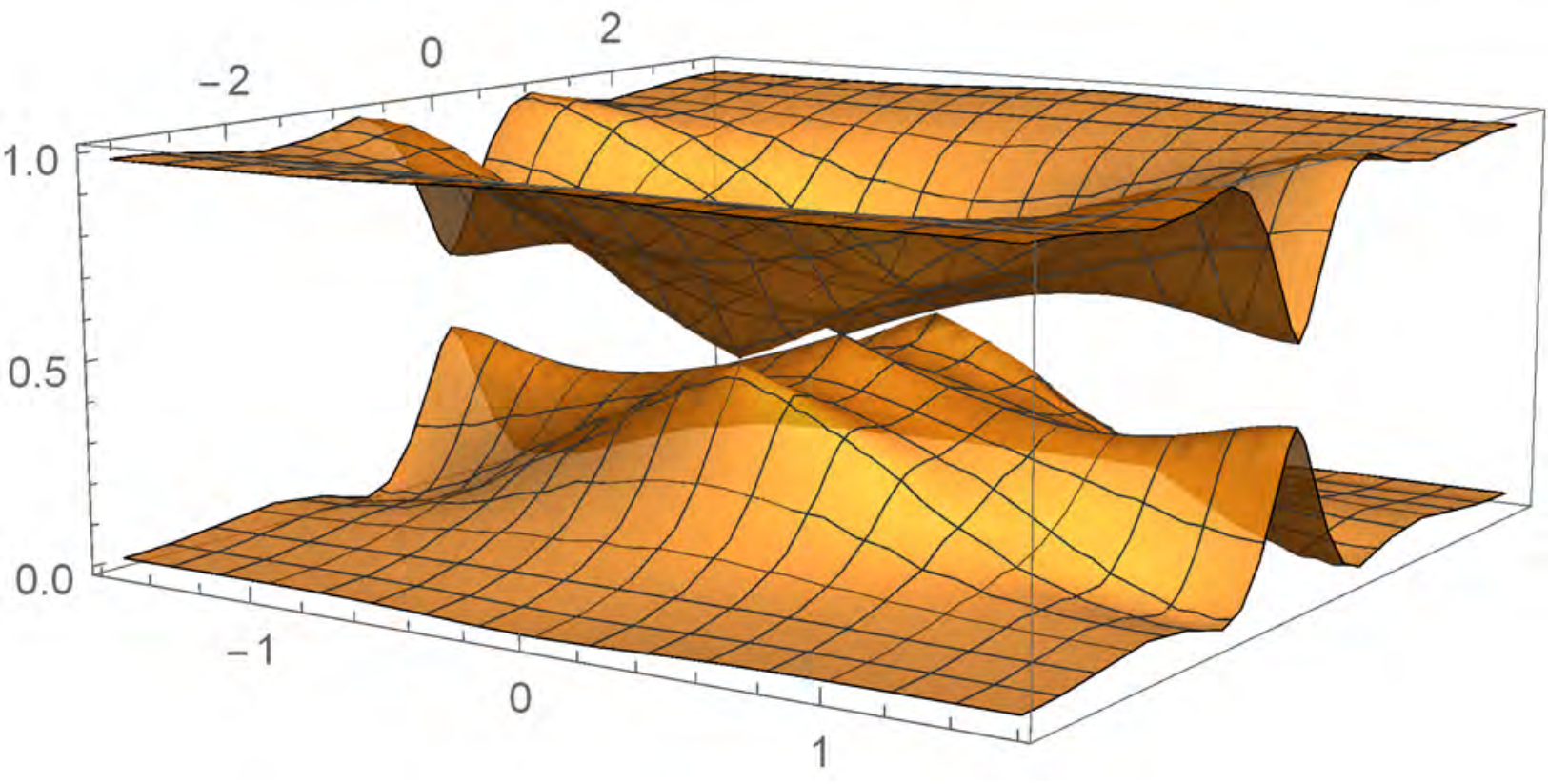}
  \end{minipage}
  \hspace{1cm} 
  \begin{minipage}[b]{0.2\textwidth}
    \includegraphics[width=\textwidth]{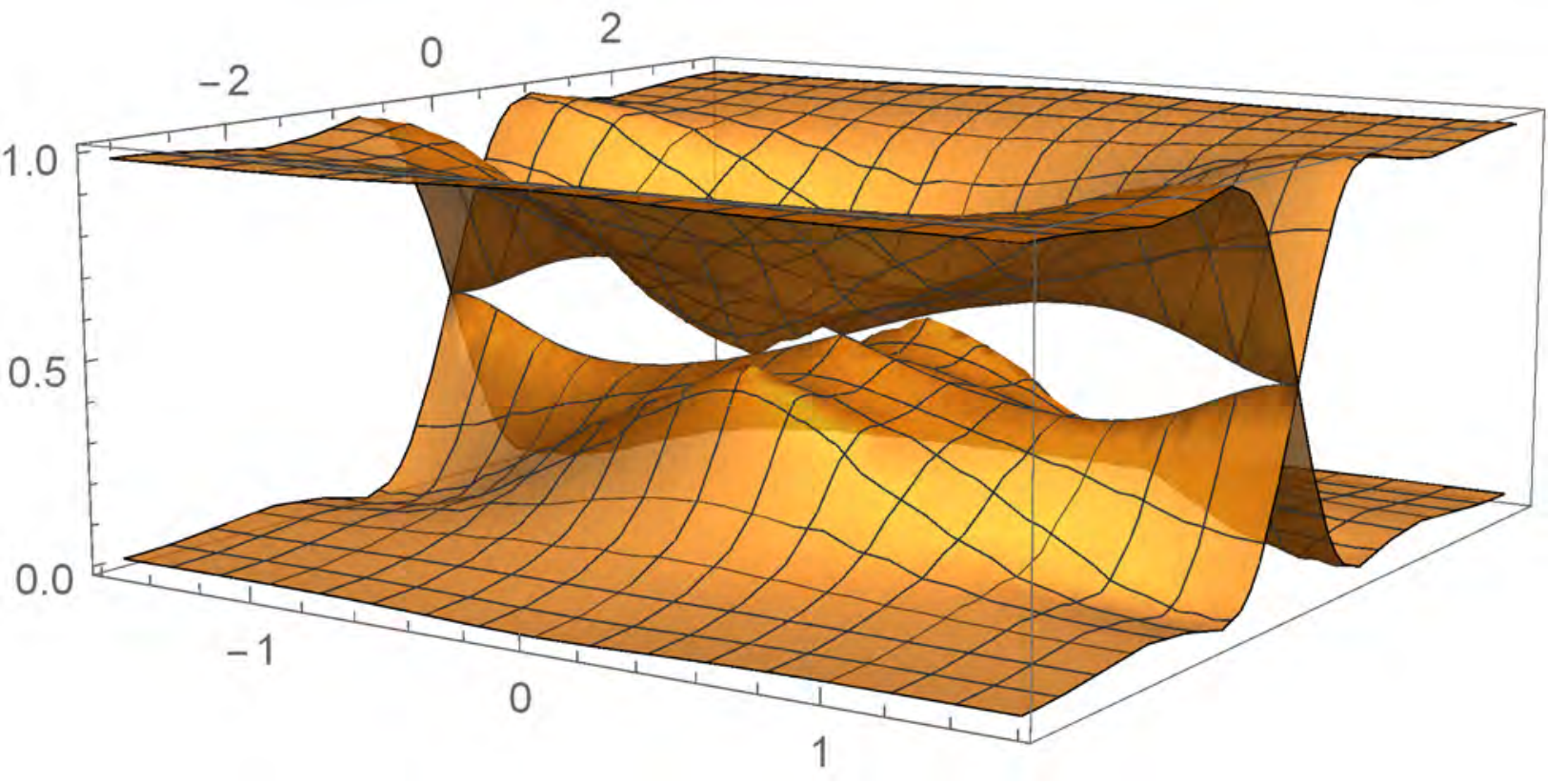}
  \end{minipage}
  \hspace{1cm} 
  \begin{minipage}[b]{0.2\textwidth}
    \includegraphics[width=\textwidth]{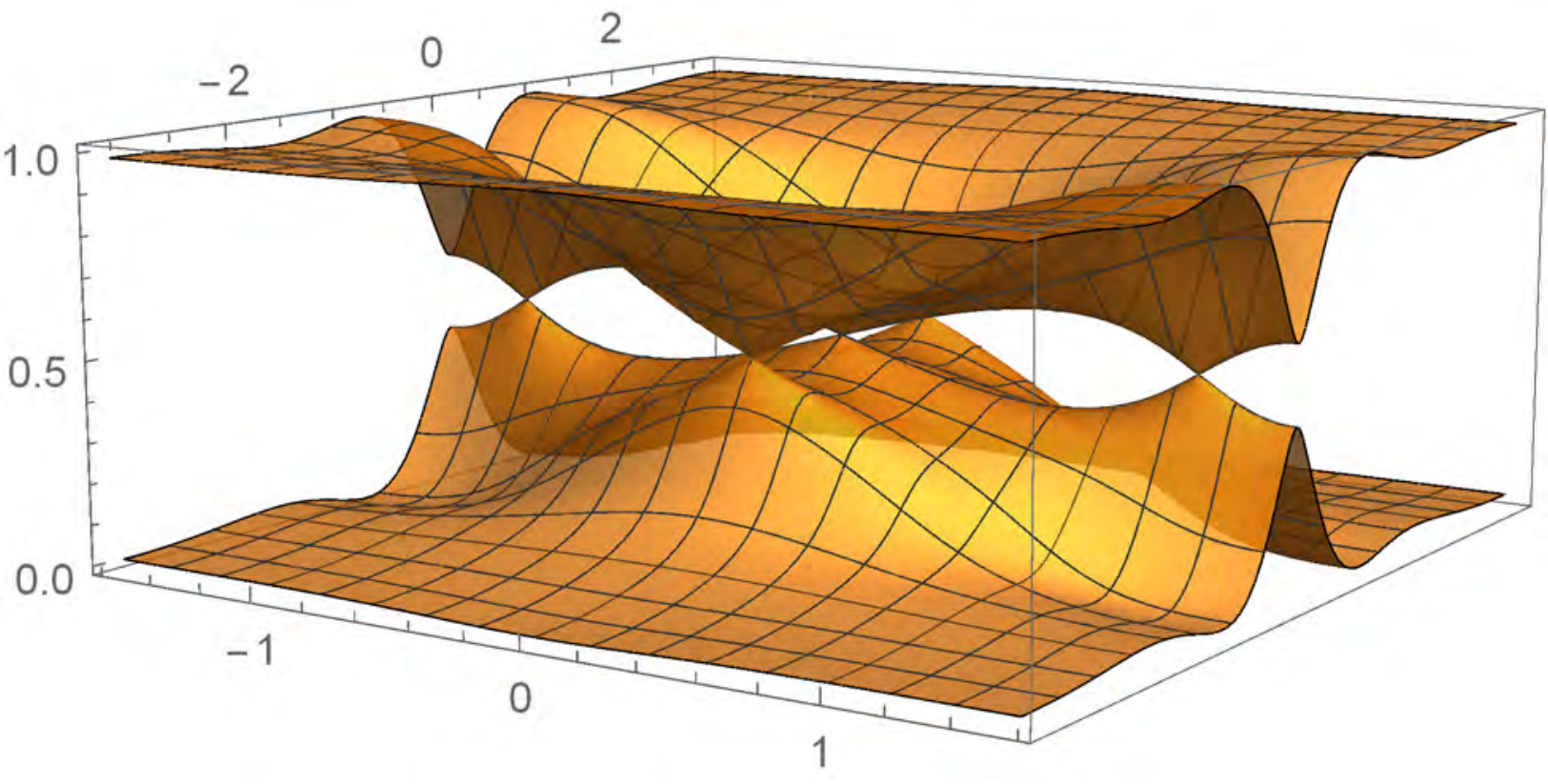}
  \end{minipage}

  \vspace{0.65cm}

  \begin{minipage}[b]{0.2\textwidth}
    \includegraphics[width=\textwidth]{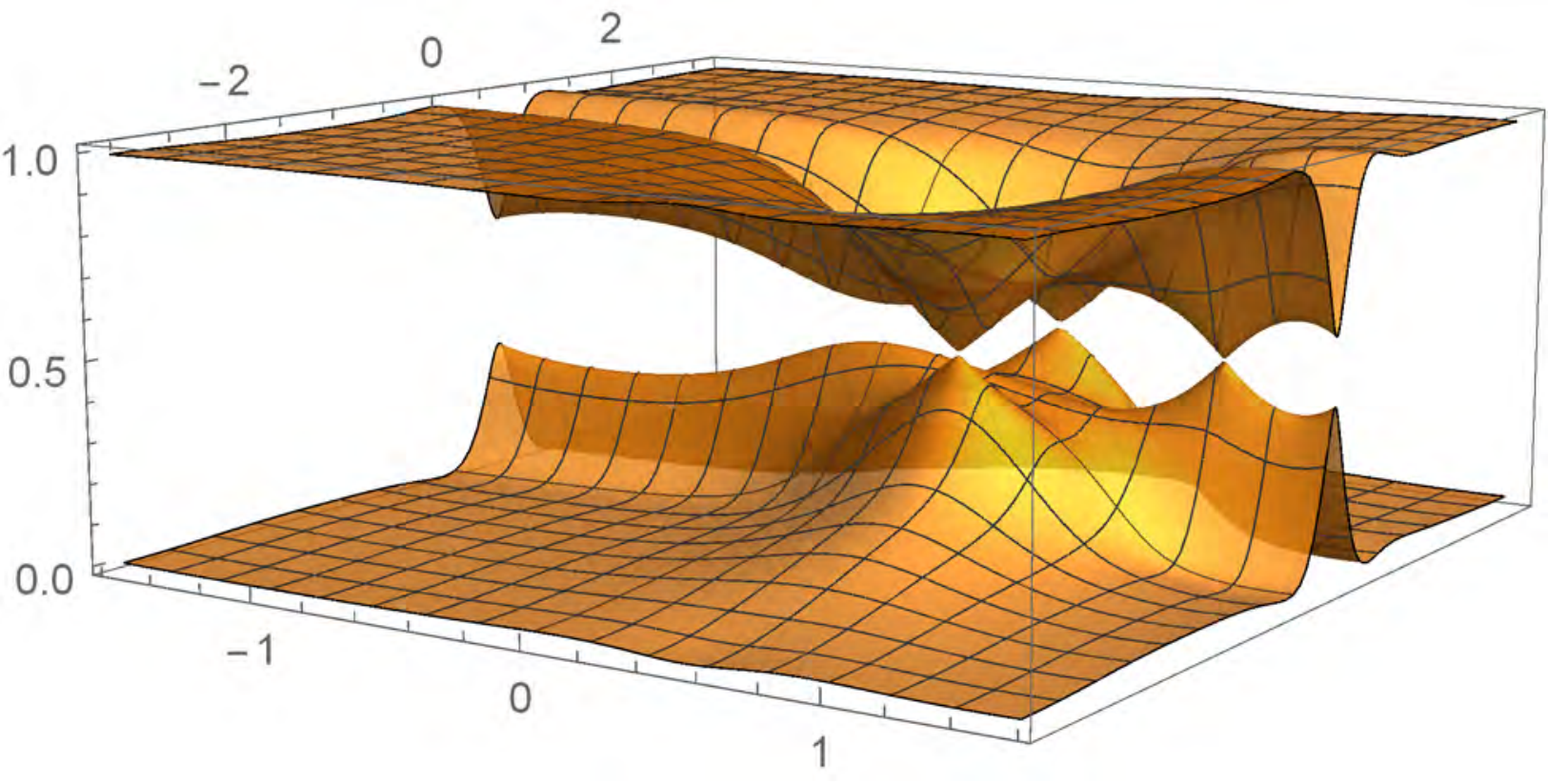}
  \end{minipage}
  \hspace{1cm} 
  \begin{minipage}[b]{0.2\textwidth}
    \includegraphics[width=\textwidth]{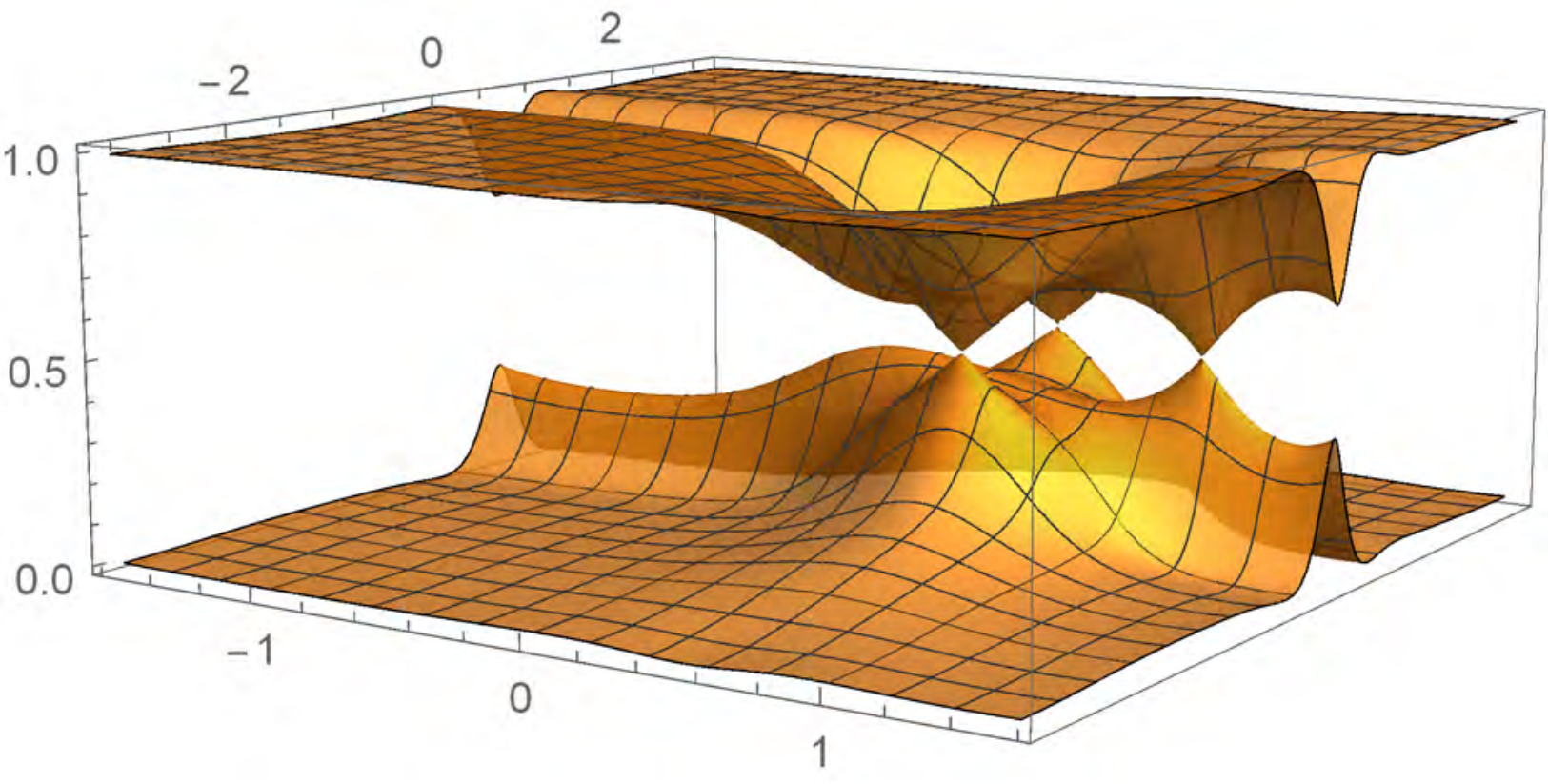}
  \end{minipage}
  \hspace{1cm} 
  \begin{minipage}[b]{0.2\textwidth}
    \includegraphics[width=\textwidth]{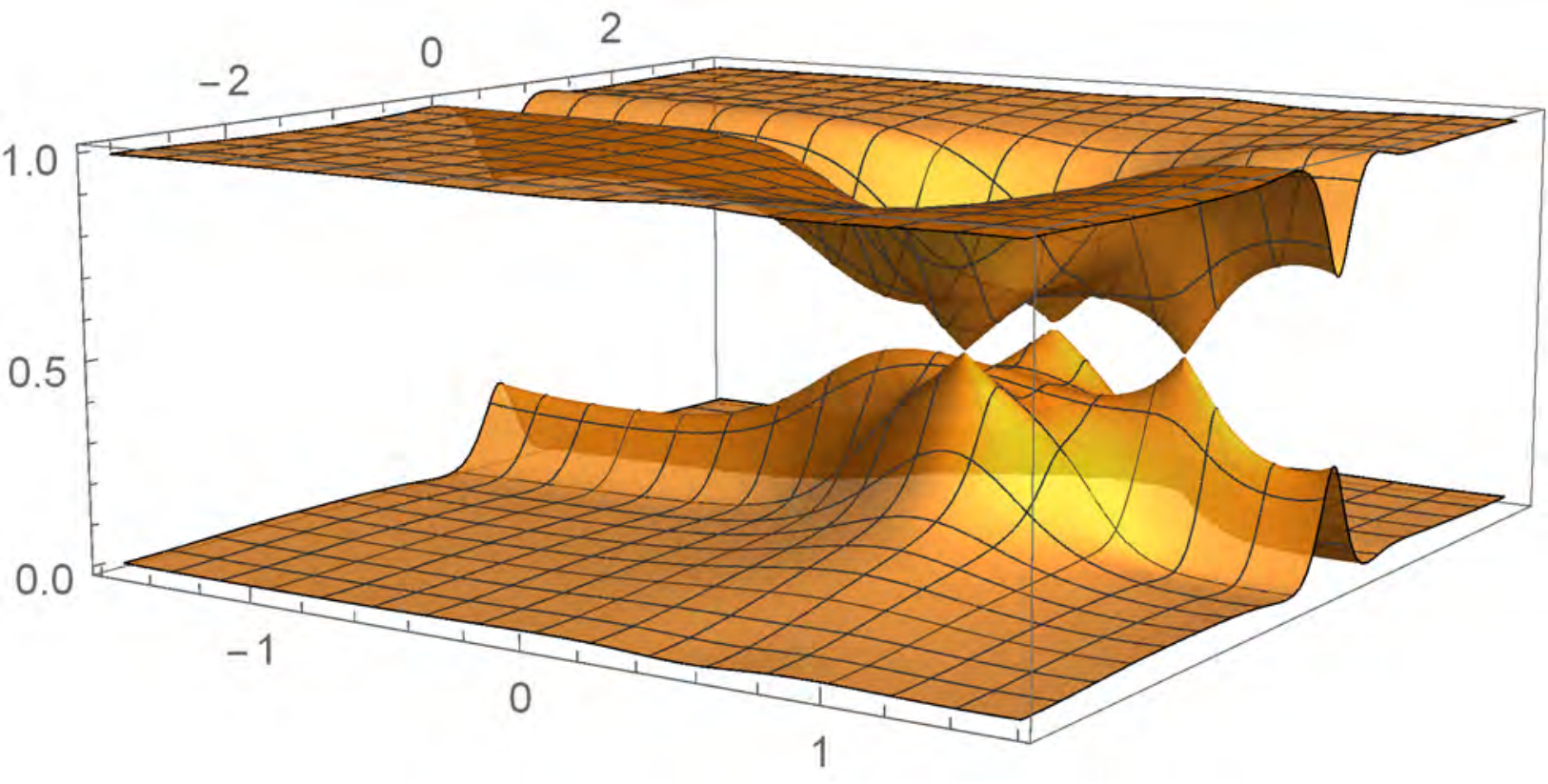}
  \end{minipage}

\caption{BES for $\bf CI_{3c}$ for $-4<\mu<0$ ($\mathtt{C}=3$). Left: $\mu=-1.8$. Middle: $\mu=-2$. Right: $\mu=-2.2$. Top: without deformation. Bottom: with deformation $a_x=b_x=1$, $a_y=b_y=1$.}\label{fig-CI3c-1}
\end{figure*}

For $-4<\mu<-2$ ($\mathtt{C}=3$) there are three Dirac nodal points at $\mathbf{q}_\mathrm{O}$, $\mathbf{q}_\mathrm{N}$, and $\mathbf{q}_\mathrm{S}$, respectively.  They remain robust under deformation \eq{deformk}.  See the left column of Fig.\ \ref{fig-CI3c-1}.

For $\mu=-2$ (still $\mathtt{C}=3$), there are four Dirac nodal points at $\mathbf{q}_\mathrm{O}$, $\mathbf{q}_\mathrm{N}$  $\mathbf{q}_\mathrm{S}$, $\mathbf{q}_\mathrm{E}=\mathbf{q}_\mathrm{W}=(\pm\pi/2,0)$ in the BES before deformation is imposed. In contrast, there are five Dirac points specified by \eq{5 nodal-1} in the BES for $-2<\mu<0$ (still $\mathtt{C}=3$) .  When we impose the deformation, only three Dirac points remain in both cases. See  the middle and right columns respectively of Fig.\ \ref{fig-CI3c-1}.

In the $\mathtt{C}=1$ phase, there are five different points in the $\bf q$-space on which the band crossings may occur, and thus \eq{not master} is violated.
\begin{eqnarray}
&&\mathbf{q}_\mathbf{A} := (0,\pm\pi)\;, \quad
\mathbf{q}_\mathrm{U,D} := \left(0, \pm \pi \mp \cos^{-1}\frac{\mu}{4}\right), \nn
\\ \label{5 nodal-2}
&&\mathbf{q}_\mathrm{R,L} := \left(\pm \cos^{-1}\sqrt{\frac{-2+\mu}{2}},\pm\pi\right)\;.
\end{eqnarray}

\begin{figure*}

\centering

  \begin{minipage}[b]{0.2\textwidth}
    \includegraphics[width=\textwidth]{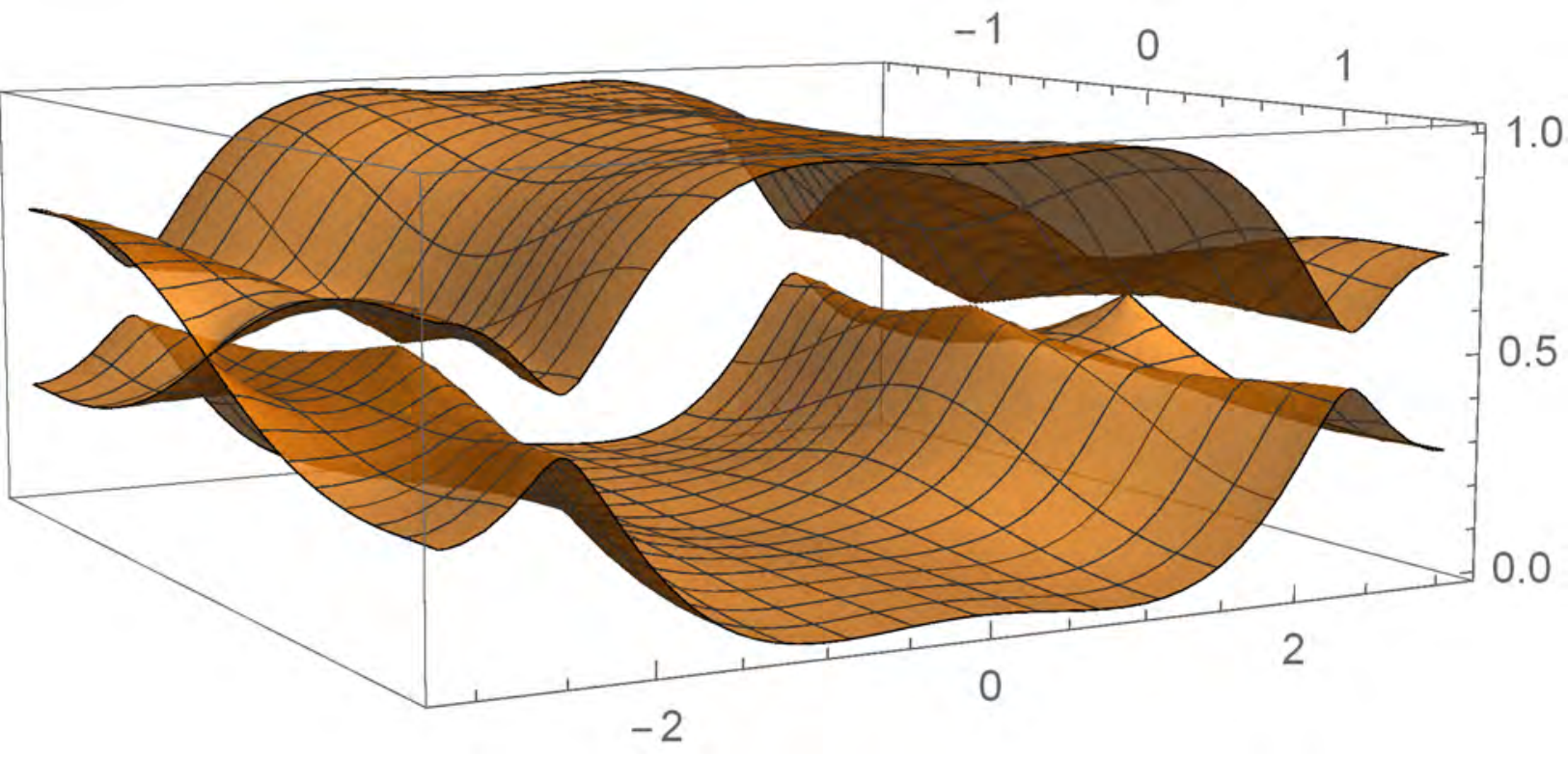}
  \end{minipage}
  \hspace{1cm} 
  \begin{minipage}[b]{0.2\textwidth}
    \includegraphics[width=\textwidth]{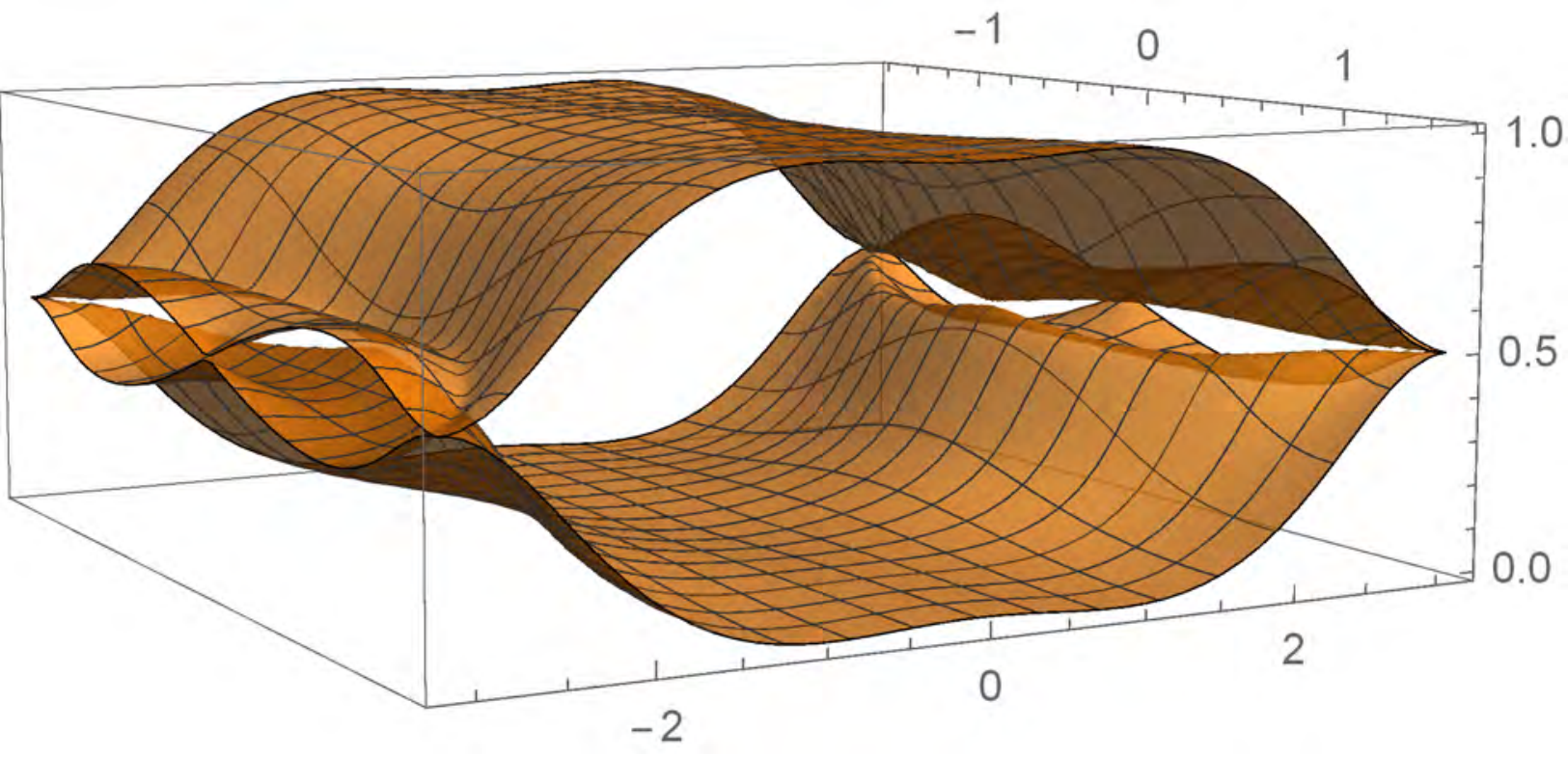}
  \end{minipage}
  \hspace{1cm} 
  \begin{minipage}[b]{0.2\textwidth}
    \includegraphics[width=\textwidth]{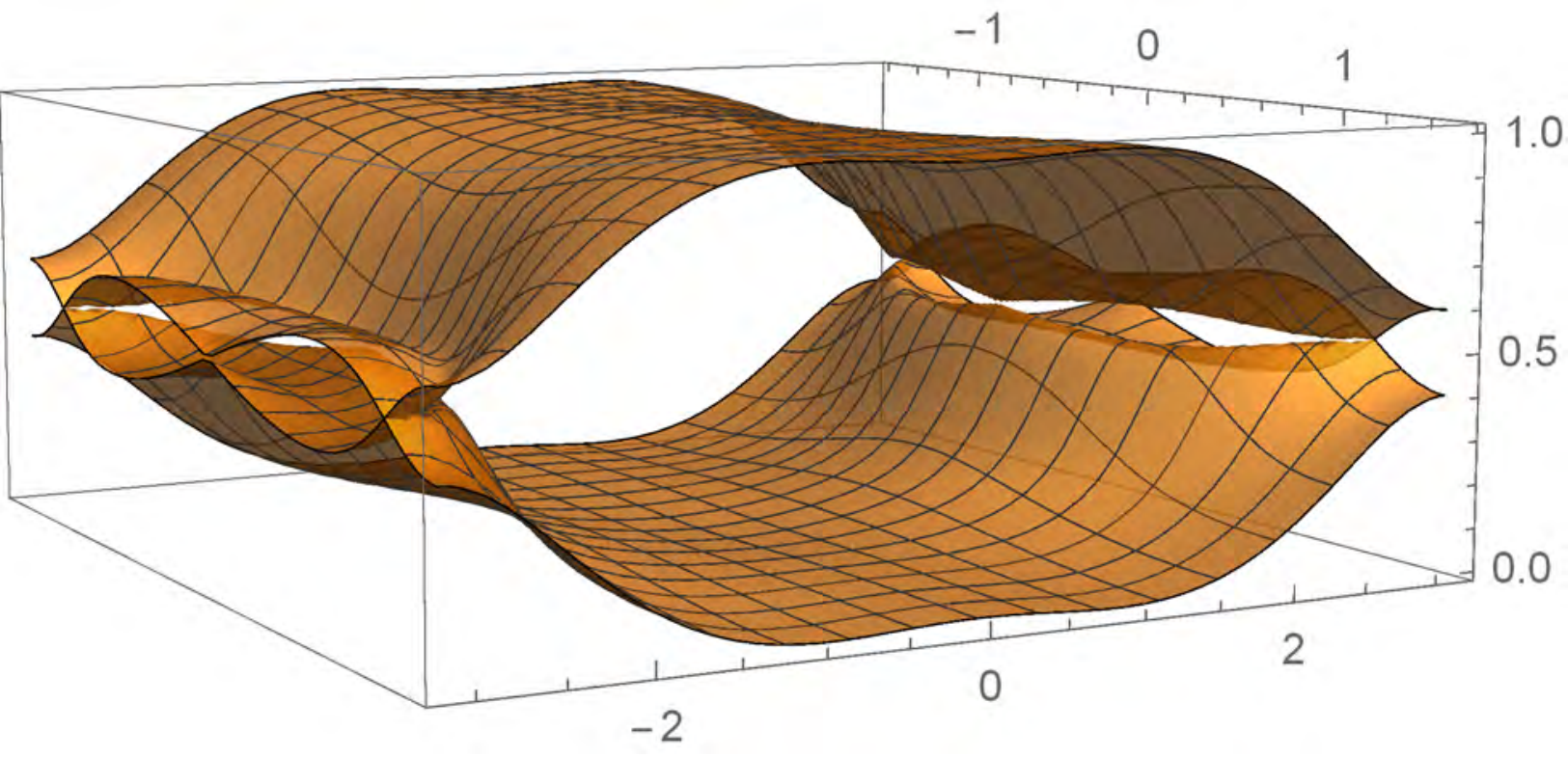}
  \end{minipage}

  \vspace{0.65cm}

  \begin{minipage}[b]{0.2\textwidth}
    \includegraphics[width=\textwidth]{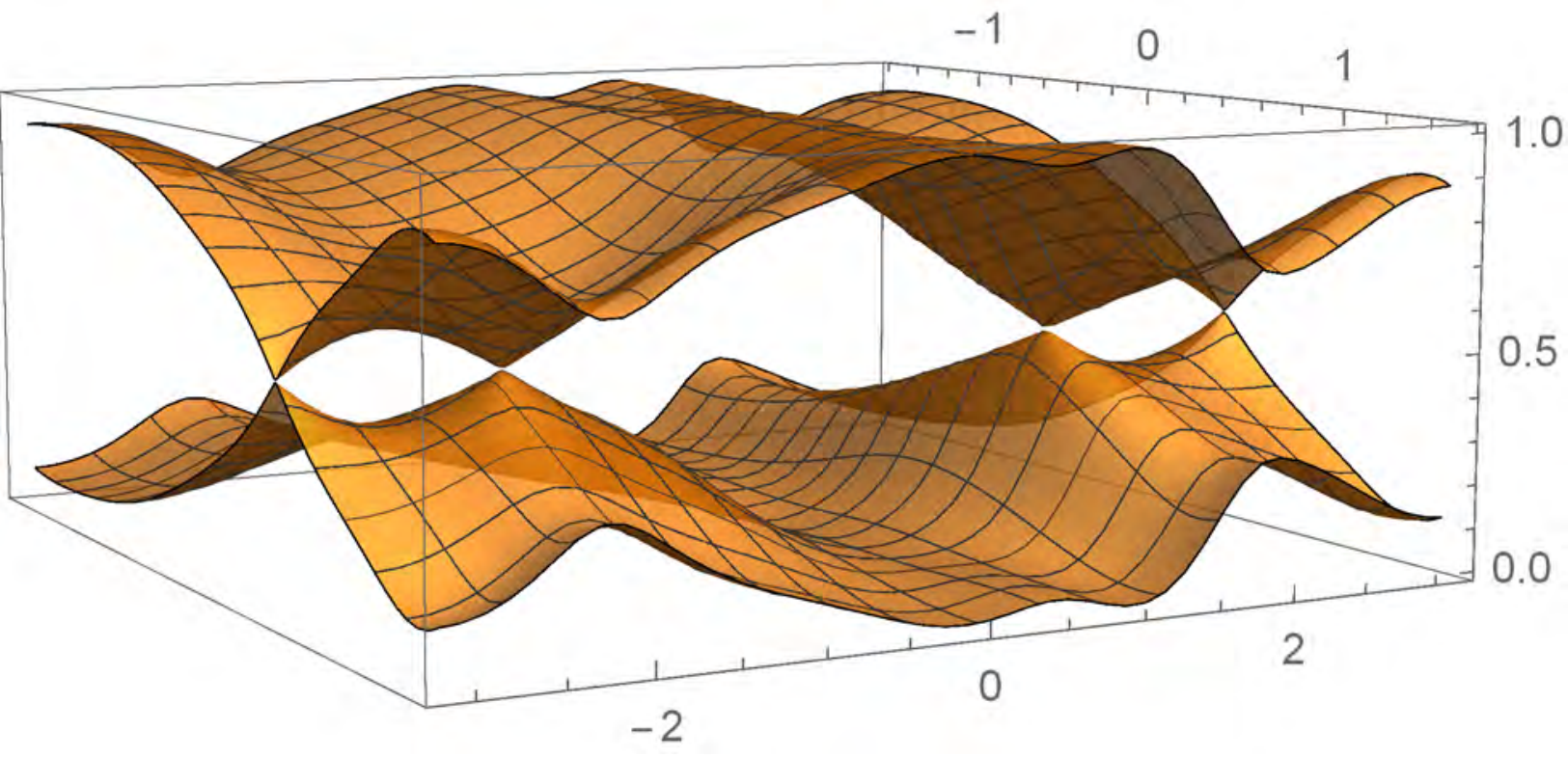}
  \end{minipage}
  \hspace{1cm} 
  \begin{minipage}[b]{0.2\textwidth}
    \includegraphics[width=\textwidth]{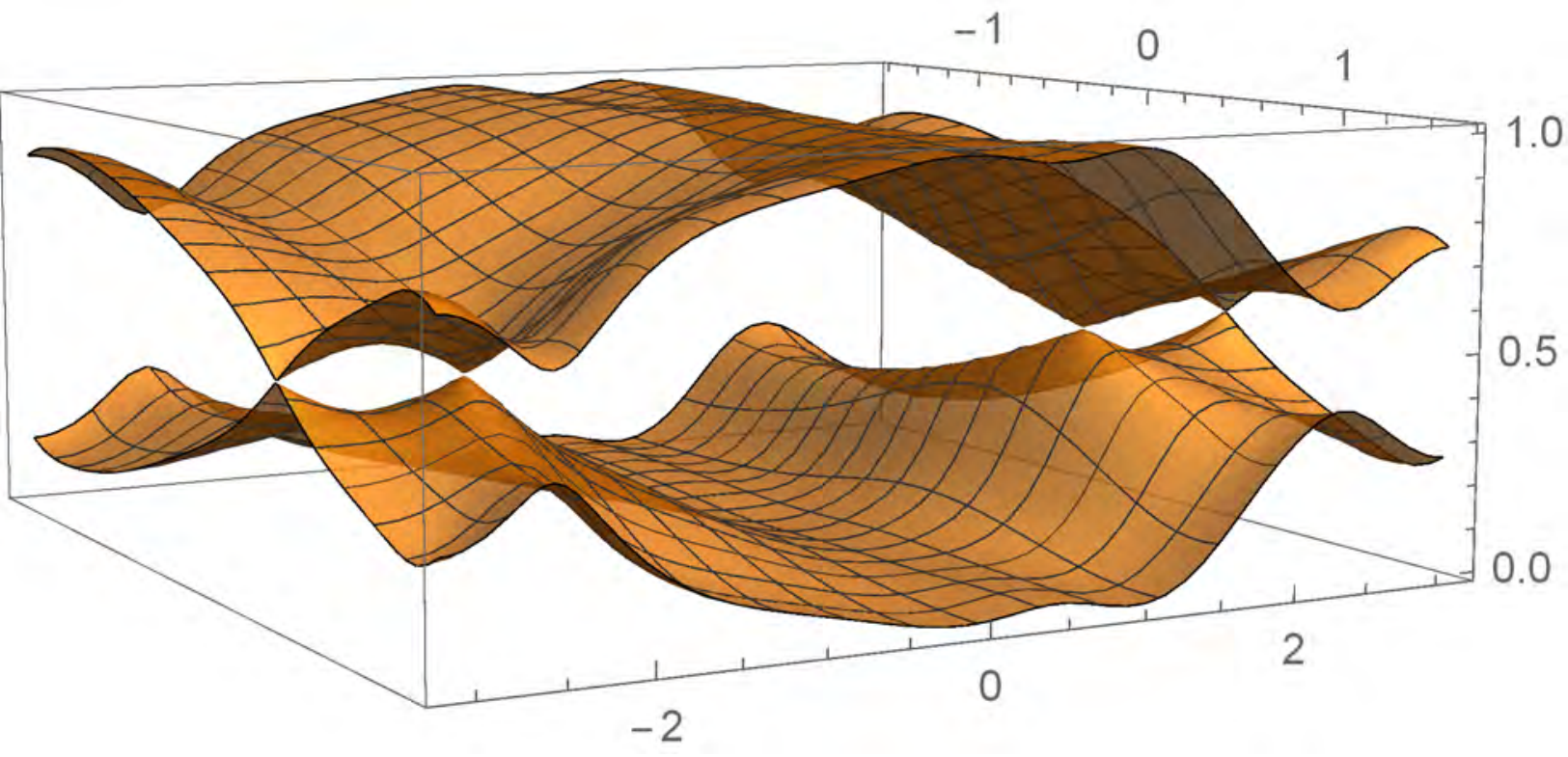}
  \end{minipage}
  \hspace{1cm} 
  \begin{minipage}[b]{0.2\textwidth}
    \includegraphics[width=\textwidth]{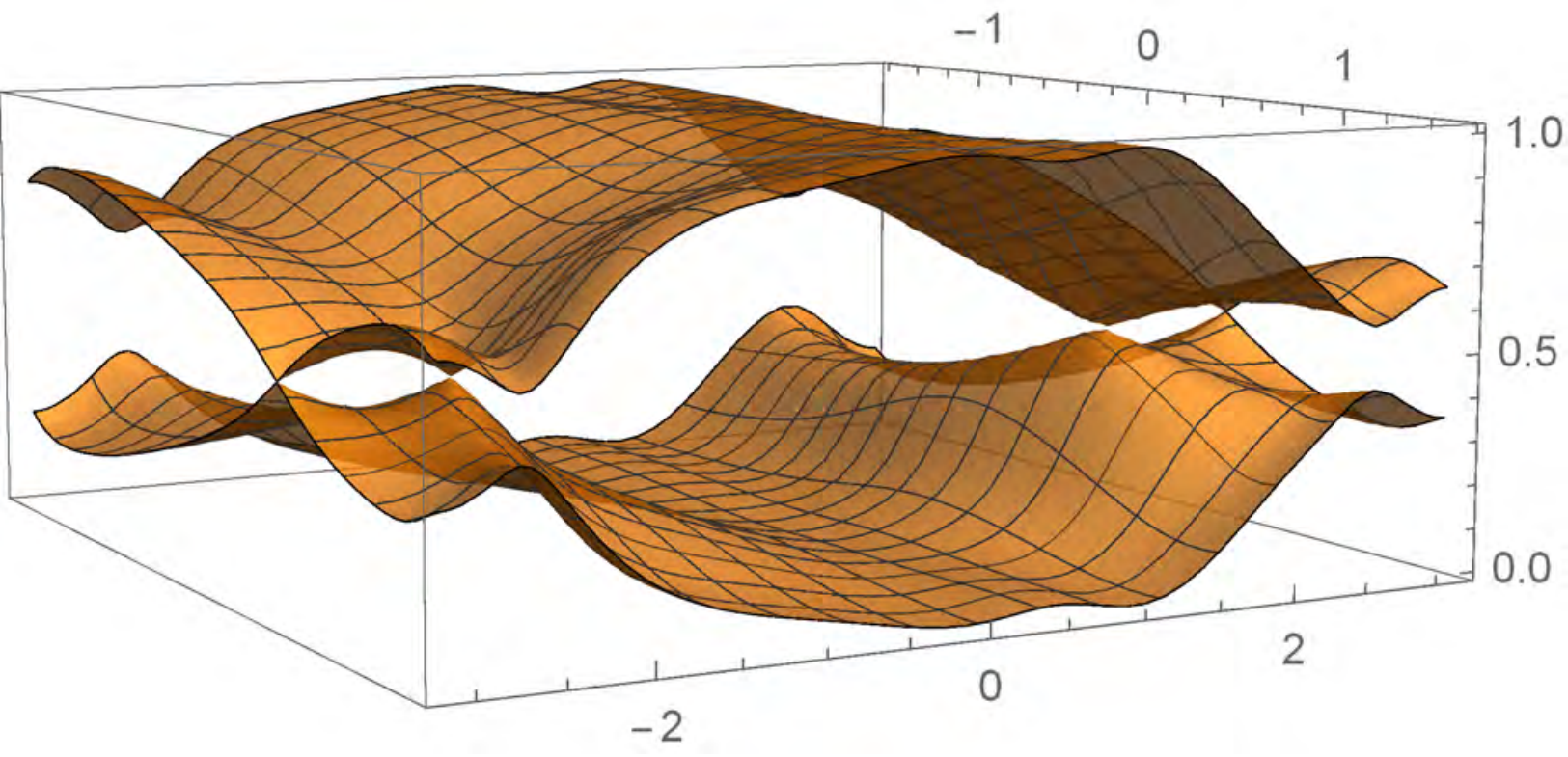}
  \end{minipage}

\caption{BES for $\bf CI_{3c}$ for $0<\mu<4$ ($\mathtt{C}=1$). Left: $\mu=1.5$. Middle: $\mu=2$. Right: $\mu=2.2$. Top: without deformation. Bottom: with deformation $a_x=b_x=1$, $a_y=b_y=1$.}\label{fig-CI3c-2}
\end{figure*}

For $2<\mu<4$ ($\mathtt{C}=1$), the BES has three Dirac nodal points at $\mathbf{q}_\mathrm{A}$, $\mathbf{q}_\mathrm{N}$, and $\mathbf{q}_\mathrm{S}$, which are robust against deformation \eq{deformk}.   See the left column of Fig.\ \ref{fig-CI3c-2}.

For $\mu=2$ (still $\mathtt{C}=1$), the BES has four nodal points at $\mathbf{q}_\mathrm{A}$, $\mathbf{q}_\mathrm{U}$, $\mathbf{q}_\mathrm{D}$, and $\mathbf{q}_\mathrm{R}\equiv\mathbf{q}_\mathrm{L}=(\pm\pi/2,\pi)$. With deformation \eq{deformk} imposed, the band crossing at $\mathbf{q}_\mathrm{R}\equiv\mathbf{q}_\mathrm{L}$ is lifted while the other three Dirac points remain robust although their locations are shifted.    See the middle column of Fig.\ \ref{fig-CI3c-2}.

For $0<\mu<2$ (still $\mathtt{C}=1$), the BES has five nodal points at \eq{5 nodal-2}. Similarly, the band crossings at $\mathbf{q}_\mathrm{R}$ and $\mathbf{q}_\mathrm{L}$ are lifted and the other three Dirac points remain robust under deformation.  See the right column of Fig.\ \ref{fig-CI3c-2}.

It is obvious that here the number of robust nodal points in the BES no longer matches the Chern number of the system, and thus eq.\ \eq{not master} does not apply in general.  Therefore, in addition to counting the order of the nodal points, we must also introduce the notion of vorticity to describe them properly. We will defer the detailed discussion about the vorticity of a nodal point to the next section. Suffice to say, with this modification we obtain the relation in \eq{master rel}, which always holds for all the cases considered in this work.

We now switch to the checkerboard partition $\bf CI_{3b}$. In this case, there is an additional x-y symmetry compared to the stripe partition.  First, we note again that there is no band crossing in the trivial phase $\mathtt{C}=0$ ($|\mu|>4$).

\begin{figure*}

\centering

  \begin{minipage}[b]{0.2\textwidth}
    \includegraphics[width=\textwidth]{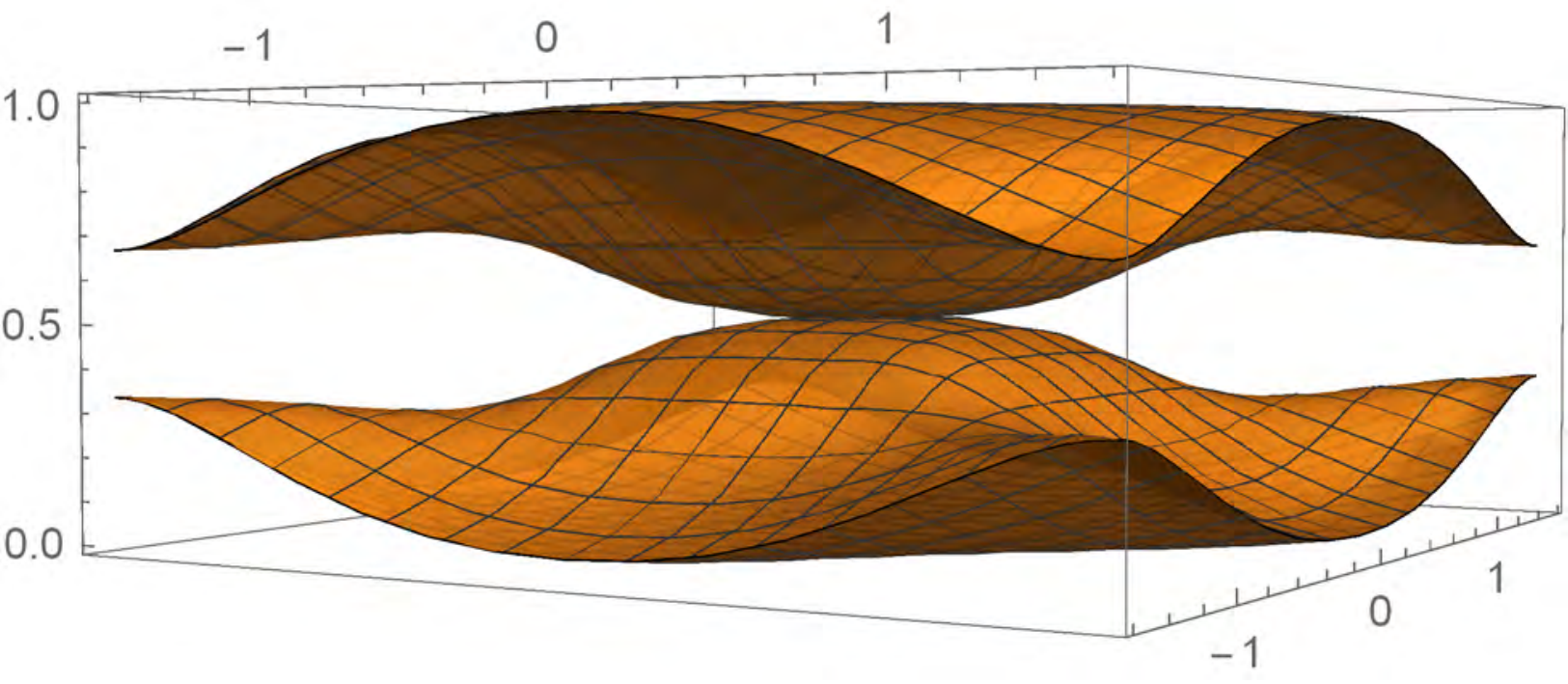}
  \end{minipage}
  \hspace{1cm} 
  \begin{minipage}[b]{0.2\textwidth}
    \includegraphics[width=\textwidth]{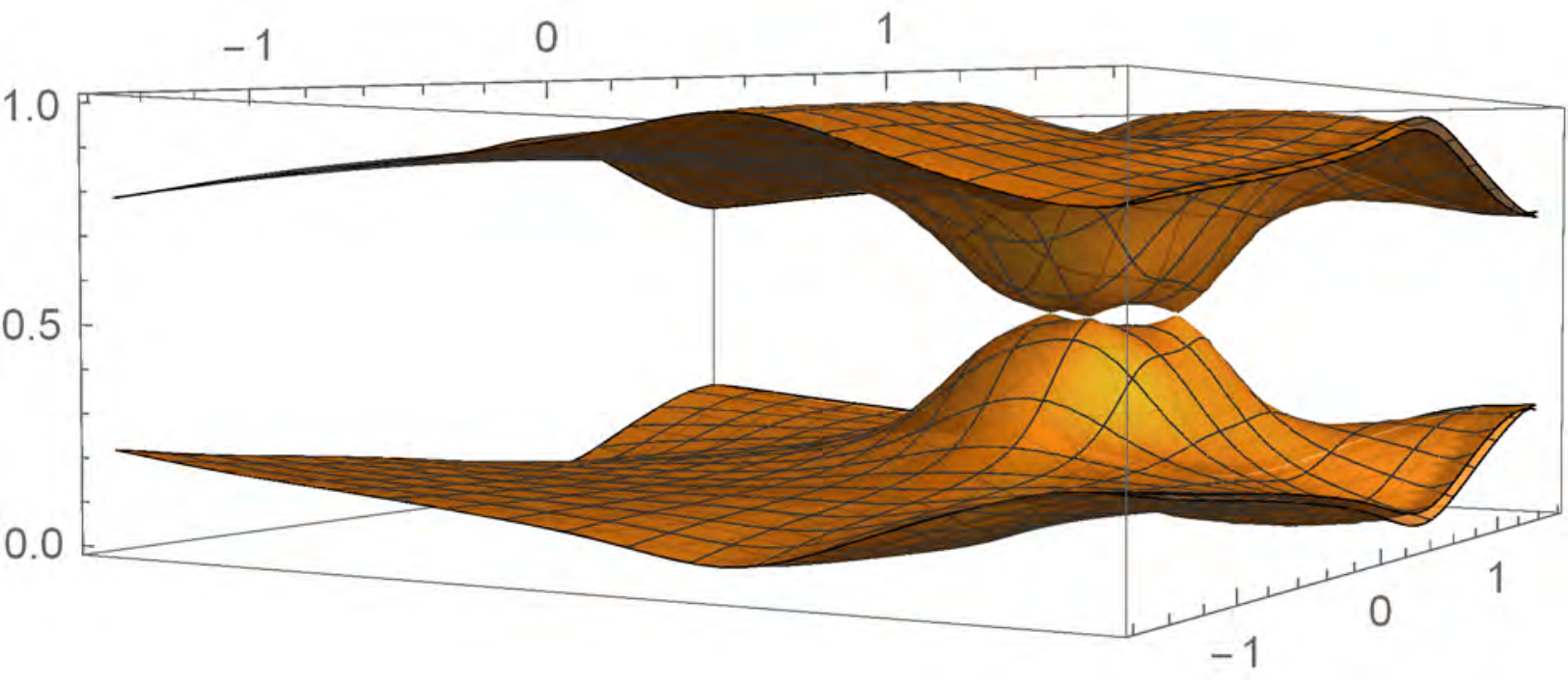}
  \end{minipage}
  \hspace{1cm} 
  \begin{minipage}[b]{0.2\textwidth}
    \includegraphics[width=\textwidth]{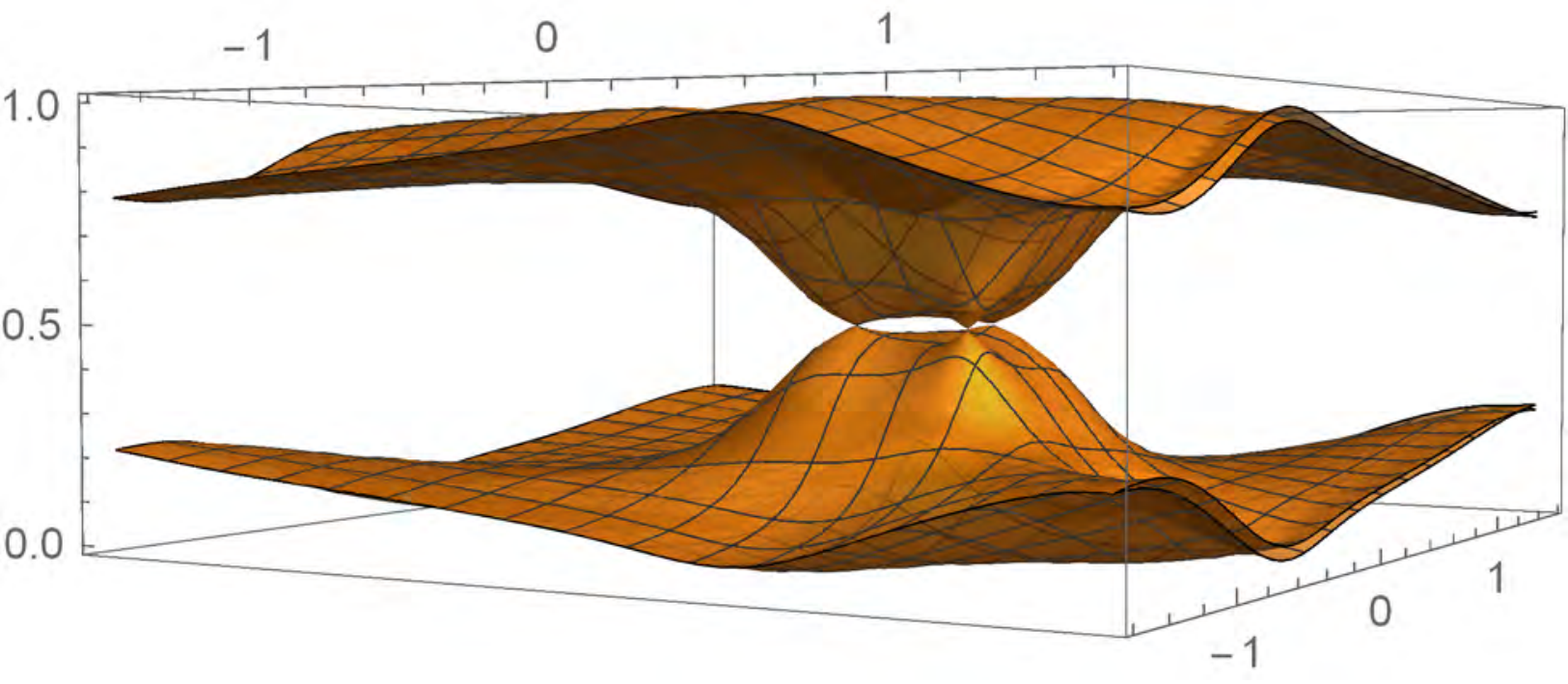}
  \end{minipage}

\caption{BES for $\bf CI_{3b}$ for $-4<\mu<0$ ($\mathtt{C}=3$). Left: without deformation. Middle: with deformation $a_x=b_x=1$, $a_y=b_y=1$. Right:  with deformation $a_x=b_x=1$, $a_y=1$, $b_y=-1$.}\label{fig-CI3b-1}
\end{figure*}

For the topological phase with $\mathtt{C}=3$ ($-4<\mu<0$), the BES has a cubic nodal point at ${\bf q}=(0,0)$, see Appendix \ref{Appendix-dispersion} for the details of checking the cubic dispersion relation. Under deformation, the cubic nodal points splits into three Dirac nodal points, see Fig.\ \ref{fig-CI3b-1}.

\begin{figure*}

\centering

  \begin{minipage}[b]{0.2\textwidth}
    \includegraphics[width=\textwidth]{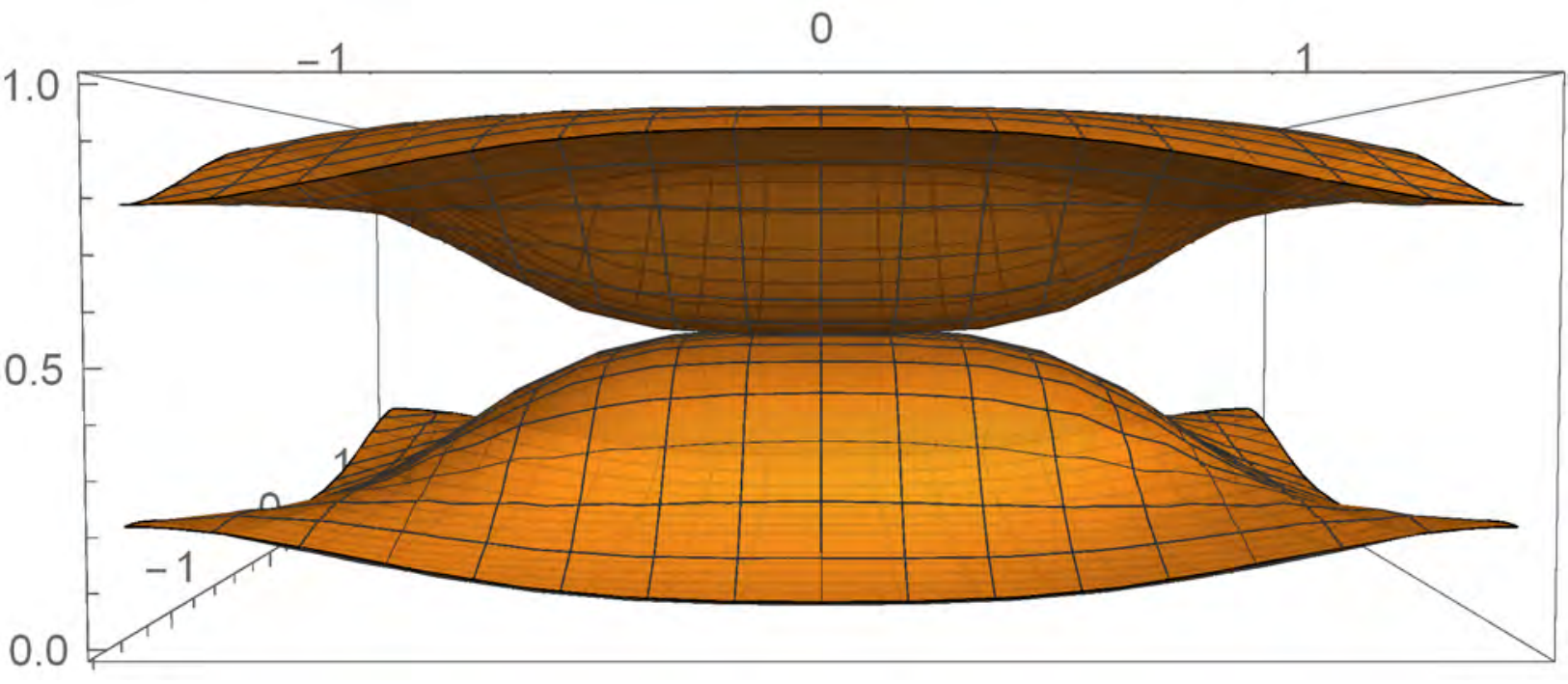}
  \end{minipage}
  \hspace{1cm} 
  \begin{minipage}[b]{0.2\textwidth}
    \includegraphics[width=\textwidth]{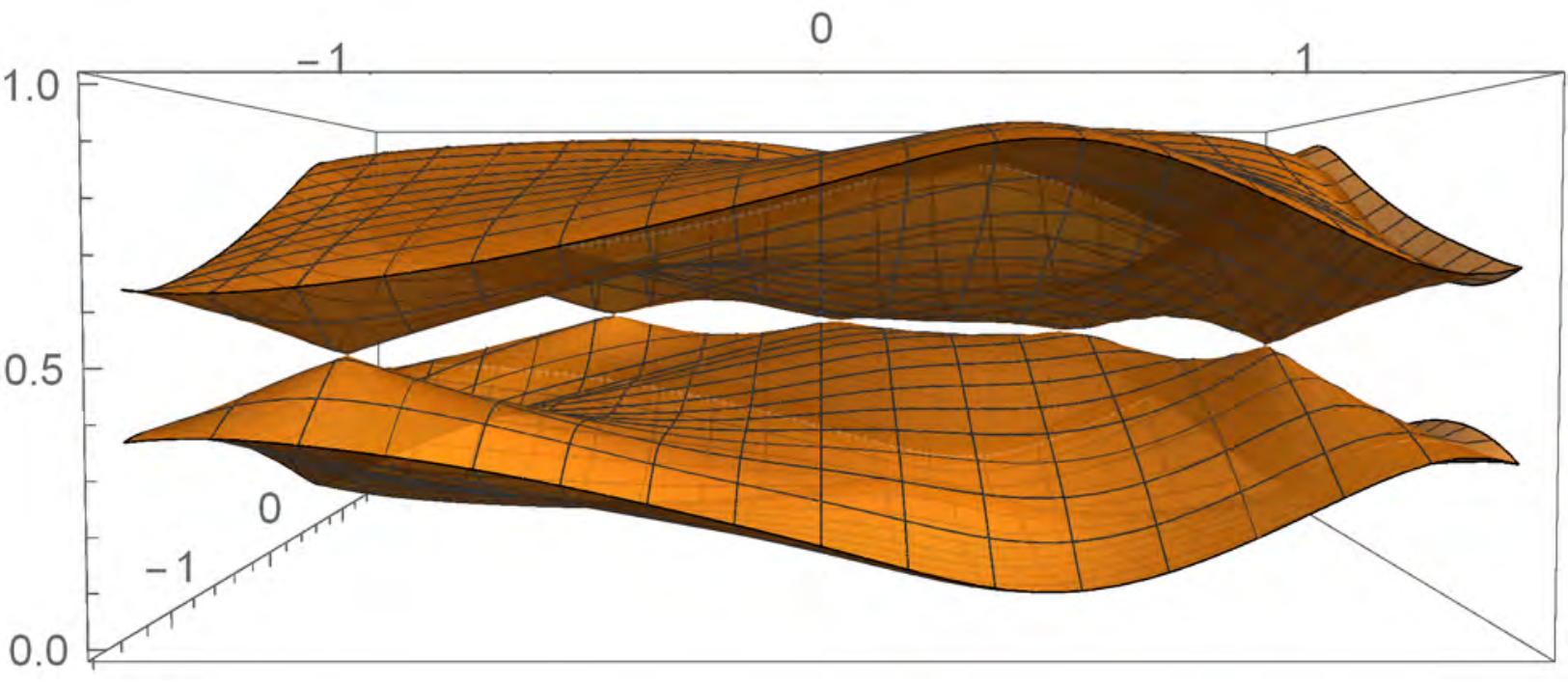}
  \end{minipage}
  \hspace{1cm} 
  \begin{minipage}[b]{0.2\textwidth}
    \includegraphics[width=\textwidth]{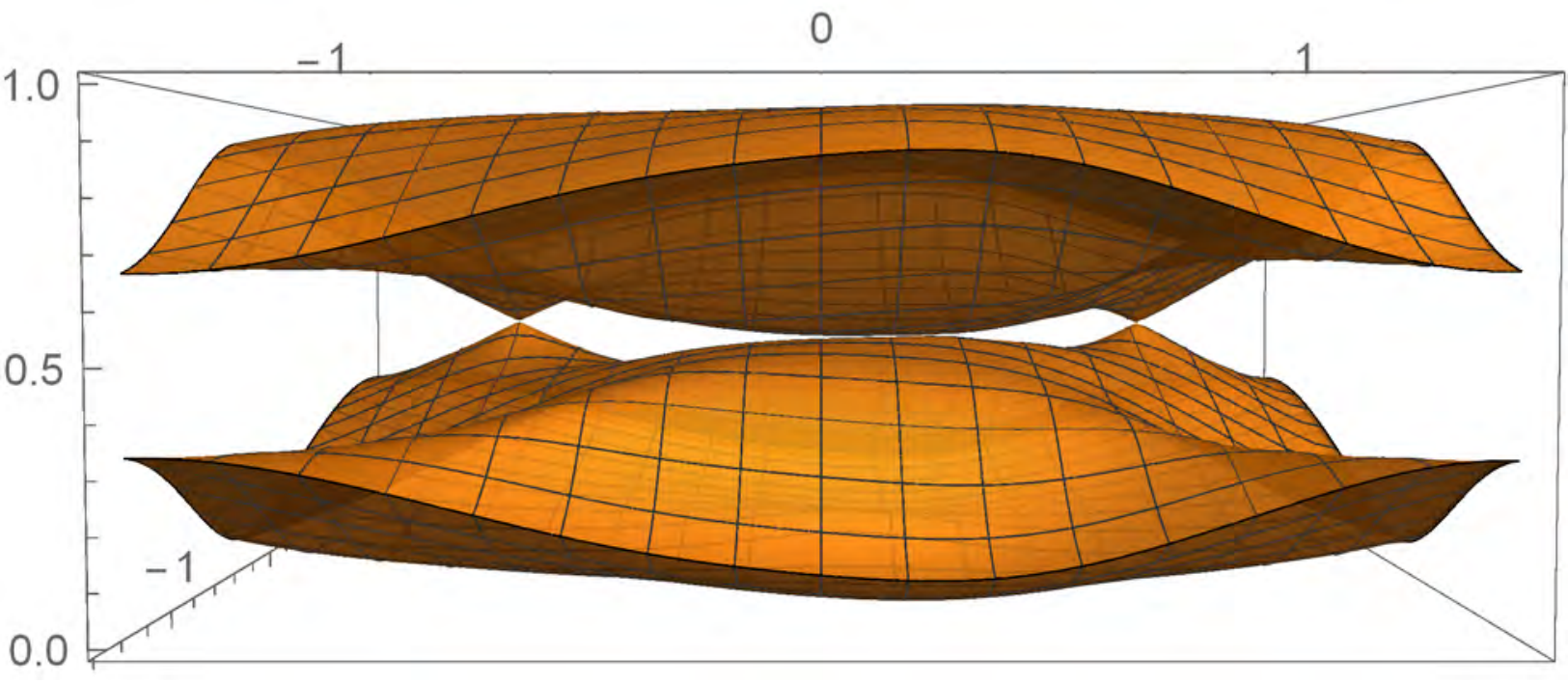}
  \end{minipage}
\caption{BES for $\bf CI_{3b}$ for $0<\mu<4$ ($\mathtt{C}=1$). Left: without deformation. Middle: with deformation $a_x=b_x=1$, $a_y=b_y=1$. Right: with deformation $a_x=b_x=0.5$, $a_y=1$, $b_y=-1$}\label{fig-CI3b-2}
\end{figure*}

For the topological phase with $\mathtt{C}=1$ ($0<\mu<4$), the BES has a cubic nodal point at ${\bf q}=(0,0)$, see Appendix \ref{Appendix-dispersion} for the verification of this. Depending on how much deformation is imposed, it either splits into three nodal points, among which there is a point with vorticity 3 and two points with vorticity -1, or it splits into five Dirac points, among which there are three points with vorticity 1 and two with vorticity -1.  See Fig.\ \ref{fig-CI3b-2}.

\subsubsection{$\bf CI_{4b}$ and $\bf CI_{4c}$}

For the model of $\bf CI_4$ we will only focus on the $\mathtt{C}=-2$ phase, since the band crossing patterns are more or less similar to the previous cases we have investigated: in the phases with $\mathtt{C}=\pm 1$ the BES exhibits a robust Dirac nodal point; while in the phase $\mathtt{C}=0$, there is no band crossing.

\begin{figure}
\centering
  \begin{minipage}[b]{0.2\textwidth}
    \includegraphics[width=\textwidth]{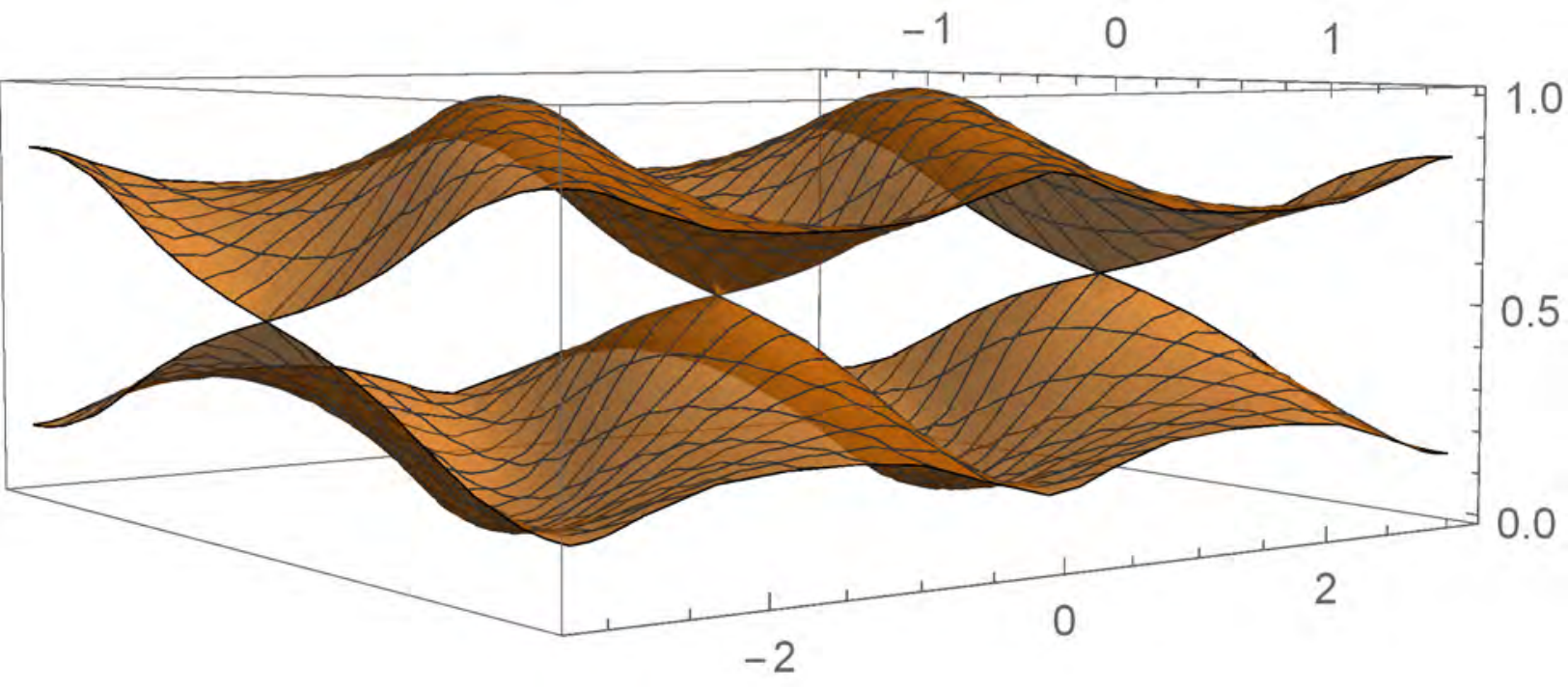}
  \end{minipage}
  \hspace{0.7cm} 
  \begin{minipage}[b]{0.2\textwidth}
    \includegraphics[width=\textwidth]{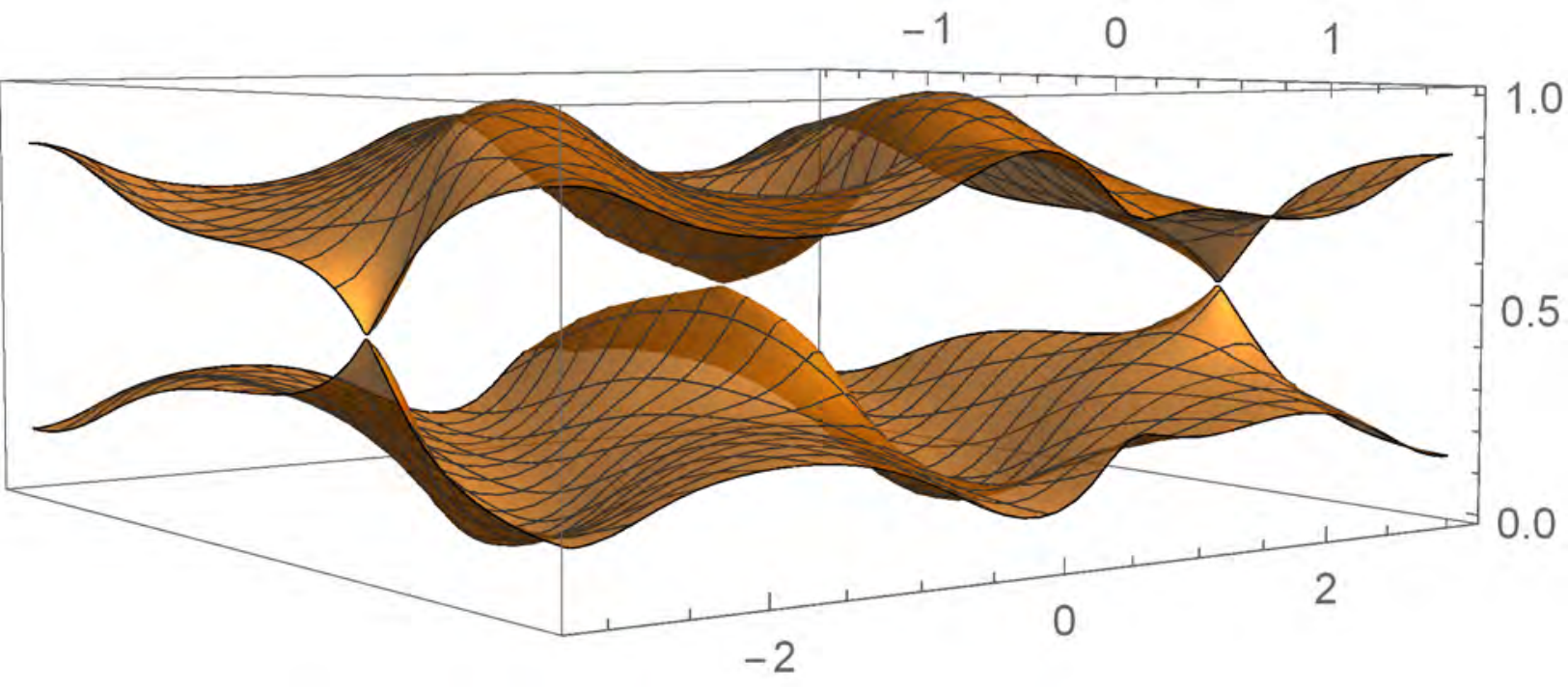}
  \end{minipage}
\caption{BES for $\bf CI_{4c}$ for $\mu=-1$, $t=0$ ($\mathtt{C}=-2$). Left:  without deformation. Right: with deformation $a_x=b_x=1=a_y=b_y=1$.}\label{fig-CI4c}
\end{figure}

We first consider the stripe partition $\bf CI_{4c}$. The BES shows two Dirac nodal points at ${\bf q}=(0,0), (0, \pi)$, which remain robust but their locations are shifted under deformation.  See Fig.\ \ref{fig-CI4c}.

\begin{figure}

\centering

  \begin{minipage}[b]{0.2\textwidth}
    \includegraphics[width=\textwidth]{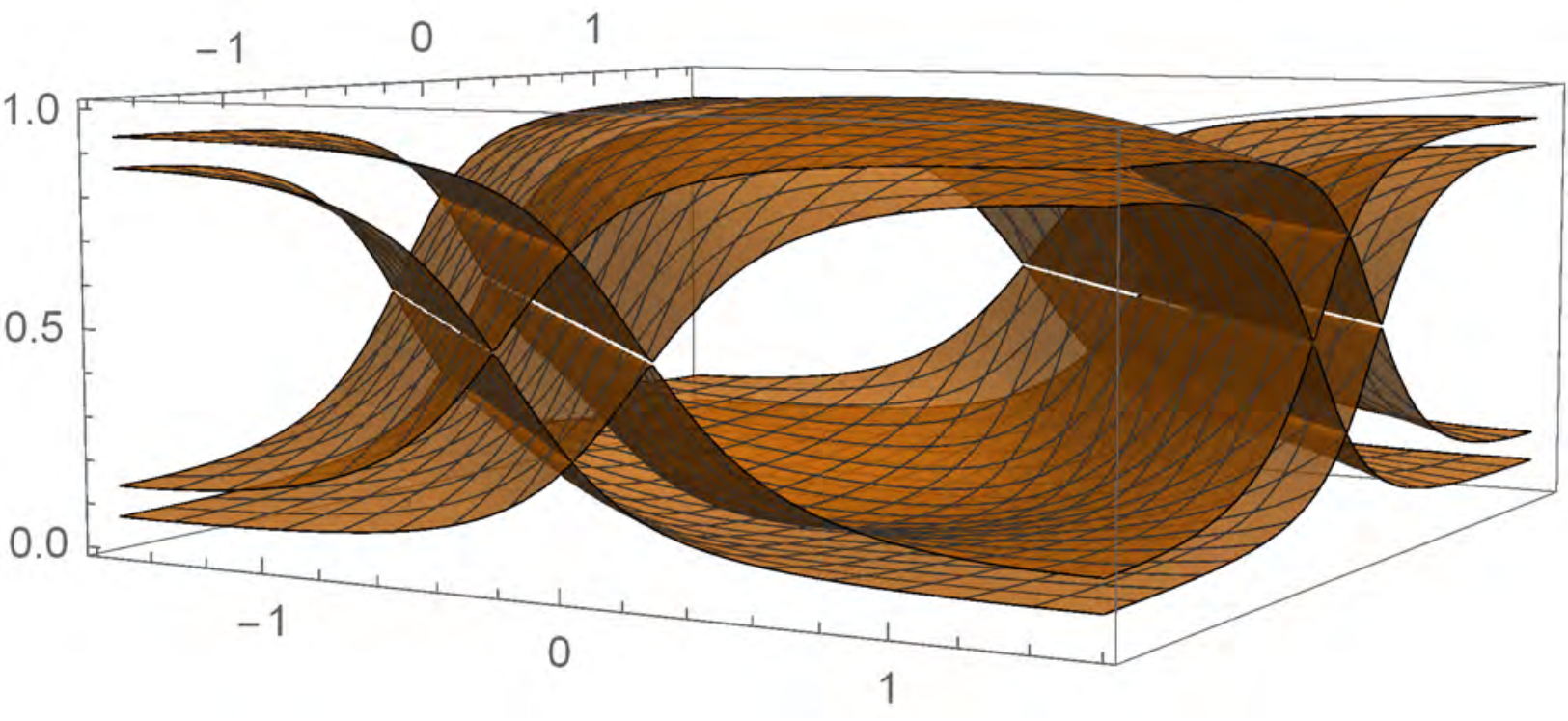}
  \end{minipage}
  \hspace{0.7cm} 
  \begin{minipage}[b]{0.2\textwidth}
    \includegraphics[width=\textwidth]{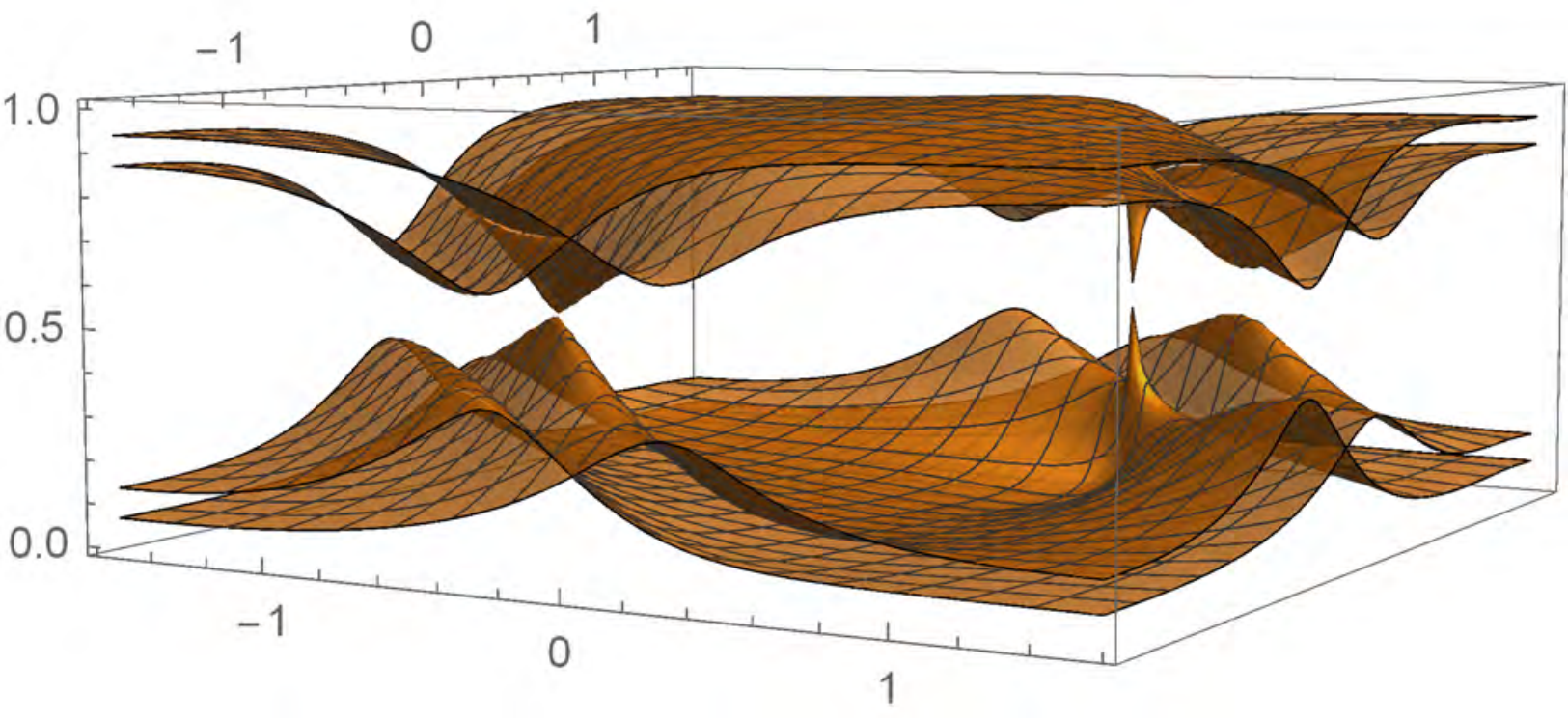}
  \end{minipage}

  \vspace{0.65cm}

  \begin{minipage}[b]{0.2\textwidth}
    \includegraphics[width=\textwidth]{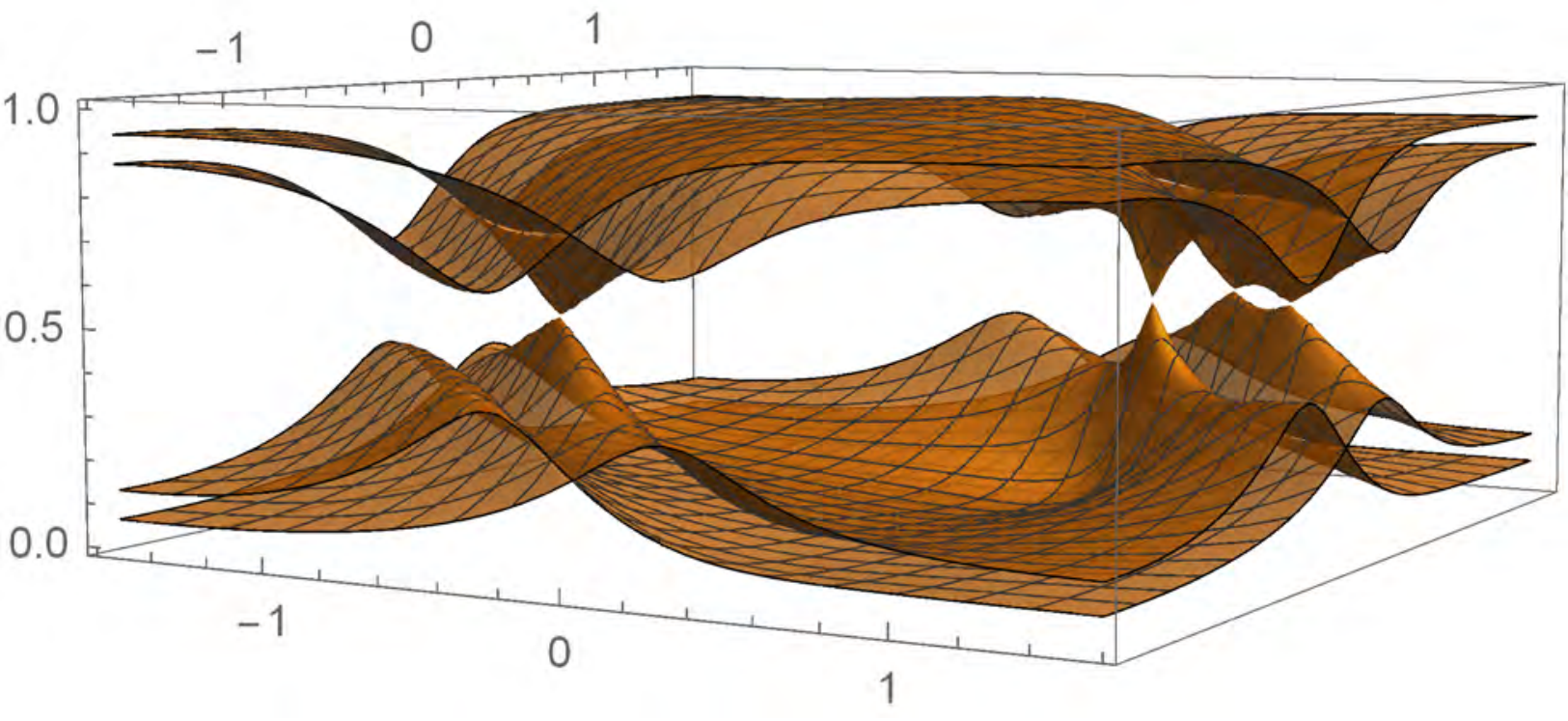}
  \end{minipage}
  \hspace{0.7cm} 
  \begin{minipage}[b]{0.2\textwidth}
    \includegraphics[width=\textwidth]{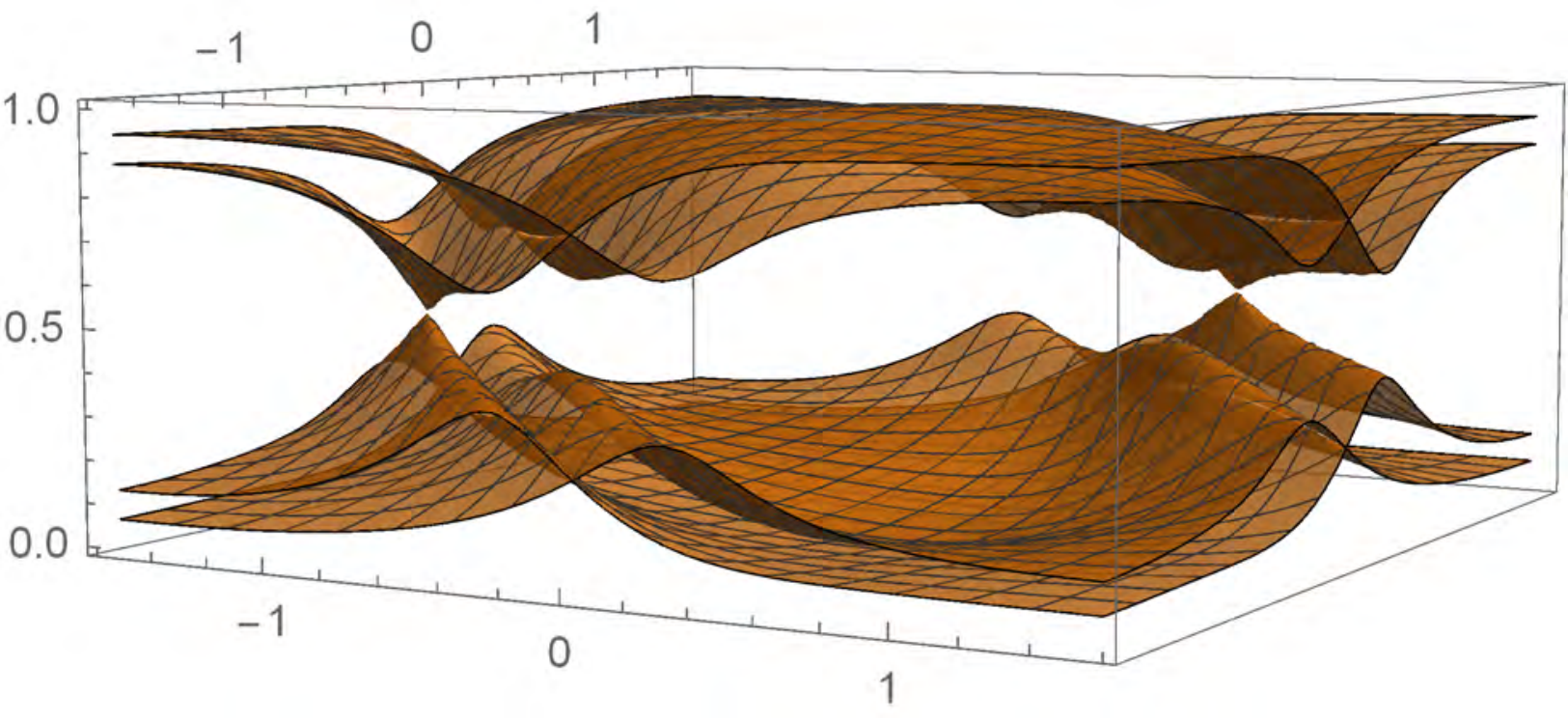}
  \end{minipage}

\caption{BES for $\bf CI_{4b}$ for $\mu=0.5$, $t=0$ ($\mathtt{C}=-2$). Upper left: without deformation.  Upper right: with deformation $a_x=b_x=1=a_y=1$, $b_y=1.5$. Lower left: with deformation $a_x=b_x=1=a_y=b_y=1$. Lower right: with deformation $a_x=b_x=1=a_y=1$, $b_y=-1$}\label{fig-CI4b}
\end{figure}

On the other hand, for the checkerboard partition $\bf CI_{4b}$, there are four nodal lines in the BES:
\be\label{nodal lines in CI4b}
q_x+q_y=\pm\cos^{-1}\frac{\mu}{2} \; \; \mbox{or} \; \; \pm\left(\pi-\cos^{-1}\frac{\mu}{2}\right),
\ee
which merge into two lines at $\mu=0$.  Under deformation, the four nodal lines reduce to four Dirac points, two Dirac points, or other more complicated configurations. See Fig.\ \ref{fig-CI4b}. Again, eq.\ \eq{master rel} holds and the notion of vorticity is important for establishing a relation between the topological characteristics in the BES and the Chern number of the underlying system.

\section{Vorticity of the nodal point and the inherited topology}\label{sec4}

From the previous study, we see nodal points may be classified by their  ``degrees''.  More specifically, they can be a linear (Dirac), quadratic, or a cubic nodal point.  In simpler cases, the sum of the  degrees of all the nodal points is equal to the absolute value of Chern number $\mathtt{C}$ of the underlying Chern insulator, see \eq{not master}.  However, there are more complicated cases for which the previous simple rule does not apply. For example, in the cases shown in Figs.\ \ref{fig-CI3c-2}, \ref{fig-CI3b-2}, and \ref{fig-CI4b}, there are more robust nodal points than $|\mathtt{C}|$, and thus it is impossible for the relation \eq{not master} to hold.  The problem can be resolved if there is a  \emph{signed} integer associated with each nodal point, which we call ``vorticity.''

Instead of trying to prove the relation \eq{master rel}, here we will just give some supporting evidence. For this purpose, we only consider the exactly solvable cases.  In partition \textbf{c} (stripe), the eigenfunctions $\psi_{1,2}(\mathbf{q})$ given in \eq{eigenvec c}--\eq{alpha3} are locally well defined everywhere except at the nodal points, where either $\chi_-(\mathbf{q})$ or $\chi_-(\mathbf{q}')$ becomes singular because they are orthogonal to each other. If $\mathbf{q}$ moves along a small circle surrounding a nodal point, the eigenfunction generally obtains a Berry phase when $\mathbf{q}$ comes back to the starting point. The reason is that the complex functions $\sqrt{\alpha_3(\mathbf{q})}$, $\sqrt{\alpha_3(\mathbf{q})^*}$ defined in \eq{alpha3} are multi-valued (if the branch cuts are to be analytically extended by using Riemann surfaces).
In the vicinity of a given nodal point, if we plot the ``phasor'' $\alpha(\mathbf{q})\equiv\abs{\alpha(\mathbf{q})}e^{i\phi(\mathbf{q})}$ as a vector field:
\begin{equation}\label{phasor for partition c}
\left(\mathrm{Re}\{\alpha(\mathbf{q})\},\mathrm{Im}\{\alpha(\mathbf{q})\}\right)
\equiv \abs{\alpha(\mathbf{q})}\left(\cos\phi(\mathbf{q}),\sin\phi(\mathbf{q})\right),
\end{equation}
we can see the corresponding vortex structure. In this way, we may associate each nodal point with an integer, \emph{vorticity}, which is the number of turning-around of the phase $\phi(\mathbf{q})$ as $\mathbf{q}$ traces around the nodal point.

In partition \textbf{b} (checkerboard), the eigenfunctions are $\psi_{1,2}(\mathbf{q})$ and $\psi_{3,4}(\mathbf{q})$ as given in \eq{eigenvec b}--\eq{alpha1 alpha2}. Similarly, the phasor field in the vicinity of a given nodal point can be defined as
\begin{subequations}\label{phasor for partition b}
\begin{eqnarray}
&&\left(\mathrm{Re}\{\alpha_1(\mathbf{q})\},\mathrm{Im}\{\alpha_1(\mathbf{q})\}\right),\nonumber\\
&& \qquad\text{if } \lambda_{1,2}(\mathbf{q})=1/2\ \text{at the nodal point},\quad\\
&&\left(\mathrm{Re}\{\alpha_2(\mathbf{q})\},\mathrm{Im}\{\alpha_2(\mathbf{q})\}\right),\nonumber\\
&& \qquad\text{if } \lambda_{3,4}(\mathbf{q})=1/2\ \text{at the nodal point}.\quad
\end{eqnarray}
\end{subequations}
The phasor field will again give the vorticity of the nodal point.

In generic A-B symmetric partitions, analytical solutions may not be available (such as in partition \textbf{e}).  As a result, we do not have the function $\alpha_i(\mathbf{q})$ which play the role of the phasor. Nevertheless, it is natural to conjecture that the notion of vorticity for each nodal point can be extended in the following way. First, we define the ``Berry connection'' for the BES as\footnote{Do not confuse it with the Berry connection defined as $\mathbf{A}:=-i \langle \chi^- | \boldsymbol{\nabla}_{\bf k} | \chi^- \rangle$ for the original Hamiltonian.}
\begin{equation}
\mathcal{A}(\mathbf{q})
:=-i\bra{\Psi(\mathbf{q})}\partial_\mathbf{q}\ket{\Psi(\mathbf{q})},
\end{equation}
where
\begin{equation}
\Psi(\mathbf{q}) = \sum_{\alpha=\uparrow,\downarrow} \chi_-^\alpha(\mathbf{q})^*\psi^\alpha(\mathbf{q})
\end{equation}
with $\psi^\alpha(\mathbf{q})$ being the normalized eigenfunction of the entanglement Hamiltonian with $\lambda\geq1/2$.
Then, the \emph{vorticity} $v$ for a given nodal point can be defined as the contour integral around it:
\begin{equation}\label{v}
v:=\frac{1}{\pi}\oint \mathcal{A}(\mathbf{q})\,d\mathbf{q}.
\end{equation}
With a moment of reflection, one sees that, in the cases of partitions \textbf{b} and \textbf{c}, the vorticity defined in \eq{v} agrees with that inferred from \eq{phasor for partition c} and \eq{phasor for partition b}. Also note that $\mathcal{A}(\mathbf{q})$ is singular at nontrivial nodal points. For generic cases, it is crucial to ask whether the definition \eq{v} always yields an integer number for $v$.  We do not have a proof, but it is tempting to speculate that the answer is affirmative provided the partition is A-B symmetric.

\begin{figure*}

\centering

  \begin{minipage}[b]{0.2\textwidth}
    \includegraphics[width=\textwidth,height=\textwidth]{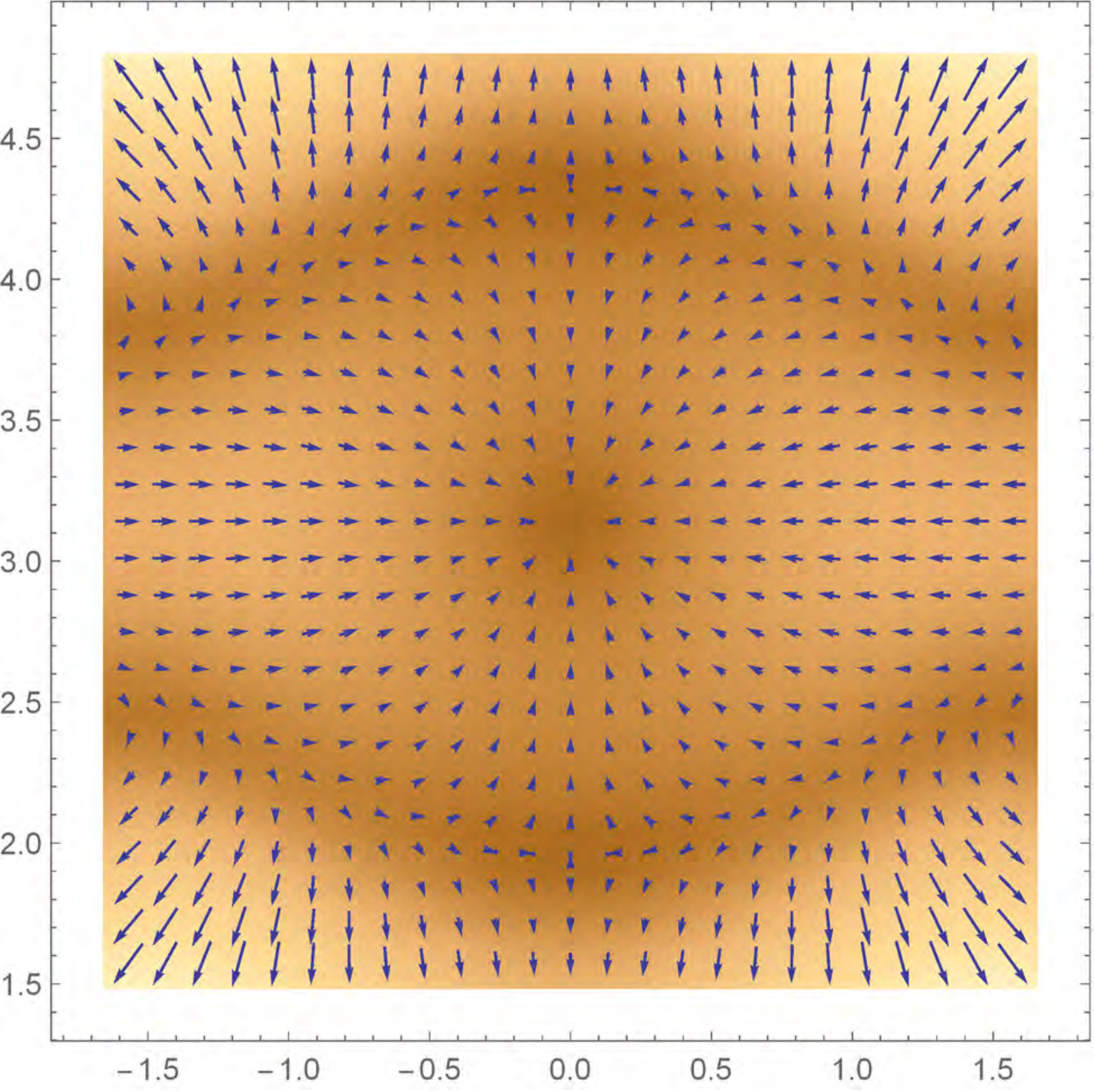}
  \end{minipage}
  \hspace{1cm} 
  \begin{minipage}[b]{0.2\textwidth}
    \includegraphics[width=\textwidth,height=\textwidth]{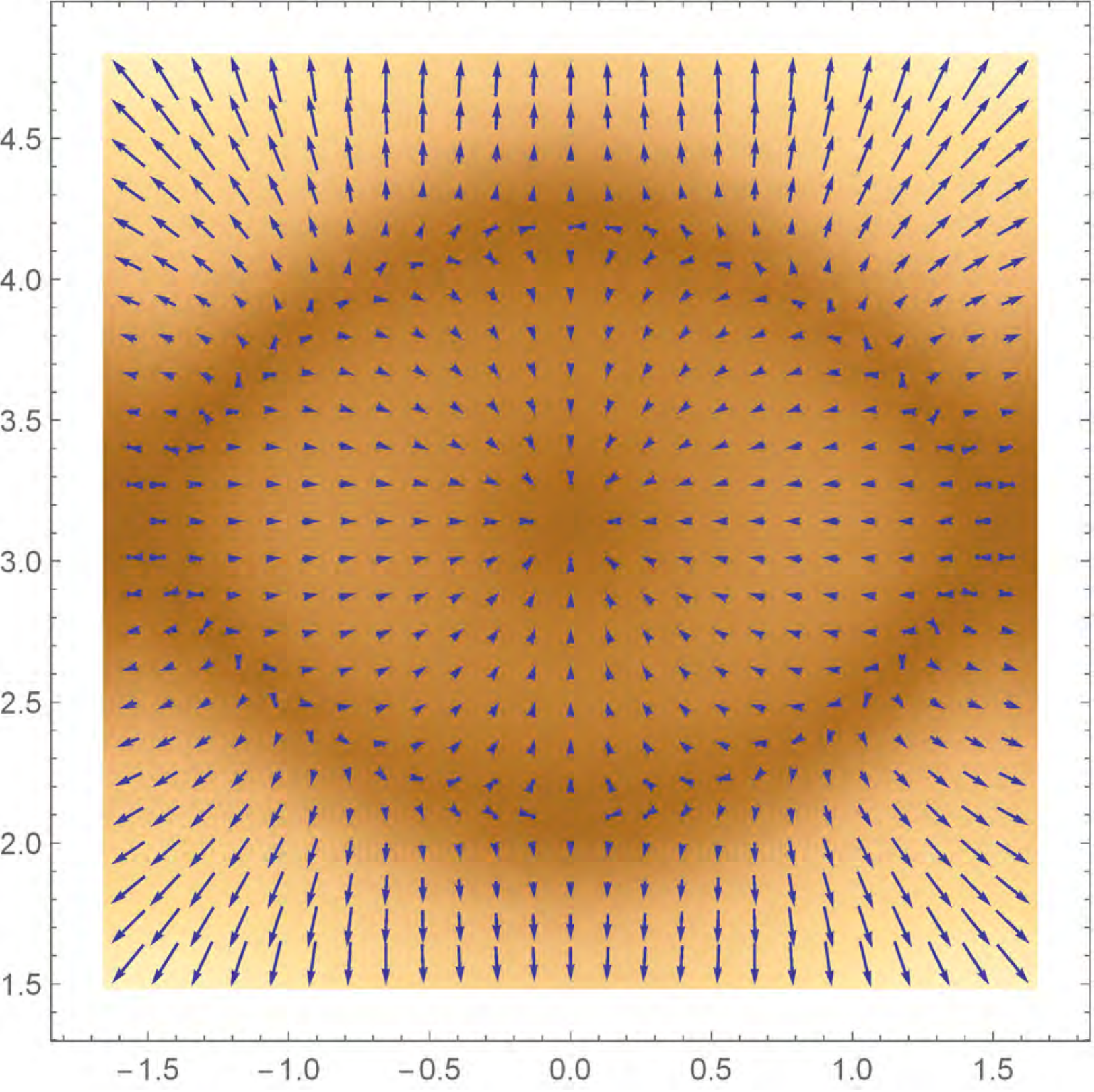}
  \end{minipage}
  \hspace{1cm} 
  \begin{minipage}[b]{0.2\textwidth}
    \includegraphics[width=\textwidth,height=\textwidth]{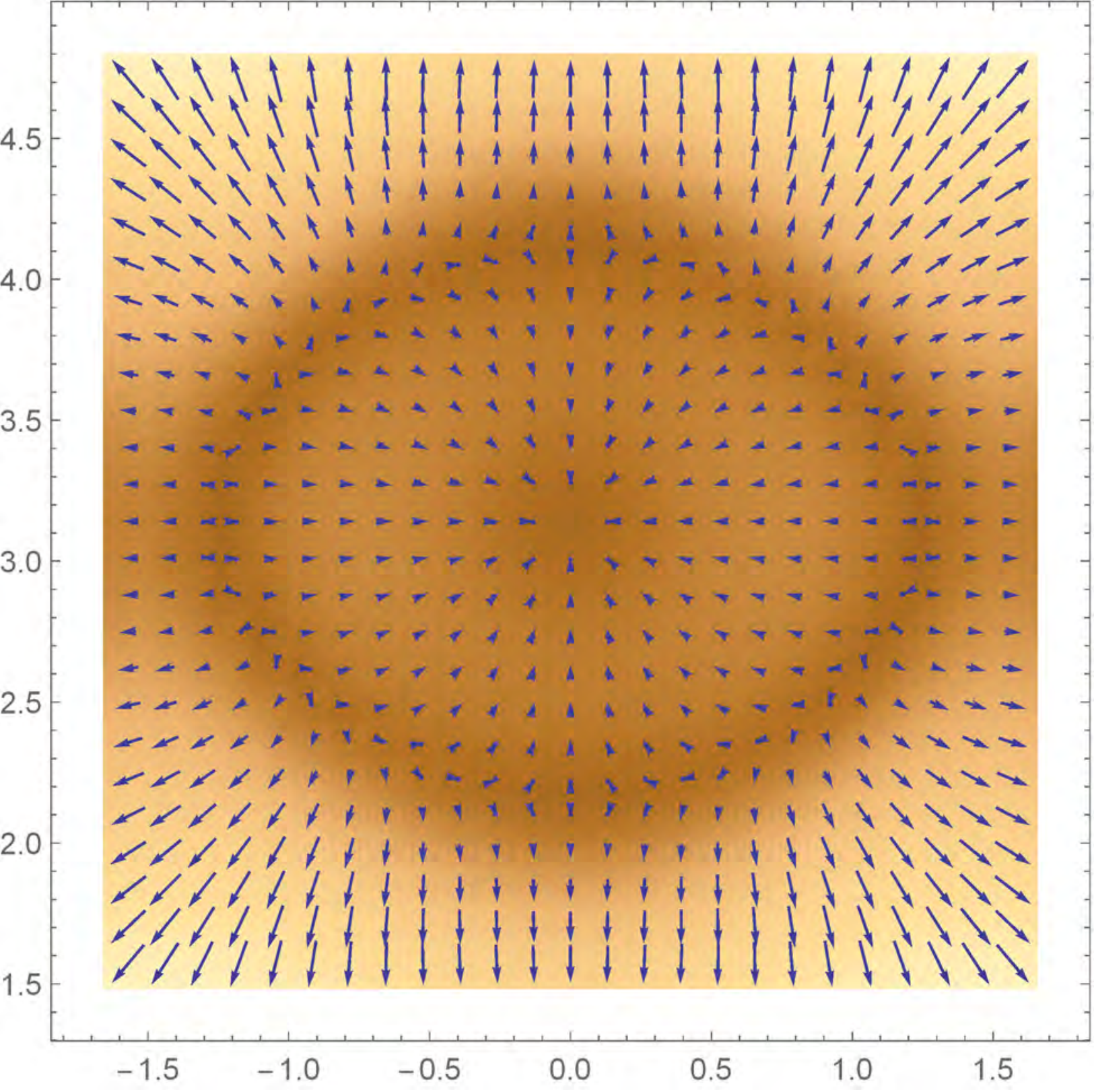}
  \end{minipage}

  \vspace{0.65cm}

  \begin{minipage}[b]{0.2\textwidth}
    \includegraphics[width=\textwidth,height=\textwidth]{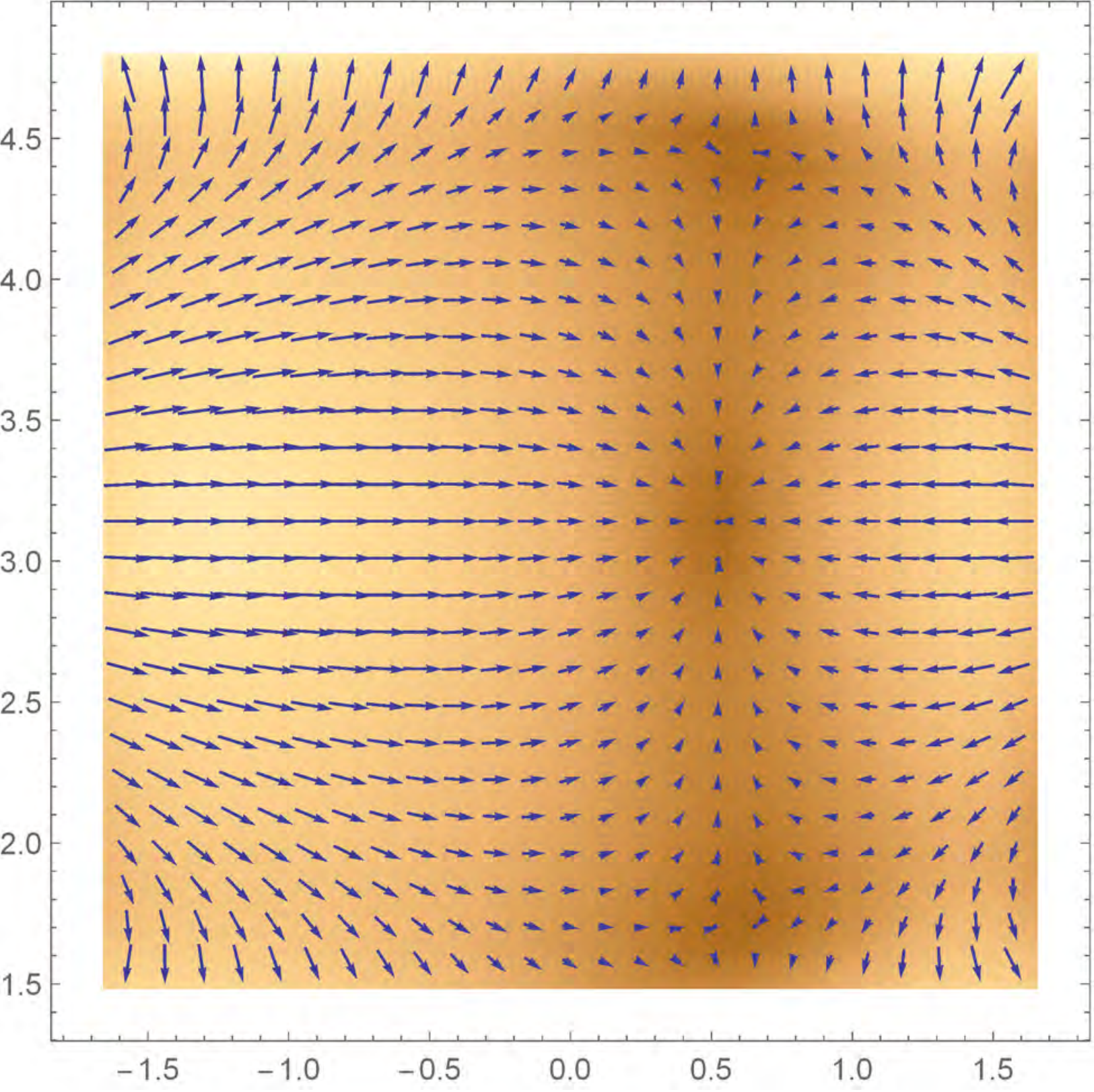}
  \end{minipage}
  \hspace{1cm} 
  \begin{minipage}[b]{0.2\textwidth}
    \includegraphics[width=\textwidth,height=\textwidth]{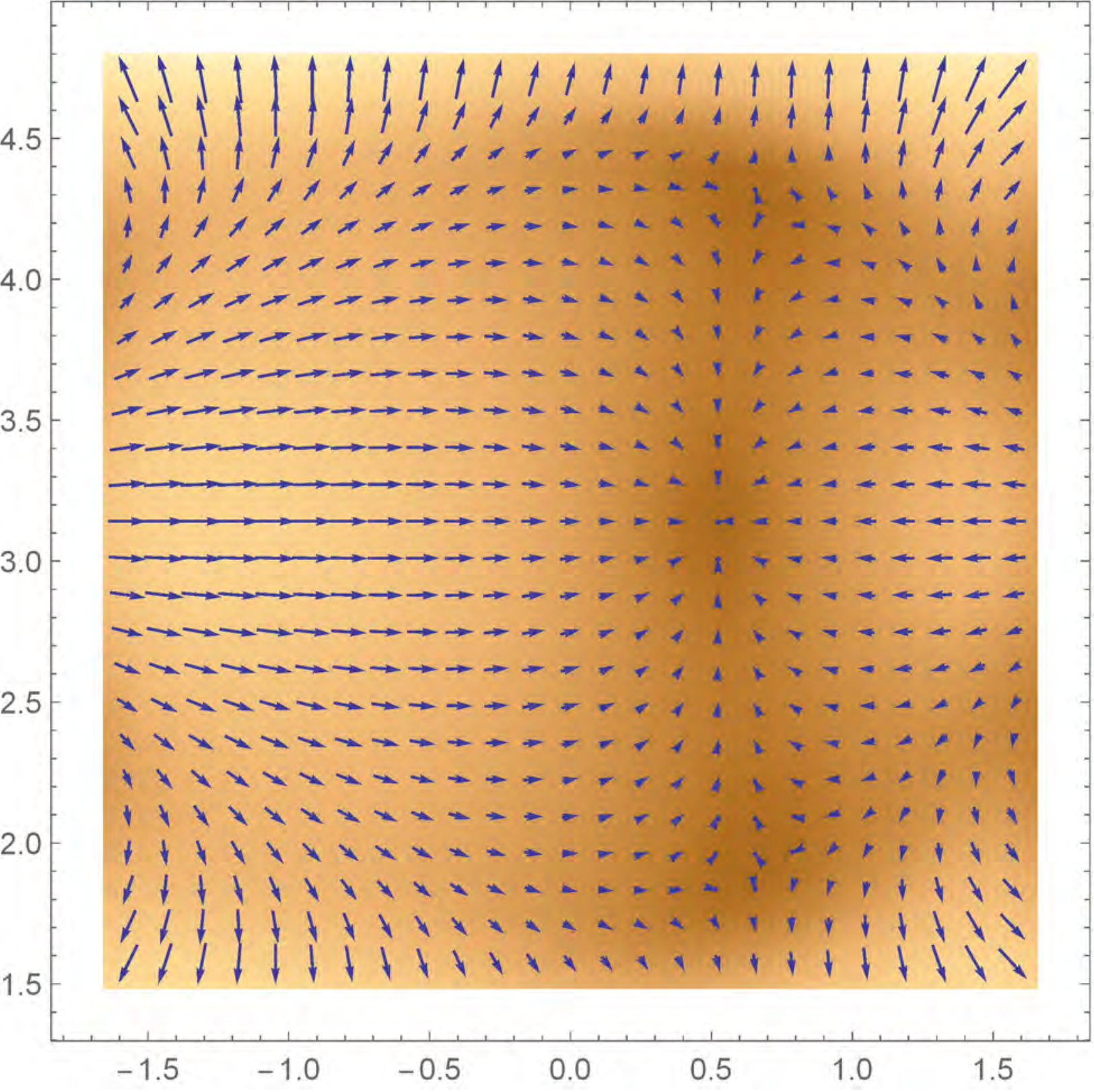}
  \end{minipage}
  \hspace{1cm} 
  \begin{minipage}[b]{0.2\textwidth}
    \includegraphics[width=\textwidth,height=\textwidth]{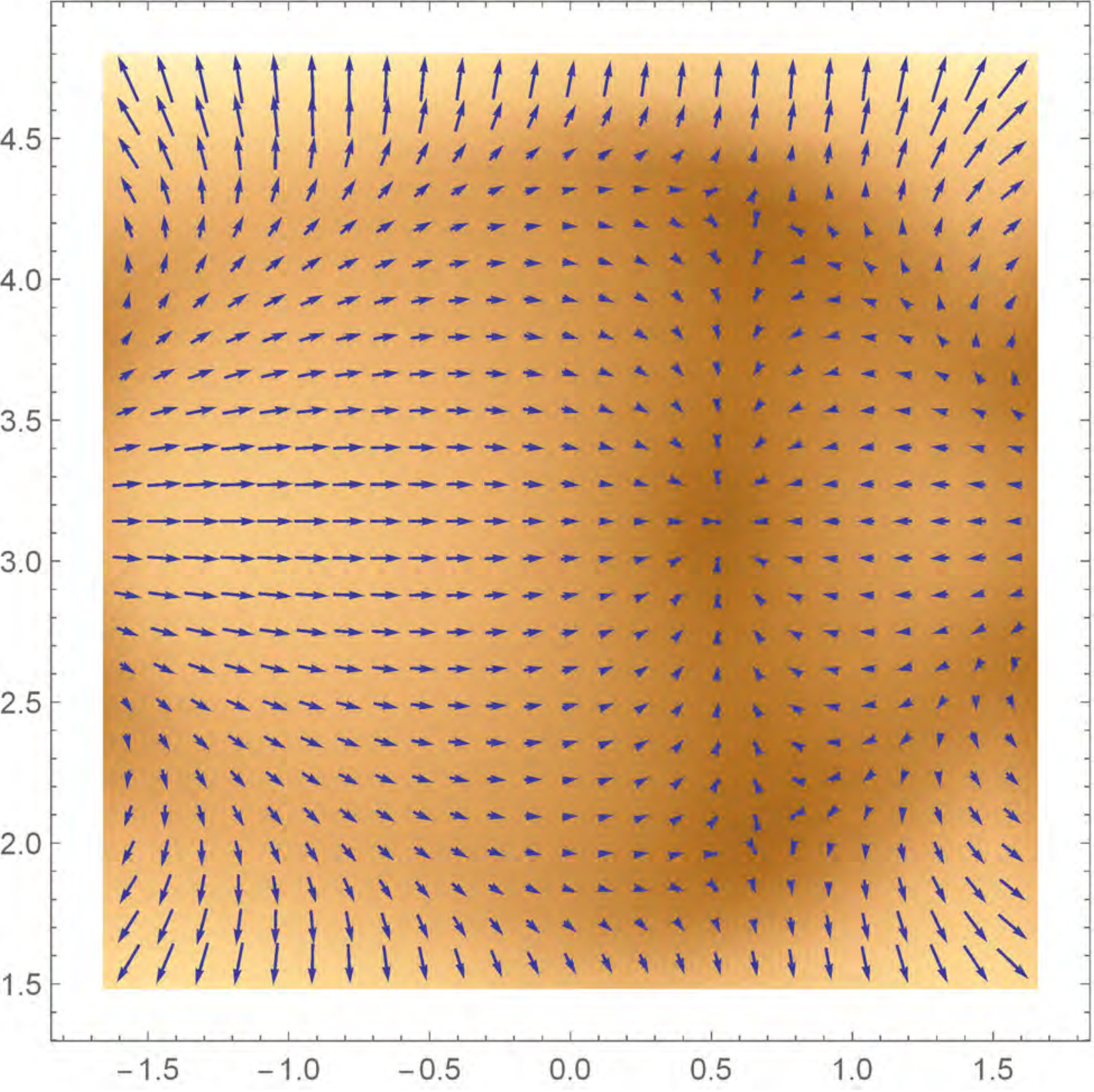}
  \end{minipage}

\caption{Vortex structure corresponding to Fig.\ \ref{fig-CI3c-2} (with $\mathtt{C}=1$), shown in $(q_x,q_y)\in(-\pi/2,\pi/2]\times[\pi/2,3\pi/2]$.  The phasor field is presented by vector arrows and its norm is also indicated by the varying background shade.  Upper left and Upper middle: two antivortices at $\mathbf{q}_\mathrm{U,D}$ with $v=-1$ and one vortex at $\mathbf{q}_\mathrm{A}$ with $v=1$; note that the nodal point at $\mathbf{q}_\mathbf{R}=\mathbf{q}_\mathrm{L}=(\pm\pi/2,\pi)$ is not a vortex (i.e.\ $v=0$). Upper right: two antivortices at $\mathbf{q}_\mathrm{U,D}$ with $v=-1$ and three vortices at $\mathbf{q}_\mathrm{O,R,L}$ with $v=1$. Bottom: all plots have two antivortices with $v=-1$ and one vortex with $v=1$.}\label{fig-CI3c-2-vor}
\end{figure*}

Generally, the phasor field $\alpha_i(\mathbf{q})$ is well defined only \emph{locally}, because $\chi_-(\mathbf{q})$ is usually regular only in a local patch. From a patch to another one, $\chi_-(\mathbf{q})$ undergoes a gauge transformation, and so does the phasor field. This corresponds to the gauge transformation for the connection vector $\mathcal{A}(\mathbf{q})$. Despite the gauge ambiguity, the contour integral \eq{v} is gauge-invariant and thus the vorticity $v$ for each nodal point is well defined.

\begin{figure*}
\centering
  \begin{minipage}[b]{0.2\textwidth}
    \includegraphics[width=\textwidth,height=\textwidth]{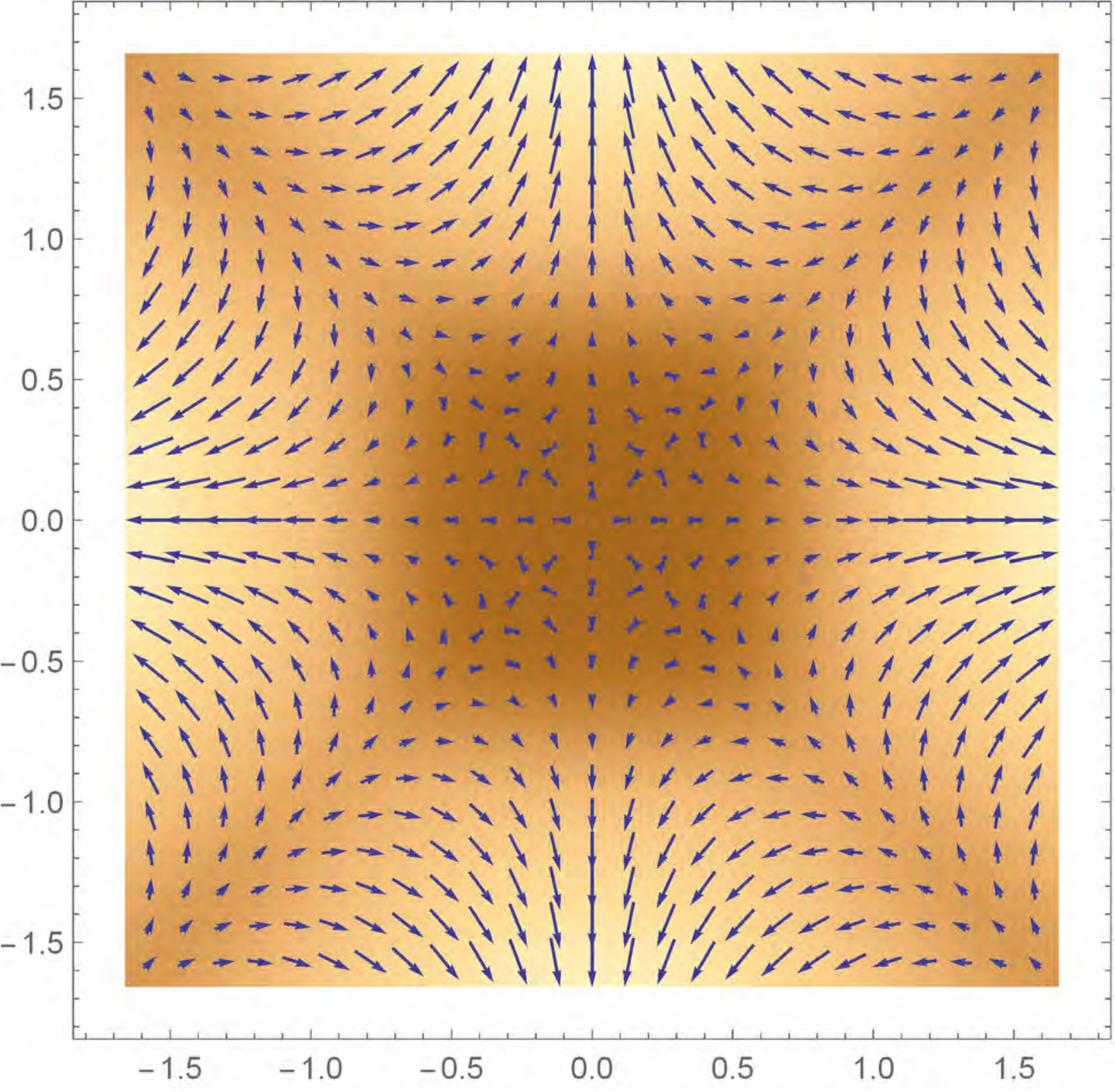}
  \end{minipage}
  \hspace{1cm} 
  \begin{minipage}[b]{0.2\textwidth}
    \includegraphics[width=\textwidth,height=\textwidth]{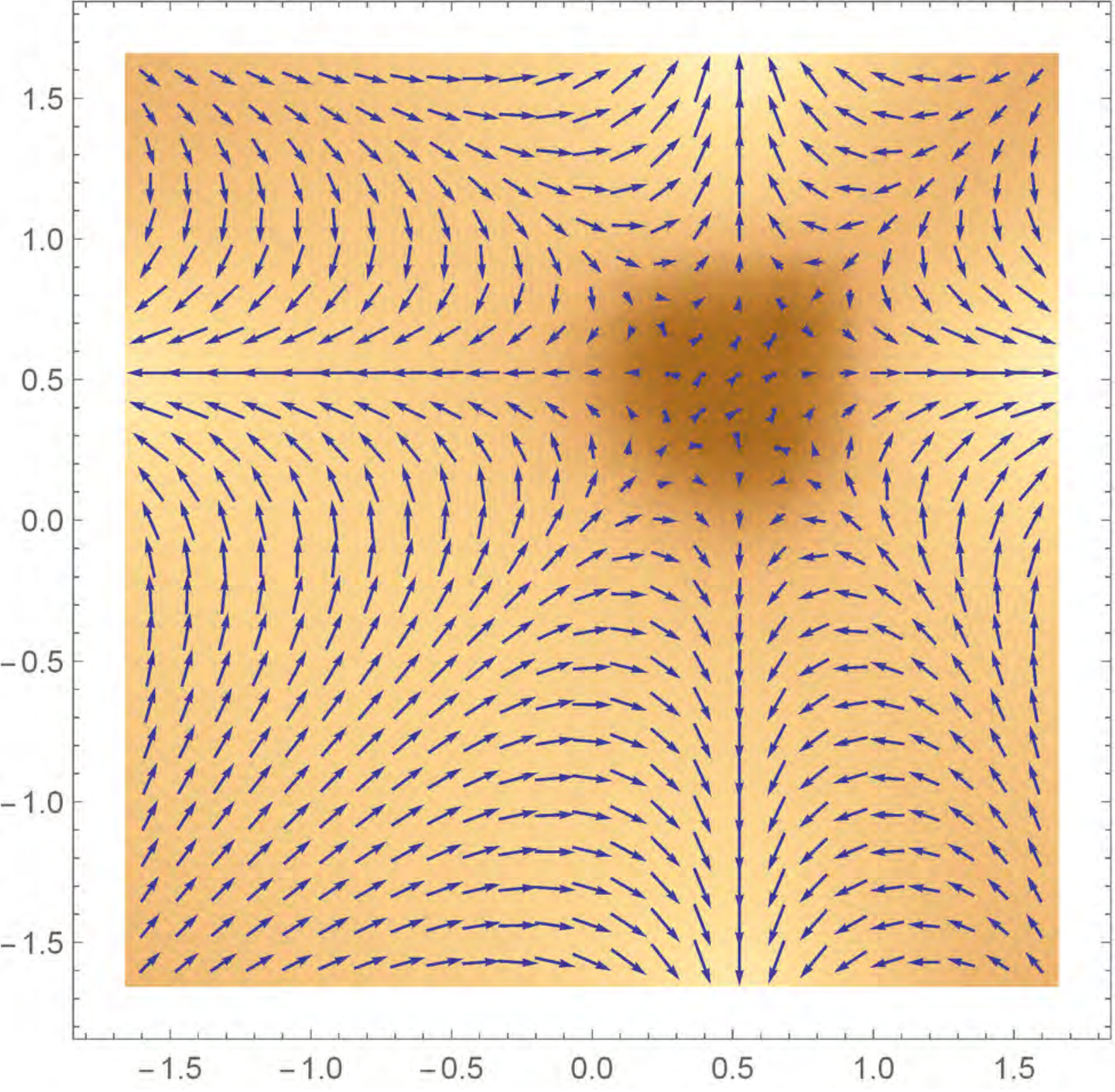}
  \end{minipage}
  \hspace{1cm} 
  \begin{minipage}[b]{0.2\textwidth}
    \includegraphics[width=\textwidth,height=\textwidth]{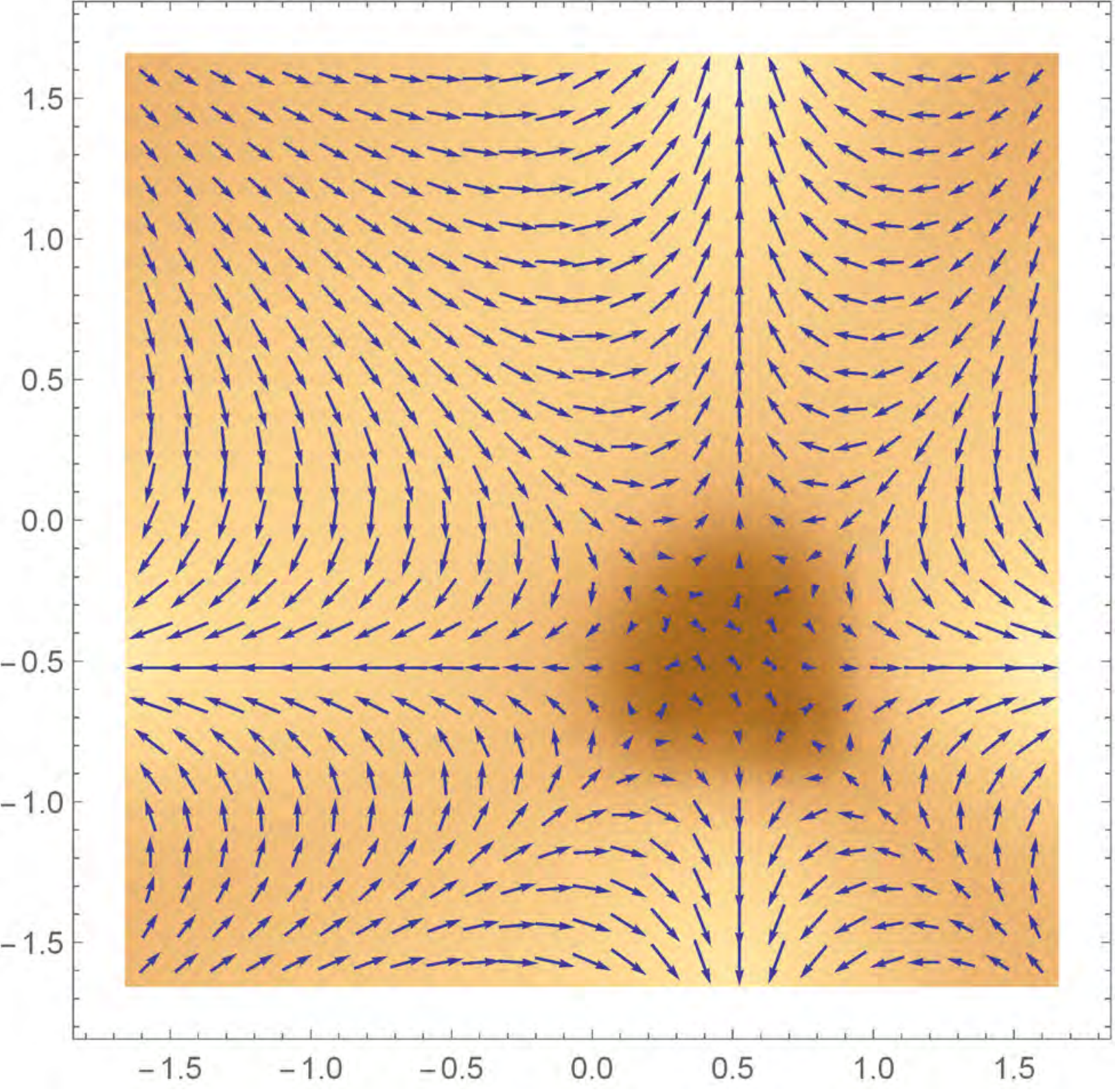}
  \end{minipage}

\caption{Vortex structure corresponding to Fig.\ \ref{fig-CI3b-1} (with $\mathtt{C}=3$).  Left: a single vortex at the origin with $v=-3$. Middle and Right: three vortices with $v=-1$.}\label{fig-CI3b-1-vor}
\end{figure*}

In Appendix \ref{Appendix-dispersion}, we study the details for the phasor vector fields in the vicinity of the corresponding Dirac, quadratic and cubic nodal points for the topological phases with $|\mathtt{C}|=1,2,3$ before imposing any deformation. The results all conform to \eq{master rel}. In particular, we would like to point out that in the case $\bf CI_{3b}$ there is a cubic nodal point at $\mathbf{q}=(0,0)$ for both $\mathtt{C}=3$ and $\mathtt{C}=1$ phases as shown in the left panels of Fig.\ \ref{fig-CI3b-1} and Fig.\ \ref{fig-CI3b-2}. However, it turns out they have different vorticities, $v=-3$ for $\mathtt{C}=3$ and $v=1$ for $\mathtt{C}=1$ so that eq.\ \eq{master rel} is still obeyed.

Furthermore, to elaborate on how the issue is resolved in more complicated cases, we investigate in details the vortex structures corresponding to Figs.\ \ref{fig-CI3c-1}, \ref{fig-CI3c-2}, \ref{fig-CI3b-1}, and \ref{fig-CI3b-2} in the following subsections.   In these cases, the number of nodal points may exceed $|\mathtt{C}|$, but nonetheless \eq{master rel} always applies.  Although our investigation is far from exhaustive, we believe the evidence gathered is enough to support the conjectured relation \eq{master rel}, which indicates that the BES indeed inherits its topological property from the underlying Chern insulator.

\begin{figure*}
\centering
  \begin{minipage}[b]{0.2\textwidth}
    \includegraphics[width=\textwidth,height=\textwidth]{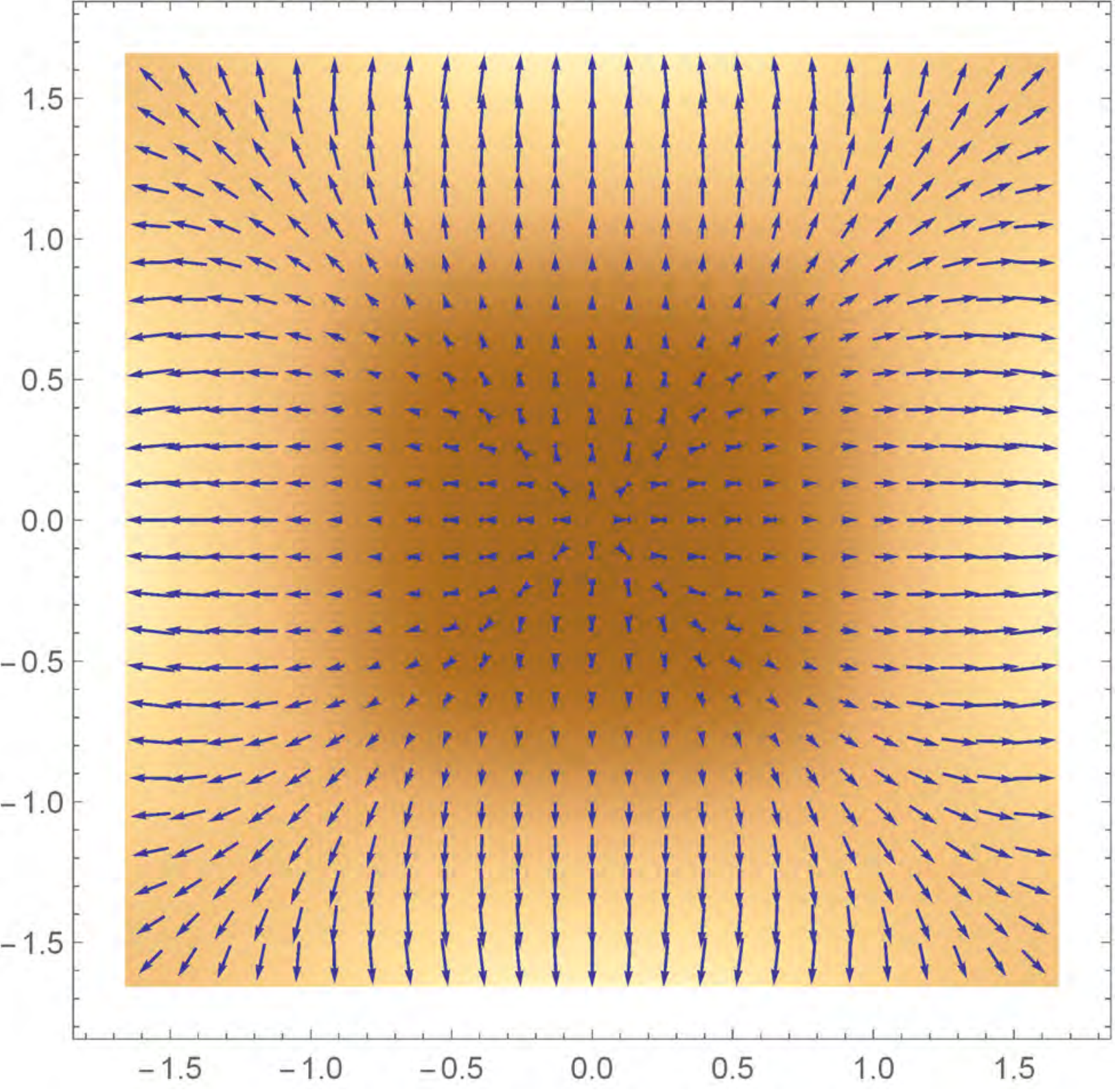}
  \end{minipage}
  \hspace{1cm} 
  \begin{minipage}[b]{0.2\textwidth}
    \includegraphics[width=\textwidth,height=\textwidth]{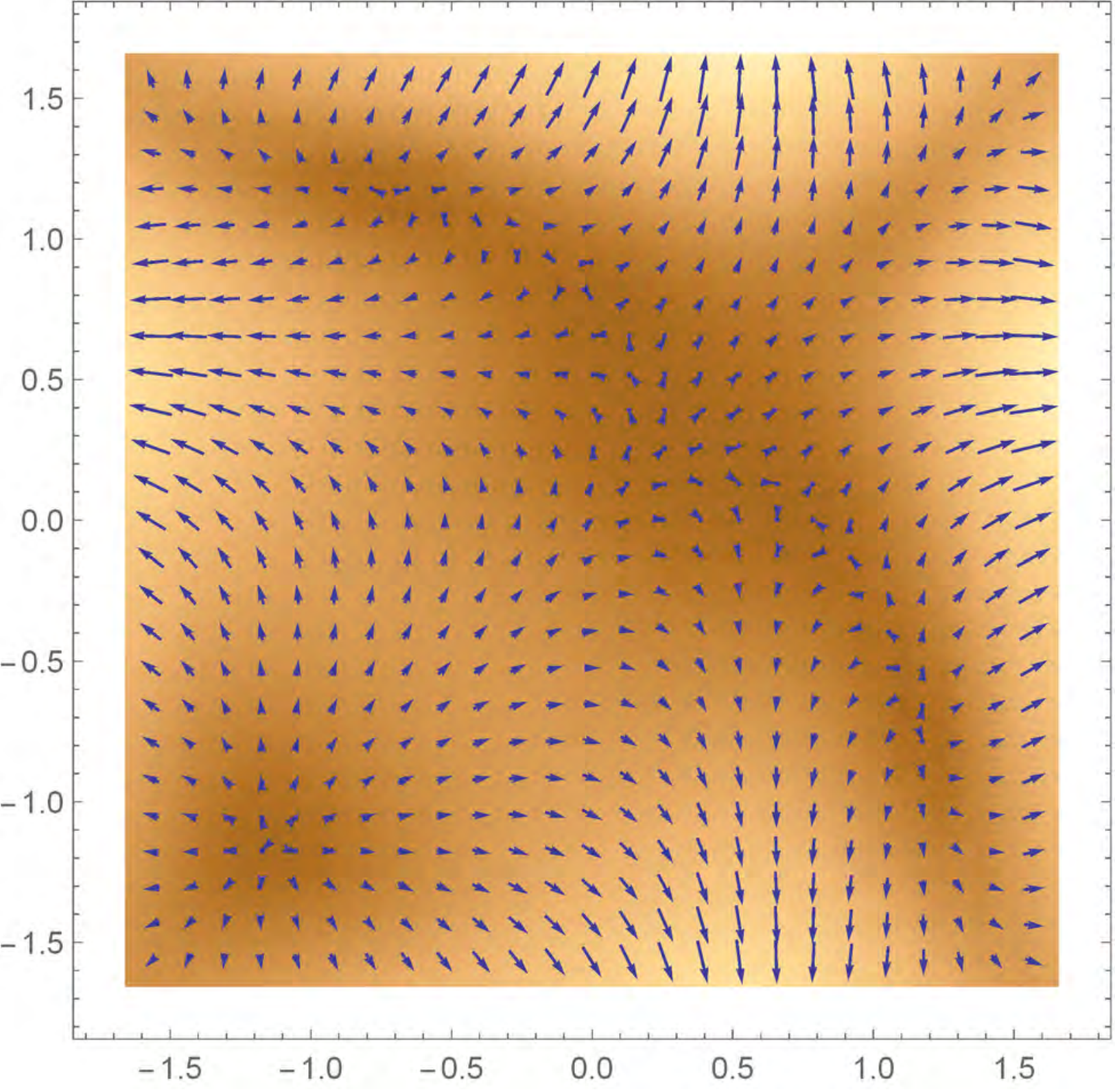}
  \end{minipage}
  \hspace{1cm} 
  \begin{minipage}[b]{0.2\textwidth}
    \includegraphics[width=\textwidth,height=\textwidth]{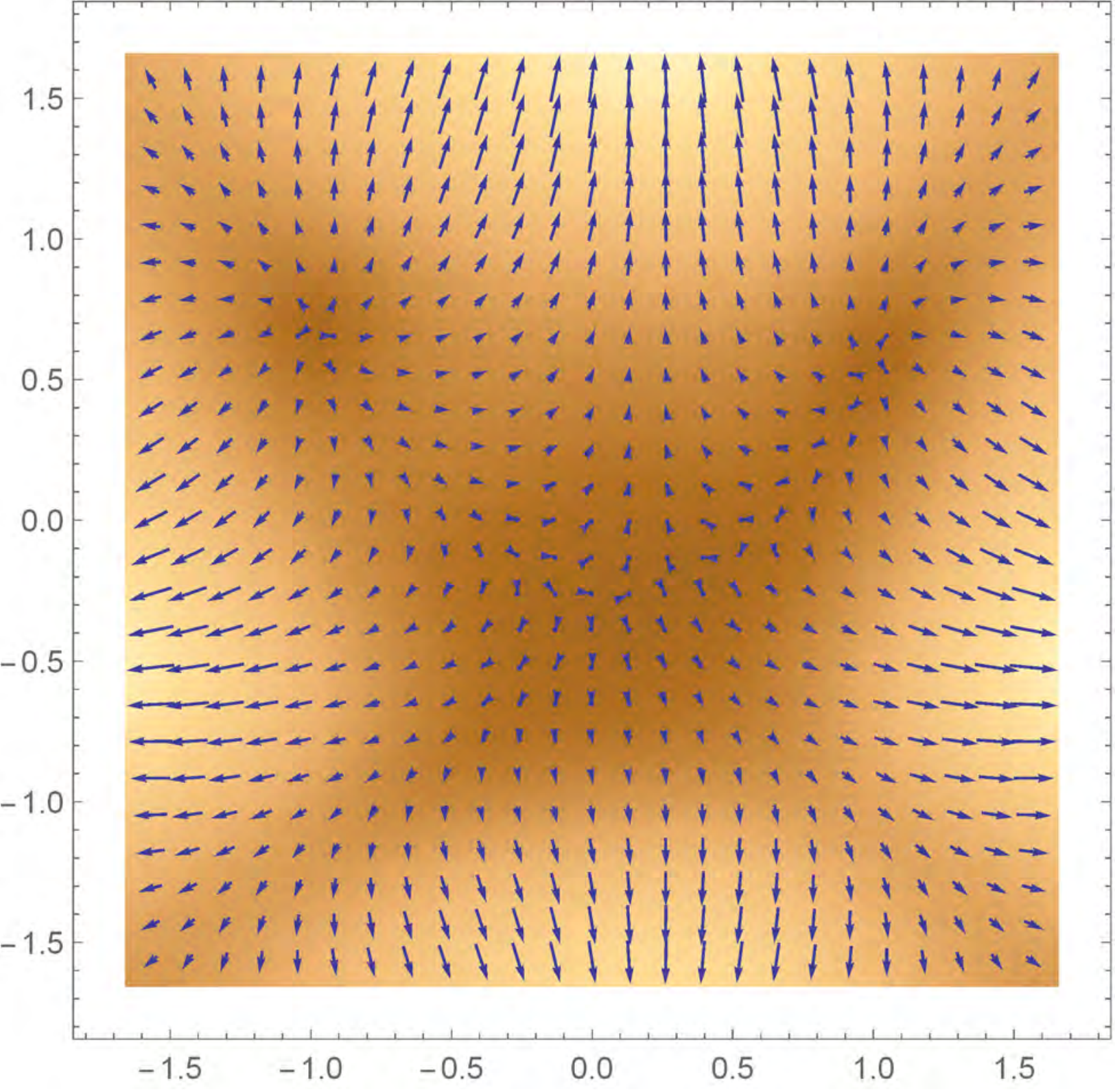}
  \end{minipage}

\caption{Vortex structure corresponding to Fig.\ \ref{fig-CI3b-2}  (with $\mathtt{C}=1$). Left: a single vortex at the origin with $v=1$. Middle: three vortices with $v=1$ and two antivortices with $v=-1$. Right: two vortices with $v=1$ and one antivortex with $v=-1$}\label{fig-CI3b-2-vor}
\end{figure*}

\begin{figure*}

\centering

  \begin{minipage}[b]{0.2\textwidth}
    \includegraphics[width=\textwidth,height=\textwidth]{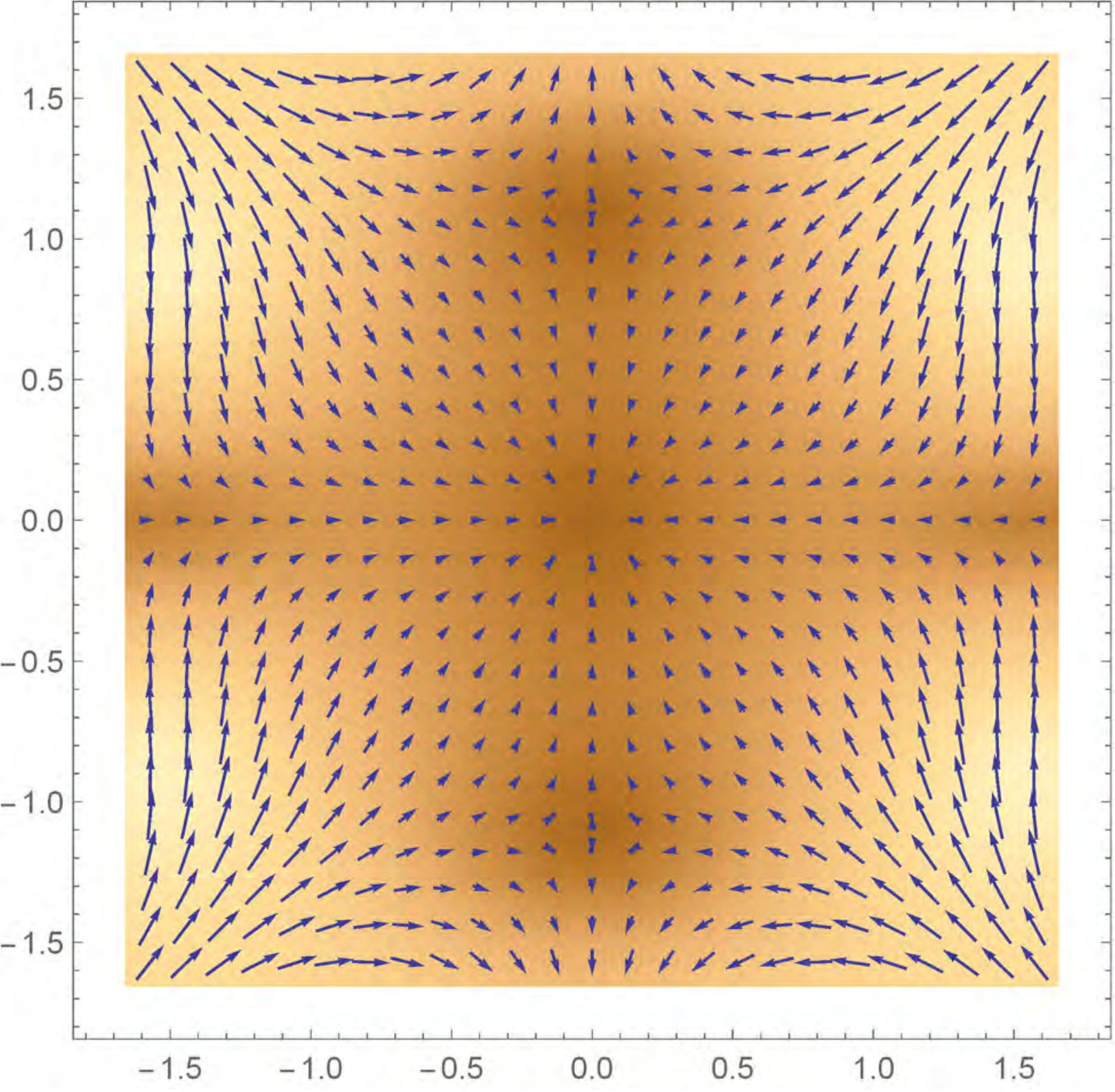}
  \end{minipage}
  \hspace{1cm} 
  \begin{minipage}[b]{0.2\textwidth}
    \includegraphics[width=\textwidth,height=\textwidth]{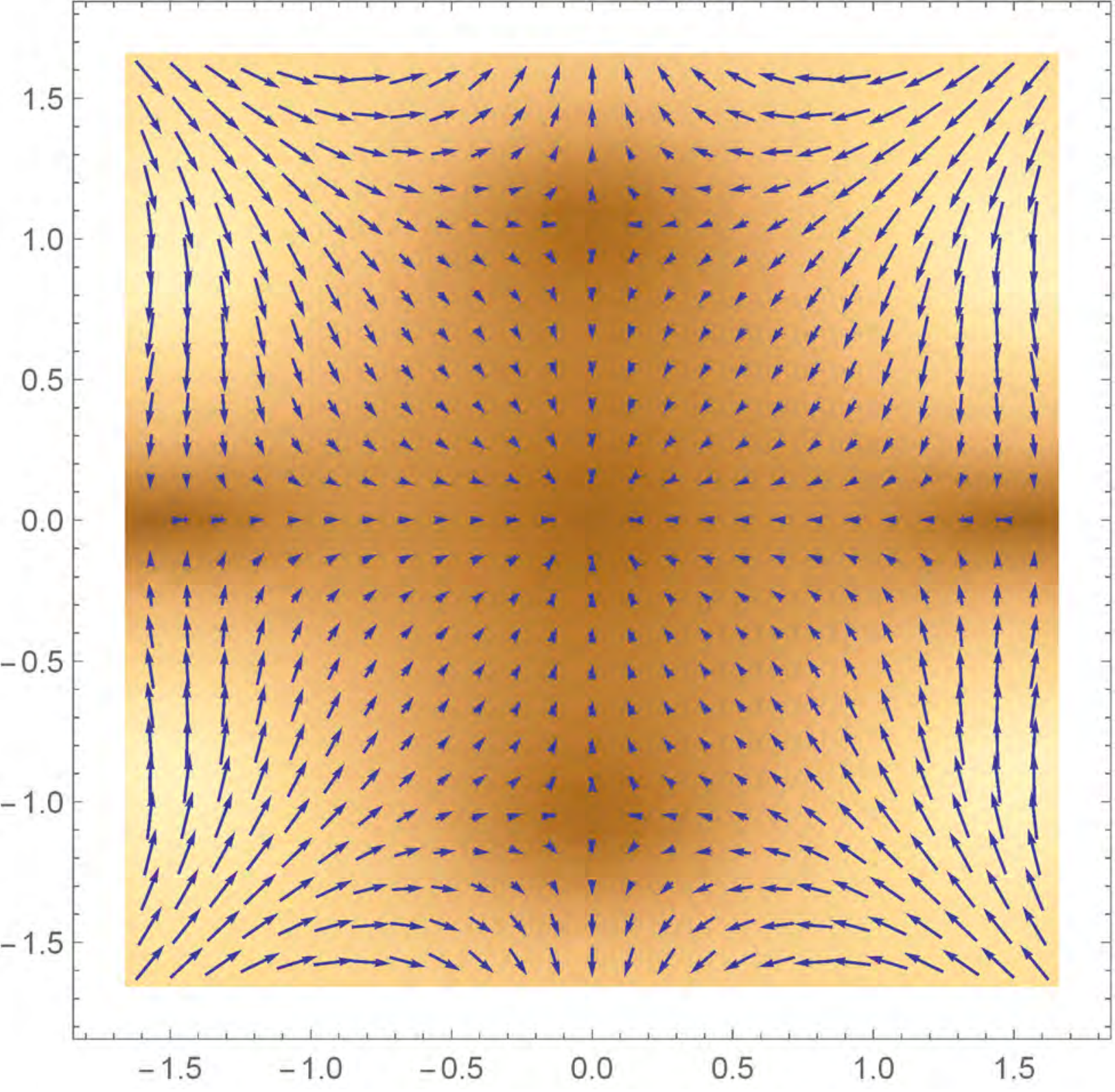}
  \end{minipage}
  \hspace{1cm} 
  \begin{minipage}[b]{0.2\textwidth}
    \includegraphics[width=\textwidth,height=\textwidth]{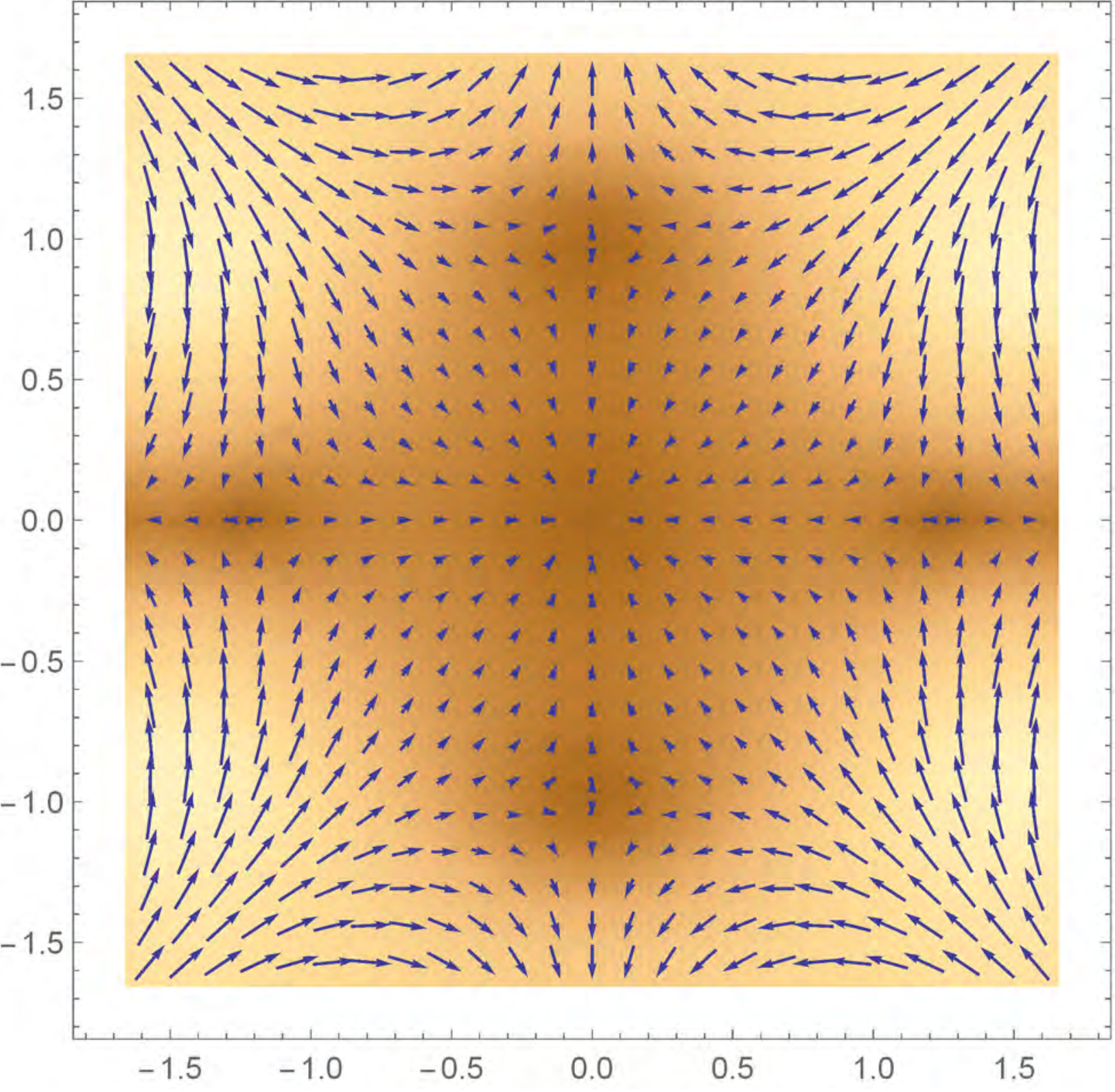}
  \end{minipage}

  \vspace{0.65cm}

  \begin{minipage}[b]{0.2\textwidth}
    \includegraphics[width=\textwidth,height=\textwidth]{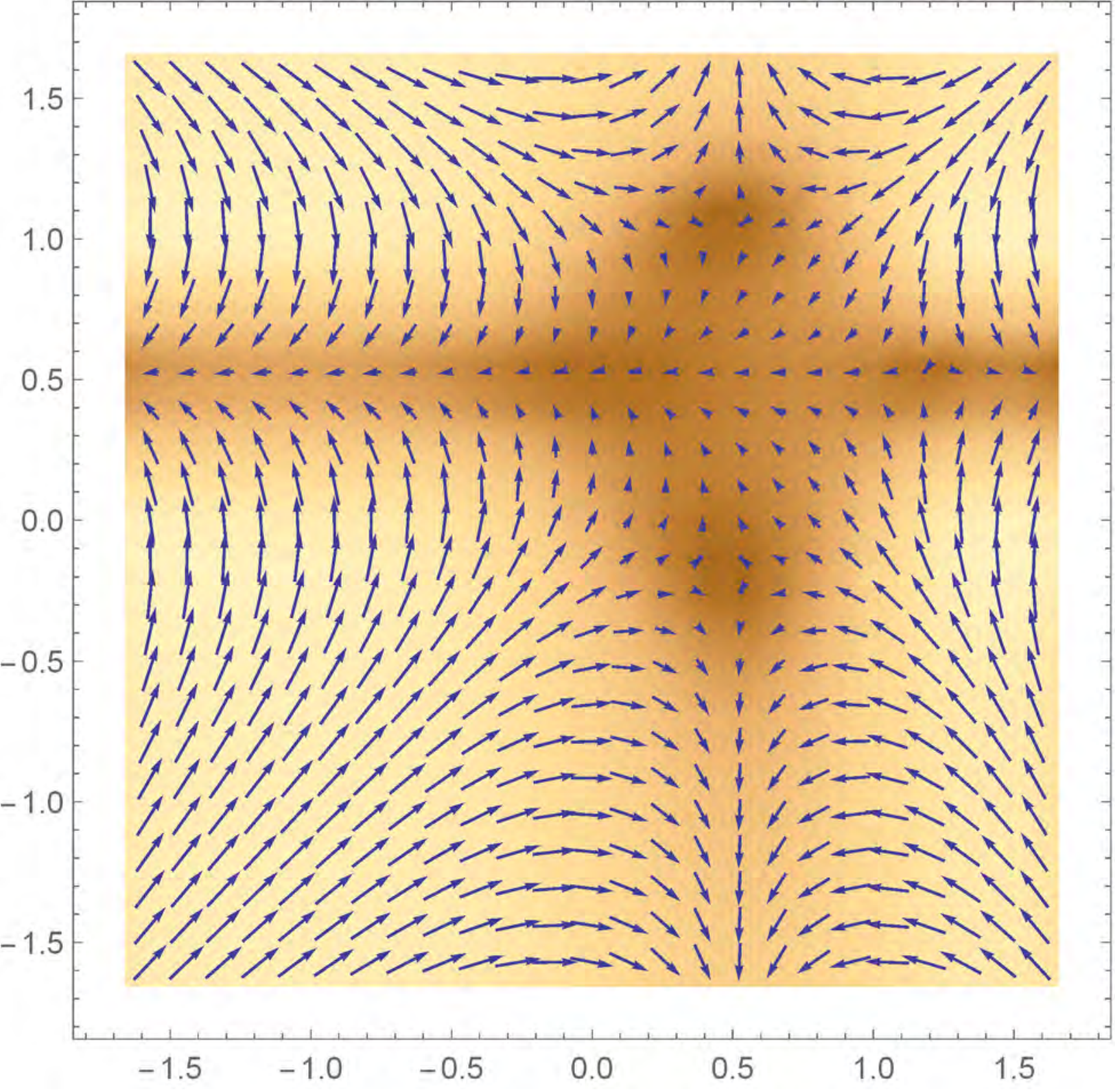}
  \end{minipage}
  \hspace{1cm} 
  \begin{minipage}[b]{0.2\textwidth}
    \includegraphics[width=\textwidth,height=\textwidth]{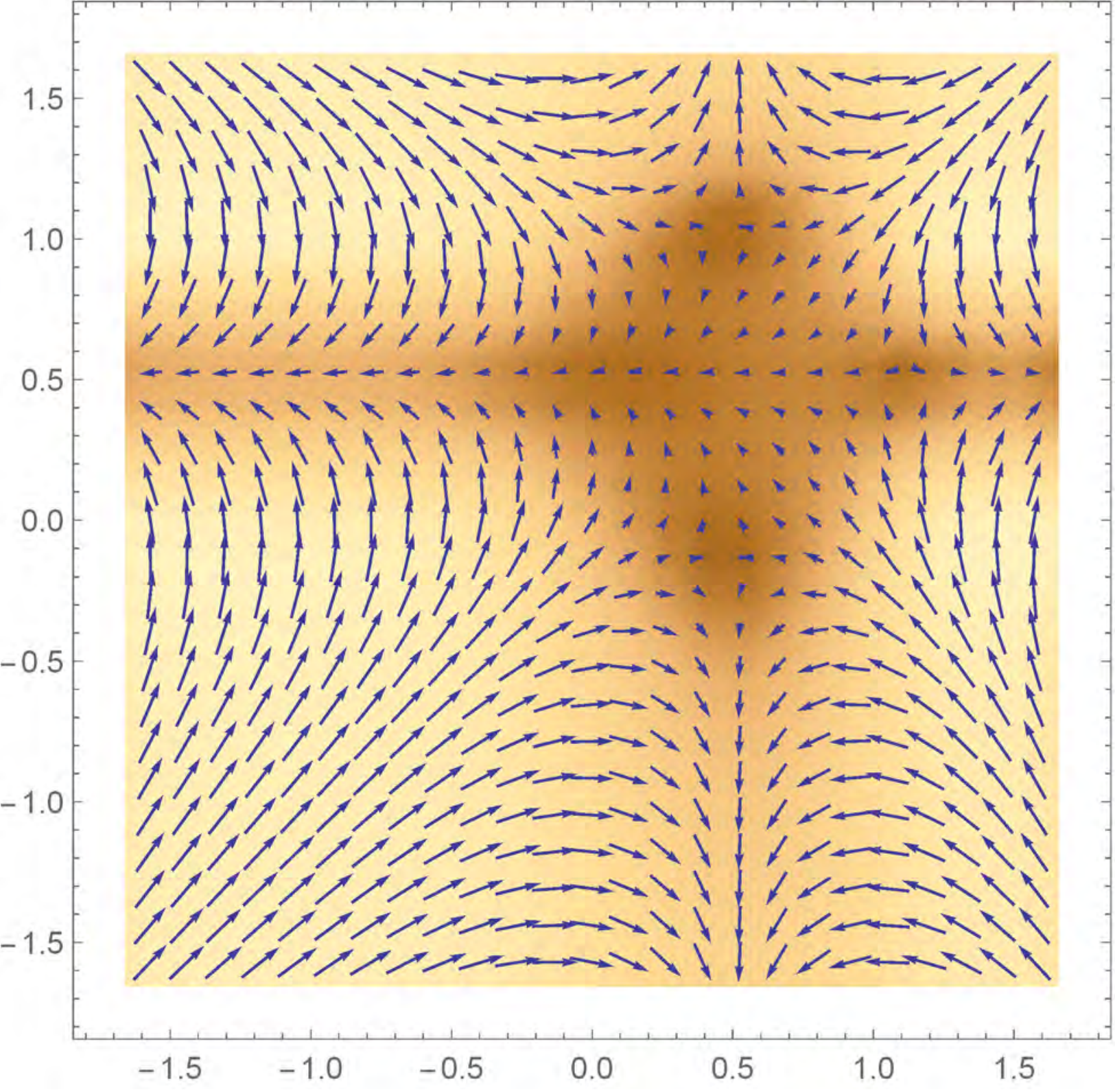}
  \end{minipage}
  \hspace{1cm} 
  \begin{minipage}[b]{0.2\textwidth}
    \includegraphics[width=\textwidth,height=\textwidth]{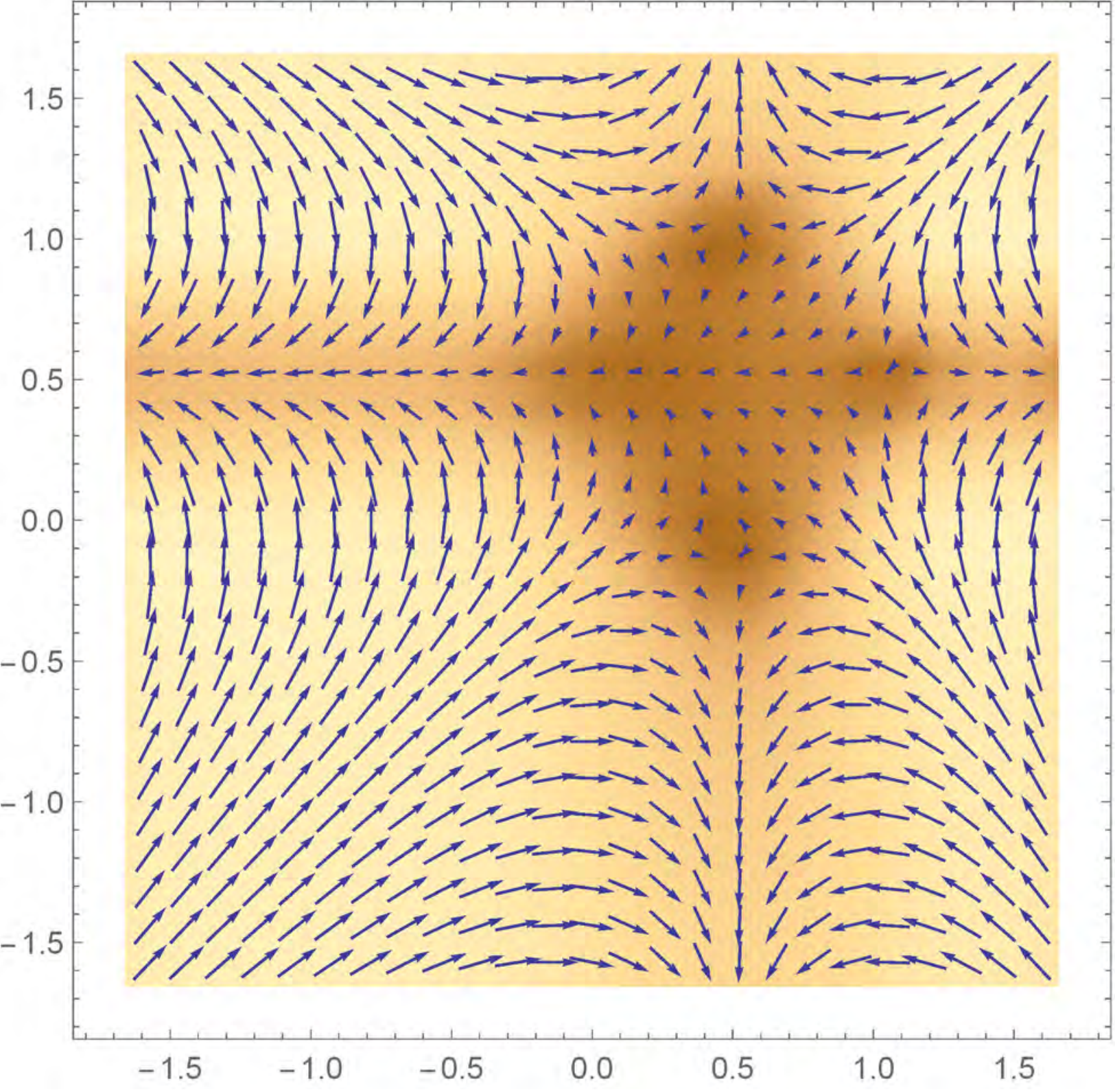}
  \end{minipage}

\caption{Vortex structure corresponding to Fig.\ \ref{fig-CI3c-1}  (with $\mathtt{C}=3$), shown in the region $(q_x,q_y)\in(-\pi/2,\pi/2]\times[-\pi/2,\pi/2]$. Upper left and Upper middle: two antivortices at $\mathbf{q}_\mathrm{N,S}$ with $v=-1$ and one vortex at $\mathbf{q}_\mathrm{O}$ with $v=1$; note that the nodal point at $\mathbf{q}_\mathbf{E}=\mathbf{q}_\mathrm{W}=(\pm\pi/2,0)$ is not a vortex (i.e.\ $v=0$). Upper right: four antivortices at $\mathbf{q}_\mathrm{N,S,E,W}$ with $v=-1$ and one vortex at $\mathbf{q}_\mathrm{O}$ with $v=1$. Bottom: all plots have three antivortices with $v=-1$.}\label{fig-CI3c-1-vor}
\end{figure*}

\subsection{$\bf CI_{3c}$ with $\mathtt{C}=1$}\label{Section-vortex CI3c c=1}
The vortex structure of $\alpha_3(\mathbf{q})$ is shown in Fig.\ \ref{fig-CI3c-2-vor} which is related to the band crossing pattern in Fig.\ \ref{fig-CI3c-2} .

In the reduced Brillouin zone $\mathbf{BZ}_\mathbf{q}$, the nodal points $\mathbf{q}_\mathrm{U}$ and $\mathbf{q}_\mathrm{D}$ have $v=-1$, while $\mathbf{q}_\mathrm{A}$,  $\mathbf{q}_\mathrm{R}$ and $\mathbf{q}_\mathrm{L}$ have $v=1$. Under deformation, $\mathbf{q}_\mathrm{R}$ annihilates with $\mathbf{q}'_\mathrm{L}:=\mathbf{q}_\mathrm{L}+(\pi,0)$, whose  $v=-1$. Likewise $\mathbf{q}_\mathrm{L}$ annihilates with $\mathbf{q}_\mathrm{L}-(\pi,0)$. See Fig.\ \ref{fig-CI3c-2-vor}. By delimiting the reduced Brillouin zone to $(-\pi/2,\pi/2]\times(0,2\pi]$, we can make the sum of the vorticities equal to 1, and thus relation \eq{master rel} is satisfied.

\subsection{$\bf CI_{3b}$ with $\mathtt{C}=3$}
The vortex structure of $\alpha_{1,2}(\mathbf{q})$, defined in \eq{alpha1 alpha2}, may be calculated numerically. It is shown in Fig.\ \ref{fig-CI3b-1-vor}, which is related to the band crossing pattern in Fig.\ \ref{fig-CI3b-1}.

Before deformation is introduced, the vortex at $\mathbf{q}=(0,0)$ has $v=-3$, which can be inferred from \eq{phasor CI3}. Under deformation, the $v=-3$ vortex splits into three $v=-1$ vortices.  From the plots in Fig.\ \ref{fig-CI3b-1-vor},  we see that the sum of the vorticities in $\mathbf{BZ}_\mathbf{q}$ is 3 and the relation \eq{master rel} again applies.

\subsection{$\bf CI_{3b}$ with $\mathtt{C}=1$}
The vortex structure of $\alpha_{1,2}(\mathbf{q})$ is shown in Fig.\ \ref{fig-CI3b-2-vor}, which is related to the band crossing pattern in Fig.\ \ref{fig-CI3b-2} .

Before we impose any deformation, the vortex at
$\mathbf{q}=(0,0)$ has $v=+1$, which can be seen from \eq{phasor CI3}. Under deformation, vortex-antivortex pairs may be created. In the middle plot, we have two additional vortex-antivortex pairs, while in the right plot, we have one additional pair.  In sum, the plots in Fig.\ \ref{fig-CI3b-2-vor} again are consistent with the relation \eq{master rel}.

\subsection{Subtlety on choosing the reduced zone}\label{Section-remarks on subtleties}

It should be emphasized that the vorticity associated with $\alpha_{i}(\mathbf{q})'s$ is defined only up to a sign ambiguity,  which gives rise to the sign ambiguity in \eq{master rel}.  In partition \textbf{c}, for example, the point $\mathbf{q}\in\mathbf{BZ}_\mathbf{q}$ is identical to $\mathbf{q}_3:=\mathbf{q}+(\pi,0)$ as far as the BES is concerned.  Thus, we have $\alpha_3(\mathbf{q}_3)\propto\alpha_3(\mathbf{q})^*$ according to \eq{eigenvec c}, and the vorticity in the reduced Brillouin zone $\mathbf{BZ}_{\mathbf{q}_3}:=\mathbf{BZ}_\mathbf{q}+(\pi,0)$ is precisely the opposite of the vorticity in the principal reduced Brillouin zone $\mathbf{BZ}_\mathbf{q}$. In partition \textbf{b}, we have a similar situation as well. The sign ambiguity is unavoidable since the total number of vortices and antivortices on a torus must be the same and the corresponding pair are always separated by $(\pi,0)$ in partition \textbf{c} and $(\pi,\pi)$ in partition \textbf{b}.  Moreover, the definition of the principal reduced Brillouin zone is mainly a matter of convention and does not have much physical significance, since the BES is known to be periodic over the reduced Brillouin zone.

Once the deformation is imposed, the boundary between the aforementioned mirror pair of reduced Brillouin zones will be deformed accordingly, since the nodal points will be shifted and vortex-antivortex pairs can be created or annihilated. This also indicates that the principal reduced zone does not have any intrinsic advantage over other reduced zones. As a result, in some particular cases one needs to choose a proper reduced zone $\widetilde{\mathbf{BZ}}_\mathbf{q}$, which is not even necessarily rectangular, to ensure that the sum of vorticities of all nodal points is identical to the Chern number of the underlying tight-binding model.

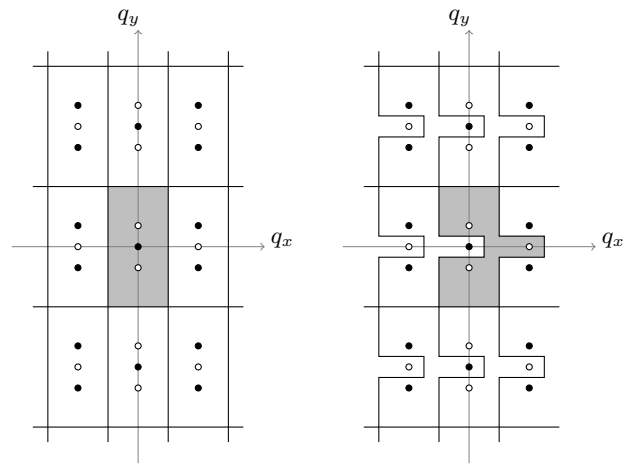
\begin{figure}
\begin{tikzpicture}

\begin{scope}[scale=0.4]

 \draw [lightgray,fill=lightgray] (-1,-2) rectangle (1,2);

 \draw [help lines,->] (-4.2,0) -- (4.2,0);
 \node at (4.8,0.2) {$q_x$};
 \draw [help lines,->] (0,-7.2) -- (0,7.2);
 \node at (-0.3,7.6) {$q_y$};

 \foreach \i in {-3,-1,1,3}{
  \draw [-] (\i,-6.5) -- (\i,6.5);
 }

 \foreach \j in {-6,-2,2,6}{
  \draw [-] (-3.5,\j) -- (3.5,\j);
 }

 \foreach \i in {0}{
  \foreach \j in {-4,0,4}{
   \begin{scope}[shift={(\i,\j)}]
    \draw [fill] (0,0) circle [radius=0.1];
    \draw [fill=white] (0,0.7) circle [radius=0.1];
    \draw [fill=white] (0,-0.7) circle [radius=0.1];
   \end{scope}
  }
 }

 \foreach \i in {-2,2}{
  \foreach \j in {-4,0,4}{
   \begin{scope}[shift={(\i,\j)}]
    \draw [fill=white] (0,0) circle [radius=0.1];
    \draw [fill] (0,0.7) circle [radius=0.1];
    \draw [fill] (0,-0.7) circle [radius=0.1];
   \end{scope}
  }
 }

\end{scope}

\begin{scope}[scale=0.4,shift={(11,0)}]

 \draw [lightgray,fill=lightgray] (1,2) -- (-1,2) -- (-1,0.35) -- (0.5,0.35) -- (0.5,-0.35) -- (-1,-0.35) -- (-1,-2) -- (1,-2) -- (1,-0.35) -- (2.5,-0.35) -- (2.5,0.35) -- (1,0.35) -- (1,2);

 \draw [help lines,->] (-4.2,0) -- (4.2,0);
 \node at (4.8,0.2) {$q_x$};
 \draw [help lines,->] (0,-7.2) -- (0,7.2);
 \node at (-0.3,7.6) {$q_y$};

 \foreach \i in {-2,0,2}{
  \foreach \j in {-4,0,4}{
   \begin{scope}[shift={(\i,\j)}]
   \draw [-] (1,2) -- (-1,2) -- (-1,0.35) -- (0.5,0.35) -- (0.5,-0.35) -- (-1,-0.35) -- (-1,-2);
   \end{scope}
  }
 }

 \draw [-] (-3.5,-6) -- (3,-6);
 \foreach \j in {-2,2,6}{
  \draw [-] (-3.5,\j) -- (-3,\j);
 }
 \foreach \i in {-3,-1,1}{
  \draw [-] (\i,6.5) -- (\i,6);
 }
 \foreach \i in {-3,-1,1}{
  \draw [-] (\i,-6.5) -- (\i,-6);
 }

 \foreach \i in {0}{
  \foreach \j in {-4,0,4}{
   \begin{scope}[shift={(\i,\j)}]
    \draw [fill] (0,0) circle [radius=0.1];
    \draw [fill=white] (0,0.7) circle [radius=0.1];
    \draw [fill=white] (0,-0.7) circle [radius=0.1];
   \end{scope}
  }
 }

 \foreach \i in {-2,2}{
  \foreach \j in {-4,0,4}{
   \begin{scope}[shift={(\i,\j)}]
    \draw [fill=white] (0,0) circle [radius=0.1];
    \draw [fill] (0,0.7) circle [radius=0.1];
    \draw [fill] (0,-0.7) circle [radius=0.1];
   \end{scope}
  }
 }

\end{scope}

\end{tikzpicture}
\caption{Vortex structure corresponding to the upper left or upper middle plot in Fig.\ \ref{fig-CI3c-1-vor}. Left: The principal reduced Brillouin zone $\mathbf{BZ}_\mathbf{q}$ and its periodic copies are shown. Within $\mathbf{BZ}_\mathbf{q}$, there are two antivortices (hollow dots) and one vortex (dot). The right and left neighboring zones have the opposite vortex configuration. Right: A properly chosen reduced Brillouin zone $\widetilde{\mathbf{BZ}}_\mathbf{q}$ and its periodic copies are shown. Within $\widetilde{\mathbf{BZ}}_\mathbf{q}$, there are three antivortices.}
\label{fig-non-principal BZ}
\end{figure}

We have encountered such examples. Specifically, in Sec.\ \ref{Section-vortex CI3c c=1}, we choose the ``non-principal'' reduced Brillouin zone $\widetilde{\mathbf{BZ}}_\mathbf{q}=(-\pi/2,\pi/2]\times(0,2\pi]$, instead of the principal one $\mathbf{BZ}_\mathbf{q}=(-\pi/2,\pi/2]\times(-\pi,\pi]$, in order to enclose the vortex at the boundary of $\mathbf{BZ}_\mathbf{q}$. A more nontrivial example occurs in the $\mathtt{C}=3$ phase of  $\bf CI_{3c}$ with its vortex structure corresponding to Fig.\ \ref{fig-CI3c-1} shown in Fig.\ \ref{fig-CI3c-1-vor}.  For all the plots in Fig.\ \ref{fig-CI3c-1-vor}, the sum of vorticities is equal to $-3$ except for the upper left and upper middle plots (the undeformed cases of $-4t<\mu<-2t$ and $\mu=-2t$), where the sum is $-1$. Even in the exceptional cases, we may still choose a non-principal reduced Brillouin zone $\widetilde{\mathbf{BZ}}_\mathbf{q}$ which encloses $\mathbf{q}_\mathrm{N}$, $\mathbf{q}_\mathrm{S}$, and $\mathbf{q}'_\mathrm{O}$ but excludes $\mathbf{q}_\mathrm{O}$.  See Fig.\ \ref{fig-non-principal BZ}. Then, within $\widetilde{\mathbf{BZ}}_\mathbf{q}$, the sum of vorticities is $-3$ as desired. Therefore, we conclude that the conjectured identity \eq{master rel} can always be satisfied by choosing a proper reduce zone $\widetilde{\mathbf{BZ}}_\mathbf{q}$.

The fact that $\widetilde{\mathbf{BZ}}_\mathbf{q}$ has to be chosen \emph{adaptively} does not mean the agreement between the both sides of \eq{master rel} is \textit{ad hoc}.
Note that the choice of $\widetilde{\mathrm{BZ}}_\mathbf{q}$ is not completely arbitrary but subject to the rather restrictive requirement that its periodic copies ``tile up'' the whole momentum space of $\mathbf{q}$.
In fact, such a relation, Eq. \eq{master rel}, can exist  only if the the sum of the positive (or, equivalently, negative) vorticities is larger than or equal to $|\mathtt{C}|$ in the combined region of the principle reduced zone $\mathrm{BZ}_\mathbf{q}$ and one of its neighboring copy.
Furthermore, the choice of $\widetilde{\mathrm{BZ}}_\mathbf{q}$ is \emph{continuous} in the sense that in response to a continuous change of tuning parameters and/or deformation we only need to \emph{continuously} update our choice of $\widetilde{\mathrm{BZ}}_\mathbf{q}$.\footnote{Under a continuous change of parameters and/or deformation, the vortices and antivortices inside the old $\widetilde{\mathrm{BZ}}_\mathbf{q}$ are continuously relocated and might go outside $\widetilde{\mathrm{BZ}}_\mathbf{q}$, and furthermore new vortex-antivortex pairs might be created somewhere. The new $\widetilde{\mathrm{BZ}}_\mathbf{q}$ can be continuously updated by enclosing \emph{both} the vortex and antivortex of the new vortex-antivortex pair, if any, as well as \emph{all} the vortices and antivortices that are already inside the old $\widetilde{\mathrm{BZ}}_\mathbf{q}$.} It is quite nontrivial that the identity \eq{master rel} \emph{always} and \emph{continuously} admits existence (albeit not uniqueness or arbitrariness) of a properly chosen reduced Brillouin zone.  Take the case corresponding to Fig.\ \ref{fig-non-principal BZ} as an example.  At first glance, it seems that one can choose a different $\widetilde{\mathrm{BZ}}_\mathbf{q}$ to yield $\pm1$ on the right-hand side of \eq{master rel}, thus violating the equality.  However, this is not \emph{always} possible under generic deformation. With proper fine-tuning of the deformation, we may make the three antivortices shown in the gray region on the right panel of Fig.\ \ref{fig-non-principal BZ} merge into a \emph{single} antivortex with $v=-3$  (this is very possible as in Fig.\ \ref{fig-CI3c-1} or Fig.\ \ref{fig-CI3c-1-vor} we already see the tendency that these three antivortices come closer to one another under deformation and change of $\mu$). In this particular situation, one can not manage to obtain $\pm1$ on the right-hand side of \eq{master rel}. The impossibility of always obtaining a number other than $\pm\mathtt{C}$ for the sum of vorticities is even more obvious if we consider the case corresponding to Fig.\ \ref{fig-CI3b-1-vor}, of which the leftmost panel gives a single vortex with $v=\pm3$.

\section{Conclusions} \label{conclusion}

In this paper we have extensively studied the BES of several Chern insulators. Our results clarify some issues. In particular, we find that robust band crossings in BES occur if and only if the system is in the topological phase of nonzero Chern number, and the A and B partitions have a dual symmetry such that the entanglement Hamiltonian possesses an emergent chiral symmetry.  Moreover, we have devised a way of imposing deformation for the entanglement Hamiltonian.  Under such deformation, only genuine topological structures remain stable.  In particular, we find that nodal lines carrying topological signature can not be lifted but reduce to nodal points under deformations. The existence of nontrivial nodal lines implies that the BES can be a more refined probe to the topological structure compared to the Chern number which is used to classify the underlying Chern insulator.

We then characterize the topological nature of the BES by the vorticities around the robust nodal points of the BES. Based on our study in this work, we propose the conjectured relation in \eq{master rel}, i.e.\ the sum of the vorticities of BES in a properly chosen reduced Brillouin zone equals the Chern number of the underlying Chern insulator.  This relation implies that entanglement Hamiltonian and thus its BES inherits certain topological order from the underlying Chern insulator.  It may also bear more refined topological signature as indicated by the existence of nontrivial nodal lines.

Inspired by the bulk/edge correspondence, there should exist some relation between the entanglement spectrum and the edge mode spectrum of topological ordered states. If we believe that the BES reflects some kind of entanglement spectrum for the extensive bulk modes, then the conjectured relation \eq{master rel} is a manifestation of this bulk/edge correspondence. Therefore, this relation deserves further study.  In particular, it would be desirable if one can come up with a better index than the vorticity used here to characterize the topology of the BES.  In addition to providing a better understanding of the relation, it may also lead to a physical mechanism which explains how and why additional pairs of vortices and antivortices are created.


\begin{figure}
\begin{tikzpicture}

\begin{scope}[scale=0.5]

 \draw [->] (-7,0) -- (7,0);
 \draw [blue!!60,->] (0,-4.5) -- (0,5);
 \draw [blue!!50] (4,-4.5) -- (4,4.5);
 \draw [blue!!50] (-4,-4.5) -- (-4,4.5);
 \node [right] at (7,0) {$\mu$};
 \node [above] at (0,5) {deformation};

 \draw [cyan] (0.8,4.5) to [out=270,in=120] (2,0) to [out=300,in=90] (3.2,-4.5);
 \draw [cyan] (-0.8,4.5) to [out=270,in=60] (-2,0) to [out=240,in=90] (-3.2,-4.5);

 \draw [fill=white] (0,0) circle [radius=0.1];
 \node [below] at (0,0) {$0$};

 \draw [fill=white] (4,0) circle [radius=0.1];
 \node [below] at (4,0) {$4$};

 \draw [fill=white] (-4,0) circle [radius=0.1];
 \node [below] at (-4,0) {$-4$};

 \draw [cyan,fill=cyan] (2,0) circle [radius=0.1];
 \draw [cyan,fill=cyan] (-2,0) circle [radius=0.1];

 \node [above] at (2,0) {$\mathtt{C}=-1$};
 \node [above] at (-2,0) {$\mathtt{C}=1$};
 \node [above] at (5.5,0) {$\mathtt{C}=0$};
 \node [above] at (-5.5,0) {$\mathtt{C}=0$};

 \node [right] at (1.2,2) {$v=\pm1$};
 \node [left] at (2.8,-2) {$v=\mp1$};
 \node [left] at (-1.2,2) {$v=\mp1$};
 \node [right] at (-2.8,-2) {$v=\pm1$};

\end{scope}

\end{tikzpicture}
\caption{The phase diagram of $\bf CI_{1c}$ in Fig.~\ref{fig-model I} enlarged with inclusion of deformation, which is schematically represented as the vertical dimension. The two curved lines represent the loci where a nontrivial nodal line appears; they separate each of the $\mathtt{C}=\pm1$ phases into two sub-phases with opposite vorticities. The two solid dots are the points of $\mu=\pm2$ with no deformation.}
\label{fig-model I (refined diagram)}
\end{figure}
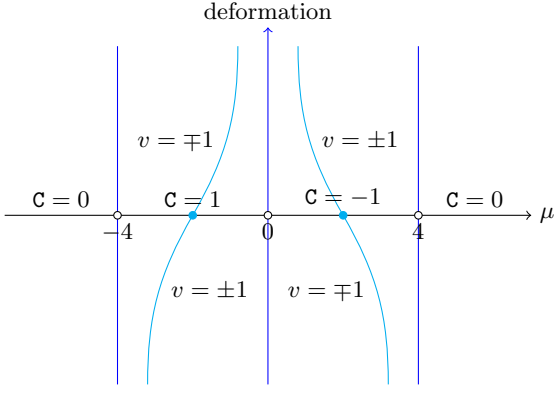

\begin{figure*}
\centering
  \begin{minipage}[b]{0.2\textwidth}
    \includegraphics[width=\textwidth]{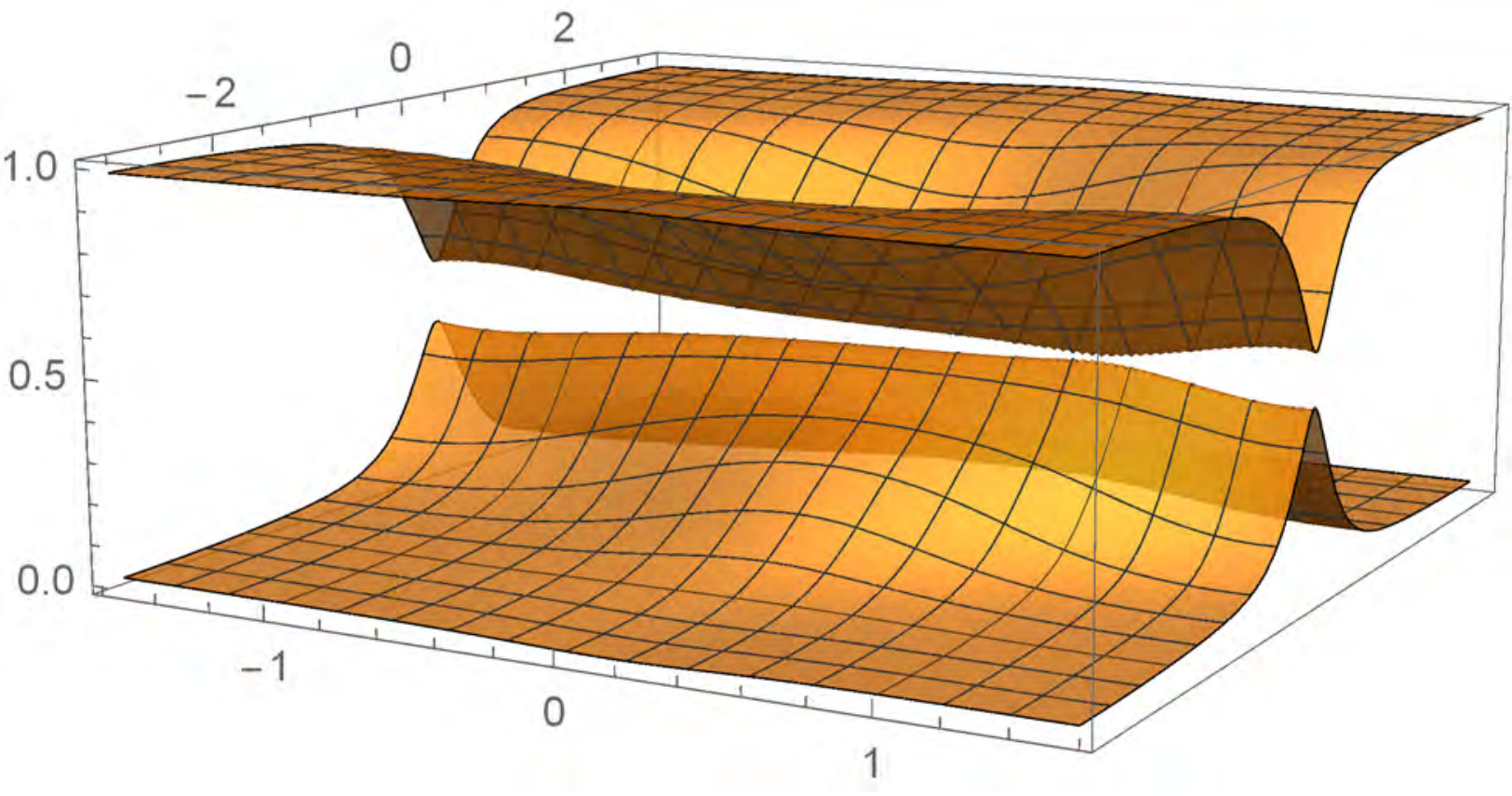}
  \end{minipage}
  \hspace{1cm} 
  \begin{minipage}[b]{0.2\textwidth}
    \includegraphics[width=\textwidth]{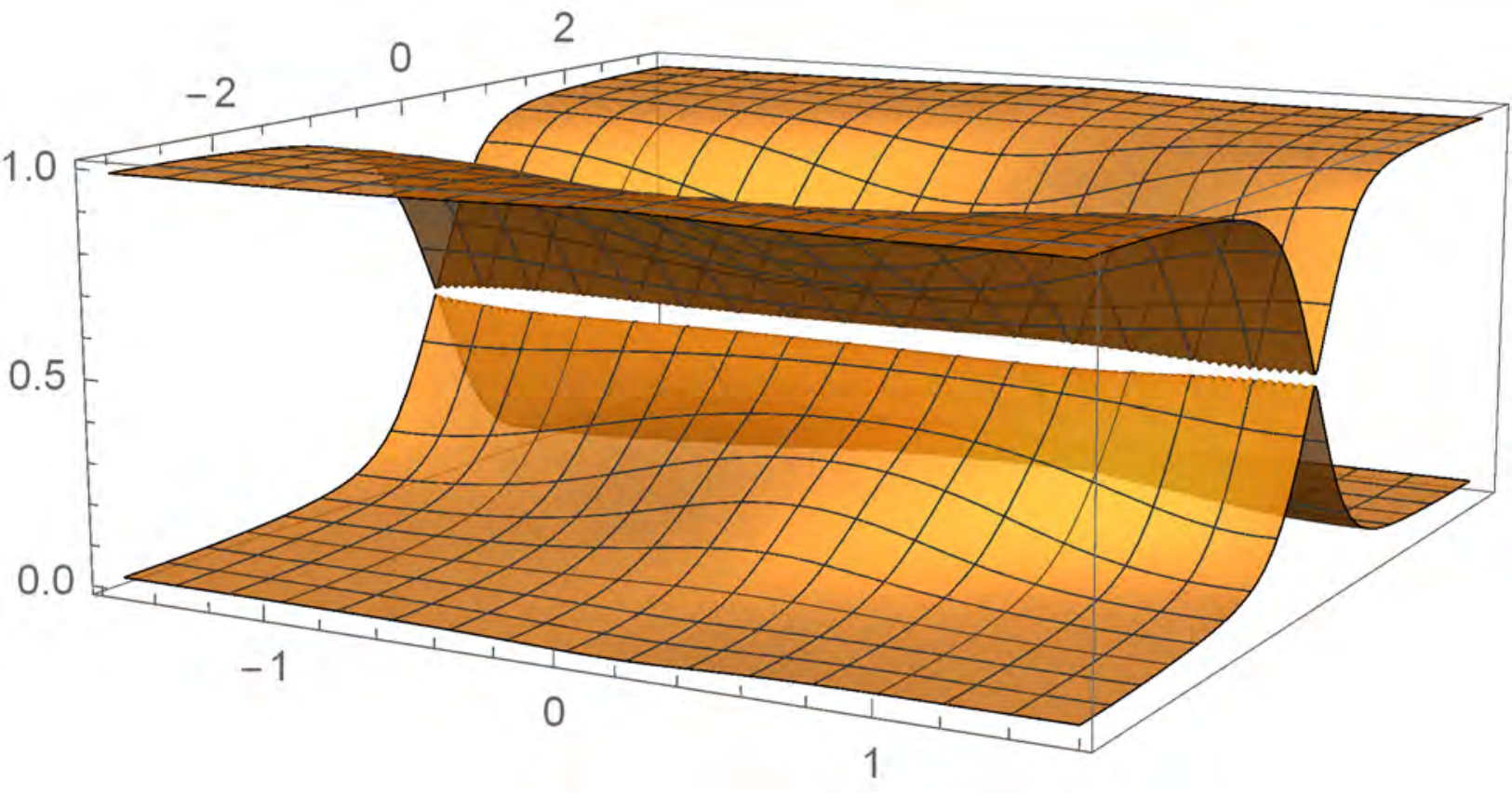}
  \end{minipage}
  \hspace{1cm} 
  \begin{minipage}[b]{0.2\textwidth}
    \includegraphics[width=\textwidth]{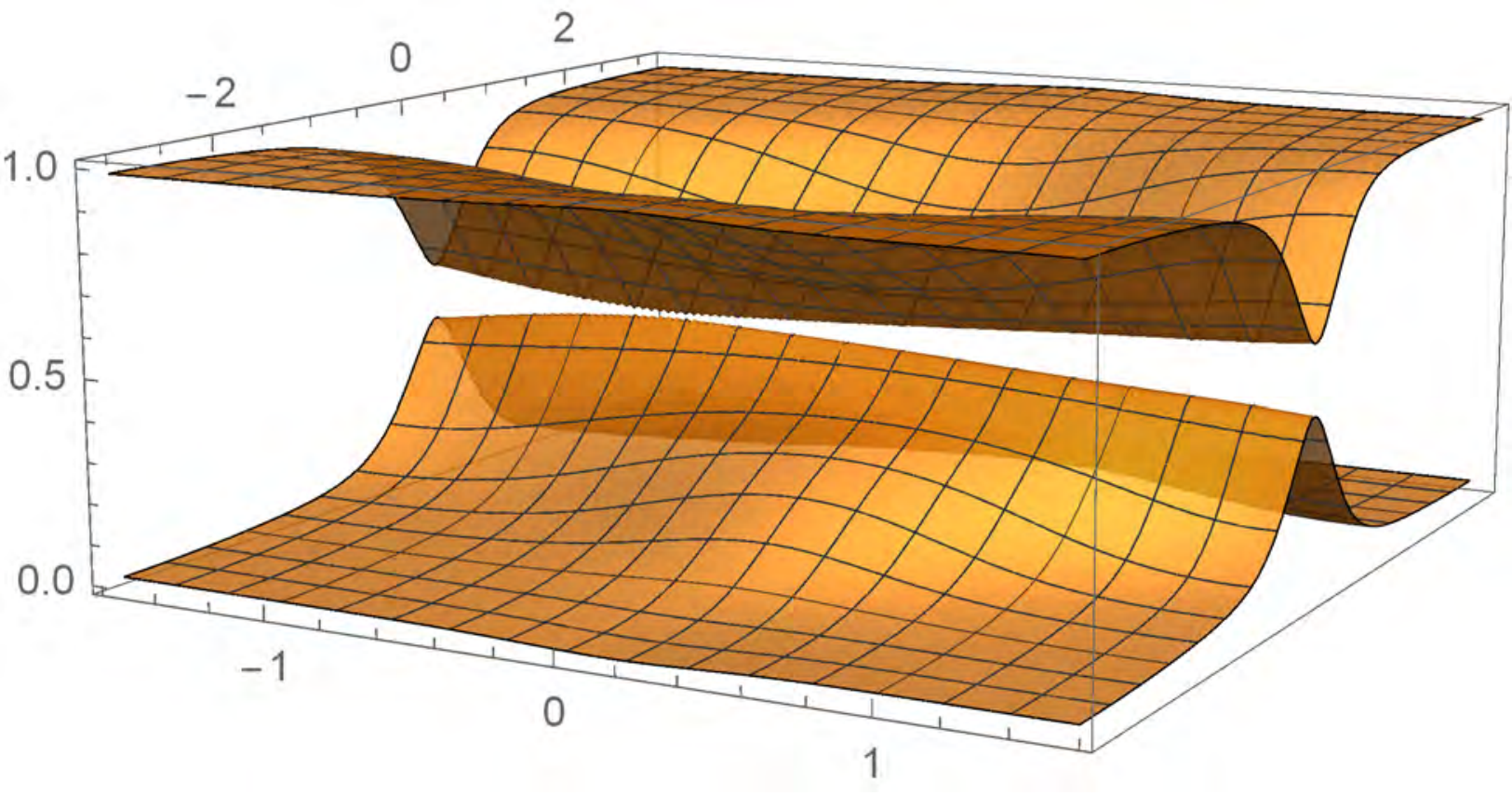}
  \end{minipage}
\caption{BES for the $\mathtt{C}=1$ phase of $\bf CI_{1c}$ imposed with deformation $a_x=b_x=a_y=b_y=0.5$. Left: $\mu=-1.7$. Middle: $\mu=-1.55$. Right: $\mu=-1.4$. With deformation fixed, by scanning the value of $\mu$, the nodal line is reproduced at the find-tuned point around $\mu\approx-1.55$.}\label{fig-CI1c-fine-tune}
\end{figure*}

\begin{acknowledgements}
We thank X.\ Wan, C.-K.\ Lu, and S.\ Ryu for helpful discussions. This work is supported in part by the Grant 104-2112-M-003-012  and 103-2811-M-003-024 of the Ministry of Science and Technology, Taiwan.
\end{acknowledgements}

\appendix

\section{Dispersion relation and the Vorticity of the nodal point in BES}\label{Appendix-dispersion}
In this Appendix we would like to analyze the eigenvectors and eigenvalues of the entanglement Hamiltonian.  In particular,  we will focus on the dispersion relation and the vorticity around the nodal points.

The eigenvectors and eigenvalues for the entanglement Hamiltonian are given in \eq{eigenvec c}-\eq{alpha3} for partition $\bf c$ (stripe) and in \eq{eigenvec b}-\eq{eigenval b} for partition $\bf b$ (checkerboard).
Expanding $\alpha_i({\bf q})$ around a nodal point, one can determine the dispersion relation to see which kind of nodal point it is: Dirac, quadratic, or cubic. We consider the phasor field in \eq{phasor for partition c} and \eq{phasor for partition b} case by case for demonstration.

The first one is for ${\bf CI_{1c}}$ with the Hamiltonian vector given in \eq{model-1}.  For $-4<\mu<0$ (except for $\mu= -2$) the nodal point is at ${\bf q}=(0,0)$. To determine the order and vorticity of the nodal point, we first expand the wavefunctions $\chi_-^s({\bf q})$ around ${\bf q}=(0,0)$ and $(\pi,0)$, and then combine them according to \eq{alpha3}.  We obtain
\be\label{alpha CI1c}
\alpha(\mathbf{q})\big|_{\mathbf{q}\approx(0,0)+\delta {\bf q}}
= -\frac{(\mu+2)\delta q_x +2 i\delta q_y}{|\mu(\mu+4)|}
+O(\delta q^2)
\ee
where $\delta {\bf q}=(\delta q_x, \delta q_y)$.
After plugging \eq{alpha CI1c} into \eq{eigenval c} to obtain the BES around the nodal point, we see that it is indeed a Dirac point as the dispersion relation is linear. Also, the phasor associated with the above $\alpha(\mathbf{q})$ around the nodal point indeed exhibits vorticity ${\it v}=1$ or ${\it v}=-1$, depending on the sign of $\mu+2$.  This sign flip around $\mu=-2$ prevents one from matching the vorticity to the Chern number $\mathtt{C}=1$. We can at best match their magnitudes, which is what we conjecture in \eq{master rel}.
Also note that in the special condition $\mu=-2$, the BES inferred from \eq{alpha CI1c} gives rise to a nodal line along $q_y=0$, as can be seen in the left plot of Fig.\ \ref{fig-CI1c-III}.

Similarly, for $0<\mu<4$ (except for $\mu= 2$) the nodal point is at ${\bf q}=(0,\pi)$, and we obtain
\be\label{alpha CI1c-II}
\alpha(\mathbf{q})\big|_{\mathbf{q}\approx(0,\pi)+\delta {\bf q}}
= \frac{(2-\mu)\delta q_x +2 i\delta q_y}{|\mu(4-\mu)|}
+O(\delta q^2).
\ee
From the above result, it is again straightforward to see it is a Dirac nodal point with vorticity ${\it v}=1$ or ${\it v}=-1$, depending on the sign of $\mu-2$. Thus, the result conforms to \eq{master rel}.
Again, note that in the special condition $\mu=2$, the BES inferred from \eq{alpha CI1c-II} gives rise to a nodal line along $q_y=0$.

Next, we consider the case ${\bf CI_{2b}}$ with the Hamiltonian vector given in \eq{model-2}.  For $|\mu|<4$ (except for $\m= 0$) there is a nodal point at ${\bf q}=(0,0)$. In the main text we claim it is a quadratic nodal point. This can by see by expanding $\alpha({\bf q})$ around ${\bf q}=(0,0)$ and $(\pi,\pi)$. The result is
\be\label{alpha CI2b}
\alpha(\mathbf{q})\big|_{\mathbf{q}\approx(0,0)+\delta {\bf q}}
 =-\frac{\mu(\delta q_x^2-\delta q_y^2)
+8 i\, \delta q_x\delta q_y}{2 |16-\mu^2|}  +O(\delta q^2).
\ee
This indeed yields a quadratic dispersion relation.  Moreover, the phasor associated with the above $\alpha(\mathbf{q})$ around the nodal point yields vorticity ${\it v}=2$ or ${\it v}=-2$ depending on the sign of $\mu$. Again it conforms to \eq{master rel} for $\mathtt{C}=2$.
Also note that in the special case $\mu=0$, the BES inferred from \eq{alpha CI2b} gives rise to two nodal line along $q_x=0$ and $q_y=0$

Finally, we consider the case $\bf CI_{3b}$. As mentioned before, in this case there is a cubic nodal point at ${\bf q}=(0,0)$ for both $\mathtt{C}=1$ and $\mathtt{C}=3$ phases. We would like to see that the vorticity can tell the difference between these two phases so that it yields the crucial supporting evidence for our conjectured relation \eq{master rel}. Similar to what we have done above, we evaluate  $\alpha({\bf q})$ according to \eq{alpha1 alpha2} by expanding around ${\bf q}=(0,0)$ and $(\pi,\pi)$. It then yields
\begin{eqnarray}\label{phasor CI3}
\alpha(\mathbf{q})\big|_{\mathbf{q}\approx(0,0)+\delta {\bf q}}
= \frac{1}{|16-\mu^2|}
\Big(2 \delta q_x(\delta q_x^2-\delta q_y^2)+\mu\delta q_x\delta q_y^2 &&
\nn\\
+ i\left(2\delta q_y(\delta q_y^2-\delta q_x^2)+\mu\delta q_y\delta q_x^2\right)
\Big)+O(\delta q^4). \qquad &&
\end{eqnarray}
This formula holds both for $-4<\mu<0$ ($\mathtt{C}=3$) and $0<\mu<4$ ($\mathtt{C}=1$). Thus, it is clear that the nodal point is cubic in both phases. However, regarding to the vorticity of the phasor field we find that it yields ${\it v}=-3$ for the $\mathtt{C}=3$ phase, and ${\it v}=1$ for the $\mathtt{C}=1$ phase. For the detailed vortex structure, see Fig.\ \ref{fig-CI3b-1-vor} and Fig.\ \ref{fig-CI3b-2-vor} in the main text.  Again, this is consistent with \eq{master rel}.

Finally, we can also obtain the vortex structures in a similar way for the other cases or around the nodal points other than the ones just discussed.

\section{Accidental vs nontrivial nodal lines}\label{Appendix-nodal lines}

There are two kinds (accidental vs nontrivial) of nodal lines in the BES we have encountered and they have very different physical significance.

On the one hand, there are nodal lines which are ``accidental'' in the sense that they are unstable and will be lifted completely under generic deformation. The nodal line appearing in the case of $\bf CI_{1d}$ with $\mu=-2$ as shown in Fig.\ \ref{fig-CI1d} is such an example. This kind of nodal lines has no topological significance at all.

On the other hand, there are nodal lines which are ``nontrivial'' in the sense that they will not be lifted completely but reduce to nodal points under generic deformation. We have encountered four such examples: (i) the nodal line along $q_y=0$ in the case of $\bf CI_{1c}$ with $\mu=-2$, which reduce to a Dirac point, as shown in Fig.\ \ref{fig-CI1c-III}, (and, similarly, the nodal line along $q_y=\pm\pi$ in the case of $\bf CI_{1c}$ with $\mu=2$, which also reduces to a Dirac point); (ii) the two nodal lines along $q_x=0$ and $q_y=0$ in the case of $\bf CI_{2b}$ with $\mu=0$, which reduce to a quadratic nodal point, as discussed in Sec.\ \ref{Section-CI2b and CI2c}; (iii) the two nodal lines along $q_y=\pm\cos^{-1}(-\mu/4)$ for the case of $\bf CI_{2c}$ with $\mu=0$, which reduce to two Dirac points, as discussed in Sec.\ \ref{Section-CI2b and CI2c}; (iv) the four nodal lines given by \eq{nodal lines in CI4b} in the $\mathtt{C}=-2$ phase of ${\bf CI_{4b}}$, which are reduced to two, four, or more Dirac points, as shown in Fig.\ \ref{fig-CI4b}.
This kind of nodal lines does exhibit some topological non-triviality as will be discussed below.

We will use the first example to illustrate the points.  Within the $\mathtt{C}=1$ phase of $\bf CI_{1c}$ (i.e., $-4<\mu<0$), the phasor field in the vicinity of $\mathbf{q}=(0,0)$ as given in \eq{alpha CI1c} yields $v=1$ for $-2<\mu<0$ while $v=-1$ for $-4<\mu<-2$.  Because the vortex structure is inherently topological, stepping over between the $v=1$ and $v=-1$ phases can be viewed as a kind of phase transition in the BES as it incurs a change of topology.

If we take (continuous) deformation into account and enlarge the phase diagram accordingly as depicted in Fig.\ \ref{fig-model I (refined diagram)}, we infer from the diagram that the aforementioned special condition (the left solid dot in Fig.\ \ref{fig-model I (refined diagram)}) can not be an isolated critical point in the phase diagram, otherwise the $-2<\mu<0$ and $-4<\mu<-2$ phases would be connected to each other via some trajectory with deformation, which is impossible as these two phases have different topological (vortical) structures. Instead, the critical point is extended to a critical line (the left curved line in Fig.\ \ref{fig-model I (refined diagram)}), which separates the $\mathtt{C}=1$ phase into two sub-phases with $v=1$ and $v=-1$ respectively. The critical line is the loci where a nontrivial nodal line appears.

In the phase diagram of Fig.\ \ref{fig-model I (refined diagram)}, any given horizontal line (i.e., with fixed deformation) always intersects the critical lines (curved lines in the diagram) at some point of $\mu$. This implies that, even in the presence of generic deformation, by fine-tuning the parameter $\mu$, we can always find a critical point of $\mu$ at which the BES exhibits a nodal line. In this sense, each of the $\mathtt{C}=1$ and  $\mathtt{C}=-1$ phases can be further differentiated into two sub-phases (with $v=1$ and $v=-1$) as far as the BES is concerned. Our numerical result indeed reassures that even with deformation imposed the nodal line can always be reproduced by fine-tuning $\mu$. See Fig.\ \ref{fig-CI1c-fine-tune}.

One can repeat the same reasoning to arrive at similar conclusions for the other three examples of nontrivial nodal lines. Particularly, for the case (iv), it should be noted that the four nodal lines given in \eq{nodal lines in CI4b} merge into two lines at $\mu=0$, which is to be viewed as the critical point of the phase transition between the $v=2$ and $v=-2$ sub-phases. Consequently, it is expected that, under generic deformation, only two nodal lines (instead of four) can always be reproduced by fine-tuning $\mu$.

It is rather surprising that a single phase in the physical Hamiltonian is further differentiated into two sub-phases,  which exhibit physical significance only in regard to the entanglement Hamiltonian, but not to the physical Hamiltonian. This suggests that the BES not only inherits the topology of the Chern insulator but, in some particular A-B symmetric partitions, also manifests a more refined topological structure which can not be seen from the spectrum of the physical Hamiltonian.


\begin{thebibliography}{plain}

\bibitem{spt} X. Chen, Z.-C. Gu, Z.-X. Liu and X.-G. Wen, Science \textbf{338}, 1604 (2012); arXiv:1106.4772v6.


\bibitem{Li} H. Li and F.~D.~M. Haldane, Phys. Rev. Lett. {\bf 101}, 010504 (2008).

\bibitem{Ryu1} S. Ryu and  Y. Hatsugai, Phys. Rev. B {\bf 73}, 245115 (2006).

\bibitem{Pollman} F. Pollmann, E. Berg, A. M. Turner, M. Oshikawa, Phys. Rev. B {\bf 81}, 064439 (2010).

\bibitem{Spain1} I.~D. Rodriguez and G. Sierra, Phys. Rev. B {\bf 80}, 153303 (2009).

\bibitem{Turner} A.~M. Turner, Y. Zhang, and A. Vishwanath, Phys. Rev. B {\bf 82}, 241102 (2010).

\bibitem{Fidkowski} L. Fidkowski, Phys. Rev. Lett. {\bf 104}, 130502 (2010).

\bibitem{Arovas} Z. Huang and D.~P. Arovas, Phys. Rev. B {\bf 86}, 245109 (2012).

\bibitem{Qi-Ludwig}  X.-L. Qi, H. Katsura, A.~W.~W. Ludwig, Phys. Rev. Lett. {\bf 108}, 196402 (2012).

\bibitem{PYChang} P.-Y.~Chang, C.~Mudry, S.~Ryu, J. Stat. Mech. (2014) P09014.


\bibitem{MIT1} T.~H. Hsieh and L.~Fu, Phys. Rev. Lett. {\bf 113}, 106801 (2014).

\bibitem{BES-1} M.~Legner and T.~Neupert, Phys. Rev. B {\bf 88}, 115114 (2013).

\bibitem{BES-2}  C.~Fang, M.~ J.~Gilbert, B.~A.~Bernevig, Phys. Rev. B {\bf 87}, 035119 (2013).

\bibitem{Wan2} W.-J. Rao, X. Wan, and G.-M. Zhang, Phys. Rev. B {\bf 90}, 075151 (2014).

\bibitem{MIT2} T.~H. Hsieh, L. Fu, and X.~L. Qi, Phys. Rev. B {\bf 90}, 085137 (2014).

\bibitem{Santos-1}  R. A. Santos,  J. Phys. A: Math. Theor. {\bf 48}, 155203 (2015).

\bibitem{Santos-2} R. A. Santos, C.-M. Jian, and R. Lundgren,  Phys. Rev. B {\bf 93}, 245101 (2016).

\bibitem{BES-QHE} C.-K. Lu, D.-W. Chiou, F.-L. Lin, Phys. Rev. B {\bf 92}, 075130  (2015).

\bibitem{Wan1} Q. Zhu, X. Wan, and G.-M. Zhang, Phys. Rev. B {\bf 90}, 235134 (2014).

\bibitem{MIT3} S. Vijay and L. Fu, arXiv:1412.4733

\bibitem{Wan3} M. Lu, R. Narayanan, X. Wan, G.-M. Zhang, arXiv:1502.02095.

\bibitem{Fukui} T. Fukui and Y. Hatsugai, J. Phys. Soc. Jpn. {\bf 83}, 113705 (2014).

\bibitem{Fukui1} T. Fukui and Y. Hatsugai, J. Phys. Soc. Jpn. {\bf 84}, 043703 (2015).



\bibitem{Haldane} F.~D.~M. Haldane, Phys. Rev. Lett. {\bf 61} 2015 (1988).


\bibitem{Kitaev-per} A.~Kitaev,  Proceedings of the L.D.Landau Memorial Conference "Advances in Theoretical Physics" (2008).

\bibitem{AIP-1} A.~P. Schnyder, S. Ryu, A. Furusaki, and A.~W.~W. Ludwig, AIP Conf. Proc. 1134, 10 (2009).

\bibitem{SPT-cls} S. Ryu, A.~P. Schnyder, A. Furusaki, and A.~W.~W. Ludwig, New J. Phys. 12, 065010 (2010).

\bibitem{Hasan2010} M.~Z. Hasan and C.~L. Kane, Rev. Mod. Phys. \textbf{82}, 3045 (2011).

\bibitem{Qi2011} X.-L. Qi and S.-C. Zhang, Rev. Mod. Phys. \textbf{83}, 1057 (2011).

\bibitem{Furusaki} A.~P. Schnyder, S. Ryu, A. Furusaki, and A.~W.~W. Ludwig, Phys. Rev. B {\bf 78}, 195125 (2008).

\bibitem{TKNN} D. J. Thouless, M. Kohmoto, M. P. Nightingale, and M. den Nijs
Phys. Rev. Lett. {\bf 49}, 405 (1982).

\bibitem{Peschel} I. Peschel, J. Phys. A: Math. Gen. {\bf 36}, L205 (2003).

\bibitem{Cheong} S.-A. Cheong and C.~L. Henley, Phys. Rev. B {\bf 69}, 075111 (2004).

\bibitem{Chung1} M.-C. Chung and I. Peschel, Phys. Rev. B {\bf 64}, 064412 (2001).

\bibitem{TI-model}
\url{http://www-personal.umich.edu/~sunkai/teaching/Fall_2013/chapter5.pdf}




\bibitem{Simon-4} D. Sticlet, F. Piechon, J.-N. Fuchs, P. Kalugin, P. Simon, Phys. Rev. B {\bf 85}, 165456 (2012).

\end{thebibliography}
\end{document}